\documentstyle[12pt]{article}
\begin{document}
\begin{titlepage}

\hfill{September 1996}

\hfill{Revised November 1996}


\hfill{UM-P-96/81}

\hfill{RCHEP-96/09}

\vskip 0.9 cm

\centerline{{\large \bf 
Studies of neutrino asymmetries generated by }}
\centerline{{\large \bf ordinary - sterile neutrino oscillations 
in the early Universe}}
\centerline{{\large \bf and implications for big bang nucleosynthesis bounds}}
\vskip 1.3 cm
\centerline{R. Foot and R. R. Volkas}

\vskip 1.0 cm
\noindent
\centerline{{\it Research Centre for High Energy Physics,}}
\centerline{{\it School of Physics, University of Melbourne,}}
\centerline{{\it Parkville, 3052 Australia. }}

\vskip 2.0cm

\centerline{Abstract}
\vskip 0.7cm
\noindent
Ordinary-sterile neutrino oscillations
can generate a significant lepton number asymmetry in the early Universe.
We study this phenomenon in detail. 
We show that the dynamics of ordinary-sterile neutrino oscillations
in the early Universe can be approximately described by a single
integro-differential equation which we derive from both
the density matrix and Hamiltonian formalisms.  This equation reduces
to a relatively simple ordinary first order differential equation if 
the system is sufficiently smooth (static limit). We study the 
conditions for which the static limit is an acceptable approximation.
We also study the effect of the thermal distribution of neutrino 
momenta on the generation of lepton number.  We apply these 
results to show that it is possible to evade (by many orders
of magnitude) the Big Bang Nucleosynthesis (BBN) bounds on the 
mixing parameters, $\delta m^2$ and $\sin^2 2\theta_0$, 
describing ordinary-sterile neutrino oscillations.
We show that the large angle or maximal vacuum oscillation
solution to the solar neutrino problem does not significantly
modify BBN for most of the parameter space of interest, provided 
that the tau and/or mu neutrinos have masses greater than about 1 eV.
We also show that the large angle or maximal ordinary-sterile neutrino 
oscillation solution to the atmospheric neutrino anomaly does
not significantly modify BBN for a range of parameters.
\end{titlepage}

\vskip 0.8cm
\noindent
{\bf I. Introduction}
\vskip 0.7cm
\noindent
There are three main experimental indications that neutrinos have
mass and oscillate. They are the solar neutrino problem\cite{sn},
the atmospheric neutrino anomaly\cite{ana} and the Los Alamos LSND 
experiment\cite{lsnd}.  It is also possible that dark matter may 
be connected to neutrino masses\cite{kt}.  The three experimental 
anomalies cannot all be explained with the three known neutrinos so 
it is possible that sterile neutrinos exist.

A potential problem with any model which contains sterile neutrinos
is that these extra states can contribute to
the energy density of the early Universe and spoil the 
reasonably successful Big Bang Nucleosynthesis (BBN) predictions.
For maximally mixed $\nu_{e}$ and $\nu_{e}'$ neutrinos and
$\nu_{\mu}$ and $\nu'_{\mu}$ (or $\nu_{\tau}$ and $\nu'_{\tau}$) 
neutrinos (where the primes denote sterile
species), the following rather stringent BBN bounds have been 
obtained\cite{B,B2,B3,B4}
{\it assuming that the lepton number asymmetry of the 
early Universe could be neglected}:
\begin{equation}
|\delta m^2_{ee'}| \stackrel{<}{\sim} 10^{-8} \ eV^2,\
|\delta m^2_{\mu \mu'}|, \ |\delta m^2_{\tau \tau'}|  
\stackrel{<}{\sim} 10^{-6}\ eV^2.
\label{negl}
\end{equation}
Observe that if valid these bounds would rule out the large angle 
$\nu_{\mu} - \nu_{\mu}'$ oscillation solution to the atmospheric 
neutrino anomaly and would restrict much of the parameter space for the 
maximal oscillation solution of the solar neutrino problem\cite{fv,pc}. 
However, these bounds do not hold if there is an appreciable 
lepton asymmetry in the early Universe for temperatures between 
$1-30$ MeV\cite{fv1}.  Remarkably, it turns out that ordinary-sterile 
neutrino oscillations can by themselves create an appreciable lepton 
number asymmetry\cite{ftv}.

The bound on the effective number of neutrinos $N_{\nu}^{eff}$ present
during nucleosynthesis is the subject of some discussion recently.
In Ref.\cite{hata}, it is argued that
the current information suggests $N_{\nu}^{eff} \simeq 2.1\pm 0.3$,
while other authors dispute this conclusion.  For example, in 
Ref.\cite{cst}, Ref.\cite{ks} and Ref.\cite{oliv},
the upper limits $N_{\nu}^{eff} < 3.9, \ 4.5, \ 4.0$ are 
respectively derived.  Thus, it may be possible that $N_{\nu}^{eff} = 4 $ 
is allowed.  In this case note that many of the BBN bounds derived
in Refs.\cite{B,B2,B3,B4}, quoted in Eq.(\ref{negl}), need not apply.  
However, for the present paper we will assume that the bound on the 
effective number of neutrinos is less than 4. 
This is useful even if it turns out that $N_{\nu}^{eff} > 4$
is allowed. For example, the large angle (or maximal) ordinary
sterile neutrino solutions to the atmospheric
and solar neutrino problems may require $N_{\nu}^{eff} \sim 5$
if they are to be solved simultaneously.
Also note that in the special case of mirror neutrinos\cite{flv},
the mirror interactions can potentially bring all three 
mirror neutrinos  (as well as the mirror photon and electron 
positron pair) into equilibrium (equivalent to about 6 additional 
neutrino species) if any one of the mirror neutrinos is brought into
equilibrium above the neutrino kinetic decoupling temperature.   

The purpose of this paper is two-fold. First, we will study the 
phenomenon of lepton number creation due to ordinary-sterile
neutrino oscillations in more detail than in the previous
studies\cite{ftv, shi}. For example, we will study the effect of 
the thermal distribution of neutrino momenta. Using these results we 
will then study the issue of whether or not the generation of lepton 
number due to ordinary-sterile neutrino oscillations can reconcile the 
large angle ordinary-sterile neutrino oscillation solutions to the solar 
neutrino problem and atmospheric neutrino anomaly with BBN.

The outline of this paper is as follows.
In section II, we discuss lepton number generation in the early
Universe by ordinary-sterile neutrino oscillations and derive 
a simple equation describing the evolution of lepton number.
We expand the analysis of Ref.\cite{ftv} and discuss in detail
the approximations behind this analysis.
In section III, we will use the density matrix formalism to derive 
a more exact equation describing
the rate of change of lepton number which is applicable even
when the system is changing rapidly (e.g. at the resonance).
In the appendix, we show how the same equation can be
derived from the Hamiltonian formalism.  Using this equation we derive
the region of parameter space where the much simpler equation
derived in section II is approximately valid. In section IV the thermal 
distribution of the neutrino momenta is considered.  
In section V we study the effect of non-negligible 
sterile neutrino number densities. We then apply these results 
to obtain the region of parameter space where large neutrino 
asymmetries are generated. We also determine 
the region of parameter space for which ordinary-sterile
neutrino oscillations (with $\delta m^2 < 0$ and for $|\delta m^2|
\stackrel{>}{\sim} 10^{-4} \ eV^2$) are consistent with
BBN.  Our work improves on previous studies\cite{B,B2,B3,B4}, because 
these studies were obtained without taking into account either the
neutrino momentum distribution or the result that 
ordinary-sterile neutrino oscillations create lepton number.
In section VI we first briefly review the large angle ordinary-sterile 
neutrino oscillation solution to the solar neutrino problem.  We 
then show that the generation of lepton number due to ordinary-sterile 
neutrino oscillations can significantly relax the BBN bounds 
for this solution to the solar neutrino problem.
We also show that the large angle or maximal $\nu_{\mu} - \nu_s$
oscillation solution to the atmospheric neutrino anomaly
is consistent with BBN for a range of parameters.
In section VII we conclude. 

\vskip 0.6cm
\noindent
{\bf II. Lepton number creation from neutrino oscillations - static
approximation}
\vskip 0.6cm
\noindent
Together with M. Thomson, we showed in
Ref.\cite{ftv} that ordinary-sterile neutrino
oscillations can create a large lepton asymmetry in the early
Universe\cite{ekmx}. A simple differential equation describing
the evolution of lepton number was derived which
seemed to work very well. We also checked our results with
the more exact density matrix formalism \cite{bm}.
Further numerical work, and analytical work based on the
density matrix formalism, has subsequently been done in Ref.\cite{shi}
which confirms our results.  

For ordinary-sterile neutrino two state mixing, the
weak-eigenstates ($\nu_{\alpha}, \nu_{s}$) will be 
linear combinations of two mass eigenstates ($\nu_a, \nu_b$):
\begin{equation}
\nu_{\alpha} = \cos\theta_0 \nu_a + \sin\theta_0 \nu_b,
\ \nu_{s} = -\sin\theta_0 \nu_a + \cos\theta_0 \nu_b.
\end{equation}
Note we will always define $\theta_0$ in such a way so
that $\cos 2\theta_0 \ge 0$ (this can always be done).
We also take the convention that 
$\delta m^2_{\alpha s}\equiv m_b^2 - m_a^2$. Hence 
with this convention $\delta m^2_{\alpha s}$ is positive 
(negative) provided that $m_b > m_a$ ($m_b < m_a$). 

In this section we will for simplicity neglect the effects of 
the thermal distribution of momentum, and assume that all of the
neutrino momenta are the same and equal to the average
momentum (i.e. $p = \langle p \rangle \simeq 3.15 T$).
In section IV we will consider the realistic case where the
neutrino spread is given by the Fermi-Dirac distribution.
Following Ref.\cite{ftv}, we can derive a simple equation for the 
rate of change of lepton number due to collisions and oscillations. 
Note that it is possible to identify two distinct contributions
to the rate of change of lepton number.
First, there are the oscillations between collisions which
affect the lepton number of the Universe because neutrinos and
anti-neutrinos oscillate with different matter oscillation
lengths and matter mixing angles in the CP asymmetric background.
Second, there are the collisions themselves which deplete
$\nu_e$ and $\bar \nu_e$ at different rates. This is because the
rates depend on the oscillation probability. The oscillation
probability for ordinary-sterile neutrino oscillations is
different to the oscillation probability for ordinary-sterile
anti-neutrino oscillations (which is again due to the CP
asymmetric background). Generally, the rate of change of 
lepton number is dominated by collisions in the region where
the collision rate is larger than the expansion rate\cite{ftv}.
[A possible exception to this is in the resonance region where
the matter mixing angle changes rapidly].
For the case of $\nu_{\alpha}-\nu_{s}$ oscillations
(where $\alpha = e, \mu, \tau$), 
the rate of change of $L_{\nu_{\alpha}}$ due to collisions is
governed by the rate equation,
\begin{eqnarray}
&\frac{dL_{\nu_{\alpha}}}{dt} = 
{-n_{\nu_{\alpha}} \over n_{\gamma}}
\Gamma(\nu_{\alpha} \to \nu_{s}) +
{n_{\bar \nu_{\alpha}} \over n_{\gamma}}
\Gamma(\bar \nu_{\alpha} \to \bar \nu_{s}) 
\nonumber \\
& + {n_{\nu_{s}} \over n_{\gamma}}
\Gamma(\nu_{s} \to \nu_{\alpha}) +
{-n_{\bar \nu_{s}} \over n_{\gamma}}
\Gamma(\bar \nu_{s} \to \bar \nu_{\alpha}), 
\label{io2}
\end{eqnarray}
where the $n$'s are number densities and $L_{\nu_{\alpha}} \equiv
(n_{\nu_{\alpha}} - n_{\bar \nu_{\alpha}})/n_{\gamma}$.
Using $\Gamma (\nu_{\alpha} \to \nu_{s}) = 
\Gamma (\nu_{s} \to \nu_{\alpha})$ and
$\Gamma (\bar \nu_{\alpha} \to \bar \nu_{s}) = 
\Gamma (\bar \nu_{s} \to \bar \nu_{\alpha})$ (will
justify this in a moment),  Eq.(\ref{io2}) simplifies to
\begin{equation}
\frac{dL_{\nu_{\alpha}}}{dt} = 
-\left[ {n_{\nu_{\alpha}} - n_{\nu_{s}} \over n_{\gamma}}
\right] \Gamma(\nu_{\alpha} \to \nu_{s}) +
\left[ {n_{\bar \nu_{\alpha}} - n_{\bar \nu_{s}} \over n_{\gamma}}
\right] \Gamma(\bar \nu_{\alpha} \to \bar \nu_{s}). 
\end{equation}
This equation can be re-written in the form
\begin{equation}
\frac{dL_{\nu_{\alpha}}}{dt} \simeq 
\left({\cal N}^+_{\nu_{\alpha}} - {\cal N}^+_{\nu_s}\right)
\left[- \Gamma(\nu_{\alpha}
\to \nu_{s}) + \Gamma(\bar \nu_{\alpha} \to 
\bar \nu_{s})\right] - 
\left( {\cal N}^-_{\nu_{\alpha}} - {\cal N}^-_{\nu_{s}}\right)
\left[ \Gamma(\nu_{\alpha}
\to \nu_{s}) + \Gamma(\bar \nu_{\alpha} \to 
\bar \nu_{s})\right],
\label{332}
\end{equation}
where
\begin{equation}
{\cal N}^{\pm}_{\nu_{\alpha}} \equiv 
{n_{\nu_{\alpha}} \pm n_{\bar \nu_{\alpha}} \over
2n_{\gamma}},\ 
{\cal N}^{\pm}_{\nu_{s}} \equiv 
{n_{\nu_{s}} \pm n_{\bar \nu_{s}} \over
2n_{\gamma}}.
\end{equation} 
Observe that ordinary-sterile neutrino oscillations do not
change the total particle number, from which it follows that
\begin{equation}
{\cal N}_{\nu_{\alpha}}^- + {\cal N}_{\nu_{s}}^- = 0.
\label{333}
\end{equation}
Using Eqs.(\ref{332}-\ref{333}),
the rate of change of $L_{\nu_{\alpha}}$ due to collisions is given by
\begin{eqnarray}
&\frac{dL_{\nu_{\alpha}}}{dt} \simeq 
\left({3 \over 8} - {\cal N}^+_{\nu_s}\right) \left[- \Gamma(\nu_{\alpha}
\to \nu_{s}) + \Gamma(\bar \nu_{\alpha} \to 
\bar \nu_{s})\right] \nonumber \\
& - L_{\nu_{\alpha}}
\left[ \Gamma(\nu_{\alpha}
\to \nu_{s}) + \Gamma(\bar \nu_{\alpha} \to 
\bar \nu_{s})\right] + {\cal O}(L^2_{\nu_{\alpha}}),
\label{io}
\end{eqnarray}
where we have used $n_{\nu_{\alpha}} + n_{\bar \nu_{\alpha}} 
\simeq 3n_{\gamma}/4 + {\cal O}(L_{\nu_{\alpha}}^2)$. 
We will assume for the present that negligible sterile neutrinos
are produced, i.e. $n_{\nu_{s}}, n_{\bar \nu_{s}} 
\ll n_{\nu_{\alpha}}, n_{\bar \nu_{\alpha}}$,
and hence ${\cal N}^+_{\nu_s} \ll 1$.

In order to work out the reaction rates, we can invoke a simple
physical picture\cite{pict,kain,thomo}.
The oscillations of the neutrino between collisions produce a superposition 
of states. The collisions are assumed to collapse the wavefunction into 
either a pure weak eigenstate neutrino or a pure sterile eigenstate 
neutrino.  In other words, we assume that the collisions are measurements 
(in the quantum mechanical sense) of whether the state is a sterile or
weak eigenstate. The rate of the measurements is expected to be
the collision frequency $\Gamma_{\nu_{\alpha}}$.

Actually it happens that the above picture is not completely correct.
It turns out that it does lead to an accurate description only 
if the rate of measurement is taken to be {\it half} of the collision
frequency that a pure $\nu_{\alpha}$ state would experience\cite{thomo}. 
This applies to both sterile neutrinos and ordinary neutrinos.
Thus using this result the reaction rate $\Gamma(\nu_{\alpha} 
\to \nu_{s})$ is given by {\it half} the interaction rate of the
neutrino due to collisions with the background particles
multiplied by the probability that the neutrino
collapses to the sterile eigenstate \cite{B}, that is 
\begin{equation}
\Gamma(\nu_{\alpha} \to \nu_{s}) =
{\Gamma_{\nu_{\alpha}} \over 2}\langle P_{\nu_{\alpha} \to \nu_{s}} 
\rangle. 
\label{Gamma}
\end{equation}
The thermally averaged collision frequencies $\Gamma_{\nu_{\alpha}}$ are 
\begin{equation}
\Gamma_{\nu_{\alpha}} \simeq y_{\alpha}G_F^2 T^5,
\label{Gammas}
\end{equation}
where $y_e \sim 4.0, y_{\mu, \tau} \simeq 2.9$\cite{B2},
$G_F$ is the Fermi constant ($G_F \simeq 1.17\times 10^{-11}$ 
MeV$^{-2}$) and $T$ is the temperature of the Universe 
[equations analogous to Eqs.(\ref{Gamma},\ref{Gammas}) hold 
for antineutrinos].  The quantity $P_{\nu_{\alpha} \to \nu_{s}}$
is the probability that the neutrino $\nu_{\alpha}$ collapses to
the sterile state $\nu_{s}$ after a measurement is made.
The brackets $\langle ... \rangle$ denote the average over all 
measurement times.  Note that $P_{\nu_{\alpha} \to \nu_{s}} 
= P_{\nu_{s} \to \nu_{\alpha}}$, so it follows that
$\Gamma (\nu_{\alpha} \to \nu_{s}) = \Gamma (\nu_{s} \to \nu_{\alpha})$
(given that the rate of measurement is the same for ordinary and
sterile neutrinos\cite{thomo}) and similarly for the anti-neutrino rates.
In the adiabatic limit,
\begin{equation}
\langle P_{\nu_{\alpha} \to \nu_{s}}\rangle
\simeq \sin^2 2\theta_m \langle \sin^2 
\frac{\tau}{2L_m} \rangle,
\label{P}
\end{equation}
where $\tau$ is the distance (or time) between collisions. 
The quantities $\theta_m$ and $L_m$ are the matter mixing angle and 
matter oscillation length respectively.  They are related to the 
vacuum parameters $\theta_0$ and $L_0$ by \cite{nr, msw}
\begin{equation}
\sin^2 2\theta_m = {\sin^2 2\theta_0
\over 1 - 2z\cos2\theta_0 + z^2},
\label{thetam}
\end{equation}
and
\begin{equation}
L_m = {{L_0} \over {\sqrt{1 - 2z\cos2\theta_0 + z^2}}},
\label{sat9}
\end{equation}
where $1/L_0 \equiv \Delta_0^p \equiv \delta m^2/2p$.  In this equation, 
$z = 2p V_{\alpha}/\delta m^2$ where $V_{\alpha}$ 
is the effective potential due to the interactions of the neutrinos 
with matter and $p$ is the neutrino momentum.  The effective potential 
is given by
\begin{equation}
V_{\alpha}  = (- a^p + b^p)\Delta_0^p,
\label{V}
\end{equation}
where the dimensionless variables $a^p$ and $b^p$ are given by
\begin{equation}
a^p \equiv {-\sqrt{2}G_F n_{\gamma}L^{(\alpha)} \over \Delta_0^p},\  
b^p \equiv  {-\sqrt{2}G_F n_{\gamma} A_{\alpha}T^2\over \Delta_0^p M_W^2}
{p \over \langle p \rangle},
\label{ab}
\end{equation}
where $M_W$ is the W-boson mass and $A_e \simeq 55.0, \ A_{\mu,\tau}
\simeq 15.3$ (note that the ``p'' superscript serves as a reminder 
that these quantities are neutrino momentum dependent).
The function $L^{(\alpha)}$ is given by
\begin{equation}
L^{(\alpha)} = L_{\nu_{\alpha}} + L_{\nu_e} + L_{\nu_{\mu}}
+ L_{\nu_{\tau}} + \eta,
\label{Lsuper}
\end{equation}
where $\eta$ is a small asymmetry term which arises from the asymmetries 
of baryons and electrons.  It is given by\cite{nr}
\begin{equation}
\eta = ({1 \over 2} + 2\sin^2\theta_w)L_e + ({1 \over 2} -
2\sin^2 \theta_w)L_P - {1 \over 2}L_N \simeq {1 \over 2}L_N,
\label{toomanyeq}
\end{equation}
where $\sin^2 \theta_w$ is the weak mixing angle and 
we have used $L_e = L_P \simeq L_N$.  Thus $\eta$ is expected 
to be of order $10^{-10}$. 
Note that the matter mixing angle $\overline{\theta}_m$ 
and oscillation length $\overline{L}_m$ for antineutrino oscillations 
are obtained from Eqs.(\ref{thetam}-\ref{ab}) by performing
the transformation $L^{(\alpha)} \to - L^{(\alpha)}$\cite{shif}.

We denote the thermal average of the variables $a^p, b^p$
by $a \equiv \langle a^p \rangle$, $b \equiv \langle b^p
\rangle$.  From Eq.(\ref{ab}), they are given approximately by
\begin{eqnarray}
&a \simeq {-6.3\sqrt{2} T G_F n_{\gamma} L^{(\alpha)} \over \delta m^2}
\simeq -25L^{(\alpha)} \left({eV^2 \over \delta m^2}\right)
\left({T \over MeV}\right)^4, \nonumber \\
& b \simeq {-6.3\sqrt{2} T G_F n_{\gamma} A_{e} T^2 \over 
\delta m^2 M_W^2} \simeq -\left( {T \over 13 \ MeV} \right)^6 
\left( {eV^2 \over \delta m^2}\right), \ {\rm for}\
\nu_{e} - \nu_s \ {\rm oscillations},\nonumber \\
& b \simeq {-6.3 \sqrt{2} T G_F n_{\gamma} A_{\mu, \tau} T^2 \over 
\delta m^2 M_W^2} \simeq -\left( {T \over 16 \ MeV} \right)^6 
\left( {eV^2 \over \delta m^2}\right), \ {\rm for}\
\nu_{\mu,\tau} - \nu_s \ {\rm oscillations},
\label{ab3}
\end{eqnarray}
where we have used $n_{\gamma} = 2\zeta (3)T^3/\pi^2 \simeq T^3/4.1$
[$\zeta (3) \simeq 1.202$ is the Riemann zeta function of 3].
The matter mixing angles $\theta_m$, $\bar \theta_m$ expressed
in terms of the parameters $a, b$ are given by
\begin{equation}
\sin^2 2\theta_m = {s^2 \over [s^2 + (b - a - c)^2]},\
\sin^2 2\bar \theta_{m} = {s^2 \over [s^2 + (b + a - c)^2]},
\label{bel}
\end{equation}
where $s \equiv \sin 2\theta_0$, $c \equiv \cos 2\theta_0$.
A resonance occurs for neutrinos when $\theta_m = \pi/4$ and for
antineutrinos when $\overline{\theta}_m = \pi/4$, which from
Eq.(\ref{bel}) implies that $b - a = \cos 2\theta_0$ and
$b + a = \cos 2\theta_0$ respectively. In our analysis we will often
need to consider the two distinct cases of very small mixing and 
very large mixing. For small mixing, $\cos 2\theta_0 \simeq 1$ and 
the resonance conditions become $b - a \simeq 1$ and $b + a \simeq 1$. 
For large mixing, $\cos 2\theta_0 \simeq 0$ and the resonance 
conditions become $a \simeq b$ and $-a \simeq b$.

Using the above analysis, we can derive a simple equation for the 
rate of change of $L_{\nu_{\alpha}}$,
\begin{eqnarray}
&\frac{dL_{\nu_{\alpha}}}{dt} = \frac{3}{16} \Gamma_{\nu_{\alpha}}
\left[-\sin^2 2\theta_m
\langle\sin^2 \left(\frac{\tau}{2L_m} \right) \rangle
+\sin^2 2\overline{\theta}_m
\langle\sin^2\left(\frac{\tau}{2\overline{L}_m}
\right)\rangle \right] \nonumber \\
&- {L_{\nu_{\alpha}}\Gamma_{\nu_{\alpha}} \over 2}
\left[\sin^2 2\theta_m
\langle\sin^2 \left(\frac{\tau}{2L_m} \right) \rangle
+\sin^2 2\overline{\theta}_m
\langle\sin^2\left(\frac{\tau}{2\overline{L}_m}
\right)\rangle \right].
\label{dLdt1}
\end{eqnarray}
The function, $\langle \sin^2 \left({\tau \over 2L_m}\right)
\rangle$ is given by
\begin{equation}
\langle \sin^2 \left({\tau \over 2L_m}\right) \rangle = 
{1 \over \omega_0}\int^t_0 e^{-\tau/\omega_0}
\sin^2 \left({\tau \over 2L_m}\right) d\tau,
\label{tau}
\end{equation}
where $\omega_0 \equiv 2\tau_0 = 2/\Gamma_{\nu_{\alpha}}$ is twice the mean 
time between collisions (of a pure weak eigenstate) and $t$ is the age 
of the Universe (note that $t \simeq \infty$ is a good approximation 
because $\omega_0 \ll t$). Evaluating Eq.(\ref{tau}) we find
\begin{equation}
\langle \sin^2 \left({\tau \over 2L_m}\right) \rangle = 
{1 \over 2}\left( {\omega_0^2/L_m^2 \over 1 + \omega_0^2/L_m^2}
\right),
\label{tau1}
\end{equation}
where we have assumed that $\omega_0$ and $L_m$ are approximately
independent of $t$ (static approximation).
Thus, using Eqs.(\ref{tau1}, \ref{sat9}, \ref{bel}), we can rewrite 
Eq.(\ref{dLdt1}) in the form
\begin{equation}
{dL_{\nu_{\alpha}} \over dt} =
{3 \over 8} {s^2 \Gamma_{\nu_{\alpha}} a 
(c - b)\over [x + (c - b + a)^2][x + (c - b - a)^2]}
+ \Delta,
\label{eeqq}
\end{equation}
where $\Delta$ is a small correction term
\begin{equation}
\Delta = - {1 \over 2} {L_{\nu_{\alpha}} s^2 \Gamma_{\nu_{\alpha}}  
[x + (c - b)^2 + a^2] \over [x + (c - b + a)^2][x + (c - b - a)^2]}, 
\label{delta}
\end{equation}
and $x$ is given by
\begin{equation}
x = s^2 + {1 \over 4}\Gamma^2_{\nu_{\alpha}} 
\left({2p \over \delta m^2}\right)^2 \simeq s^2 + 2\times 10^{-19}
\left( {T \over {\rm MeV}}\right)^{12}\left( {eV^2 \over \delta m^2}
\right)^2,
\label{jjds}
\end{equation}
where we have assumed $p = \langle p \rangle \simeq 3.15 T$ in
deriving the last part of the above equation.
Note that the correction term Eq.(\ref{delta}) is smaller than the
main term [Eq.(\ref{eeqq})] provided that $|L_{\nu_{\alpha}}| \ll |a|$.
In the region where the correction term is larger than
the main term, its effect is to reduce $|L_{\nu_{\alpha}}|$ such
that $L_{\nu_{\alpha}} \to 0$.  From Eq.(\ref{ab3}), the 
condition $|L_{\nu_{\alpha}}| > |a|$ only occurs for
quite low temperatures,
\begin{equation}
{T \over {\rm MeV}} \stackrel{<}{\sim} {1 \over 3} 
\left({|\delta m^2| \over eV^2}\right)^{1\over 4}.
\end{equation}
From the above equation, we see that in the main region of interest
($T \stackrel{>}{\sim} 3\ {\rm MeV}$), the correction term
is much smaller than the main term provided that
$|\delta m^2| \stackrel{<}{\sim} 10^4 \ eV^2$.
Note that for very large $|\delta m^2| \stackrel{>}{\sim} 10^4 \ 
eV^2$, the correction term may be important.

Observe that Eq.(\ref{eeqq}) differs slightly from the
equation derived in Ref.\cite{ftv}. The difference is that in
Ref.\cite{ftv}, we assumed that $\omega_0^2/L_m^2 \gg 1$ (so
that $\langle \sin^2 \tau/2L_m \rangle \simeq 1/2$)
which is always true except possibly at the very center of the resonance
\cite{ftv}. Also note that in Ref.\cite{ftv} we neglected a factor
of 2 which arises because we negligently assumed that the rate
of measurement was equal to the rate of collision.

We now pause to review and comment on the assumptions 
made in deriving Eq.(\ref{eeqq}).
There are five main simplifying assumptions:
\vskip 0.3cm
\noindent
(1) We have neglected the thermal spread of the neutrino
momenta, and have replaced all momenta by their thermal
average $\langle p \rangle \simeq 3.15 T$.
\vskip 0.3cm
\noindent
(2) We have assumed that $n_{\nu_{s}}, n_{\bar \nu_{s}} 
\ll n_{\nu_{\alpha}}, n_{\bar \nu_{\alpha}}$.  If the number 
densities $n_{\nu_{s}}, n_{\bar \nu_{s}}$ are non-negligible, 
then we must multiply the first term on the right-hand side of 
Eq.(\ref{eeqq}) by the
factor  $[n_{\nu_{\alpha}} - n_{\nu_{s}}]/n_{\nu_{\alpha}}$.
\vskip 0.3cm
\noindent
(3) We have assumed that the transformation from the vacuum
parameters to matter parameters i.e., $\sin\theta_0 \to \sin\theta_m$
and $L_0 \to L_m$ diagonalizes the Hamiltonian. This is only strictly 
true in the adiabatic limit ($|d\theta_m/dt| \ll |\Delta_m|$).
In the general case \cite{msw}, 
\begin{equation}
i{d \over dt}\left(\begin{array}{c}
\nu_m^1 \\
\nu_m^2\end{array}\right) = 
\left(\begin{array}{cc}
-{\Delta_m\over 2}&-i{d\theta_m \over dt}\\
i{d\theta_m \over dt}&{\Delta_m \over 2}
\end{array}\right) \left(\begin{array}{c}
\nu^1_m\\
\nu^2_m
\end{array}\right),
\end{equation}
with
\begin{equation}
{d\theta_m \over dt} = {1\over2}{\sin2\theta_0 \over 
(b - a - \cos2\theta_0)^2 + \sin^2 2\theta_0} {d(b-a)\over dt},
\end{equation}
where $\nu^{1,2}_m$ are the instantaneous matter eigenstates and 
$\Delta_m \equiv 1/L_m$.  Expanding out $\gamma \equiv 
|(d\theta_m/dt)/\Delta_m|$ we find (neglecting $da/dt$),
\begin{eqnarray}
& \gamma \le 2(4) \times 10^{-8} \left( {\sin^2 2\theta_0 \over
10^{-6}}\right)^{1/2}\left({eV^2 \over |\delta m^2|}\right)^{1/2},
\ {\rm away \ from \ resonance,} \nonumber \\
& \gamma \simeq 2(4) \left( {10^{-5} \over \sin^2 2\theta_0} 
\right)\left({eV^2 \over |\delta m^2|}\right)^{1/2},
\ {\rm at \ the \ initial\ resonance \ where}\ b = \cos2\theta_0,
a \simeq 0, \nonumber \\
& \gamma \simeq 6(9) \times 10^{-4}T^3\left({10^{-5} \over \sin^2 2\theta_0}
\right)\left({eV^2 \over |\delta m^2|}\right)
\ {\rm at \ the \ resonance\ where }\ |b - a| = \cos2\theta_0,
\label{zyx}
\end{eqnarray}
for $\nu_e - \nu_s$ ($ \nu_{\mu, \tau} -  \nu_s$) oscillations.
However at the initial resonance where $b = \cos2\theta_0, a\simeq 0$,
$L_{\nu_{\alpha}}$ is created rapidly. The contribution to $\gamma$ 
from a rapidly changing $L_{\nu_{\alpha}}$ at this resonance is
\begin{equation}
\gamma \simeq {(0.5)(3) \times 10^2 \over \sin^2 2\theta_0}\left( {
eV^2 \over |\delta m^2|}\right)^{{2\over 3}} {dL_{\nu_{\alpha}} \over 
d(T/{\rm MeV})},
\label{zyx2}
\end{equation}
for $\nu_e - \nu_s$ ($\nu_{\mu,\tau} - \nu_s$) oscillations (and
we have assumed that $\cos2\theta_0 \sim 1$).
Thus away from the resonance the adiabatic approximation is valid 
for the parameter space of interest. (i.e. for $|\delta m^2| \stackrel
{>}{\sim} 10^{-4}\ eV^2$). However at the resonance the adiabatic
approximation may not be valid. 

\vskip 0.3cm
\noindent
(4) Equation (\ref{eeqq}) neglects flavour conversion of neutrinos
passing through the resonance (the MSW effect).  Observe that there 
is not expected to be significant flavour 
conversion at the initial resonance (where $b \simeq \cos2\theta_0$) due 
to the MSW effect (even if the system is adiabatic at this resonance) 
because the frequency of the collisions is such that $\langle \sin^2 
\tau /2L_m \rangle \ll 1$ at the center of the initial resonance, for 
most of the parameter space of interest.  Indeed,
at the center of the resonance,
\begin{eqnarray}
&{\omega_0 \over 2L_m} = 
{\sin2\theta_0 \over y_{\alpha}G_F^2T^5}{\delta m^2 \over 2p}
\simeq 4\times 10^8\sin2\theta_0
\left( {{\rm MeV} \over T}\right)^6
{\delta m^2 \over eV^2},
\nonumber \\
& \simeq 90\tan2\theta_0, 
\ {\rm \ if \ } b = \cos2\theta_0.
\label{9999}
\end{eqnarray}
Thus, for $\sin^2 2\theta_0 \stackrel{<}{\sim} 10^{-4}$, 
$\langle \sin^2 \tau /2L_m \rangle \ll 1$.
Note however that for temperatures below the initial resonance,
the MSW effect may be important if there are neutrinos
passing through the resonance. 
\vskip 0.3cm
\noindent
(5) We have assumed that the rate of change of lepton number is
dominated by collisions. There is also a contribution from 
oscillations between collisions. Oscillations between collisions
affect lepton number because the oscillations produce a superposition 
of states, where the averaged expectation value of the state 
being a weak-eigenstate is $1 - \sin^2 2\theta_m$ for neutrinos.
This probability is generally unequal to the analogous quantity for
anti-neutrinos, which is $1 - \sin^2 2\bar \theta_m$. 
It is possible to show\cite{ftv} that for temperatures
greater than a few MeV, the change in lepton number
due to the oscillations between collisions is generally smaller than 
the change due to collisions except possibly at the resonance where 
$\sin^2 2\theta_m$ is changing rapidly.
\vskip 0.4cm
The effect of the thermal spread of the neutrino momenta should be 
to make the creation and destruction of lepton number much smoother.
At any given time, only a small fraction of the neutrinos will be
at resonance (because the resonance width is much less
than the spread of neutrino momenta). Thus, the regions away from
resonance may also be important.
We will study the effect of the thermal distribution of
momenta in section IV.

The second assumption [(2) above] will be approximately valid 
for much of the parameter space of interest. This is because we are 
essentially interested in the region of parameter space where 
the sterile neutrinos do not come into equilibrium with the 
ordinary neutrinos.  We will study the effect of the sterile 
neutrino number density being non-zero in section V.
The assumptions (3) and (5), may not be valid in the resonance region.
Note that we will denote the assumptions (3) and (5) 
collectively as the static approximation because in limit where the 
system is sufficiently smooth they will be valid. 

Clearly a more exact treatment of the resonance
is desirable, since assumptions
(3) and (5) may not be valid there. In section III
we will develop a more exact treatment of the resonance region by 
examining the appropriate equations from the density matrix.
As we will show in section III, this treatment leads to
the following equation for the rate of change of lepton number
\begin{equation}
{dL_{\nu_{\alpha}} \over dt} \simeq
{3\beta^2 \over 8}\int^t_0 e^{-\tau/\omega_0} 
\sin\left[\int^t_{t-\tau} \lambda^+ dt'\right]
\sin\left[\int^t_{t-\tau} \lambda^- dt''\right] d\tau,
\label{bob}
\end{equation}
where
\begin{equation}
\beta = {\delta m^2 \over 2p} \sin 2\theta_0,\
\lambda^+ = {\delta m^2 \over 2p}(\cos2\theta_0 - b),\
\lambda^- = {\delta m^2 \over 2p}a.
\end{equation}
This equation is valid given the assumptions (1), (2) and (4) but does
not require assumptions (3) and (5) [above]. This equation is an 
integro-differential equation and 
although compact cannot be solved analytically except in various limits.
Note that the static limit corresponds to taking $\lambda^{\pm}$ as
constant (i.e. independent of $t'$). In this limit Eq.(\ref{bob}) 
reduces approximately to Eq.(\ref{eeqq}) as expected.  In the 
appendix we show that Eq.(\ref{bob}) can also be obtained using 
the Hamiltonian formalism provided that the rate of measurement is 
taken to be half the collision frequency.

Qualitatively, it turns out that the simplified equation, 
Eq.(\ref{eeqq}), gives a reasonable description of the creation of 
lepton number as the Universe evolves.  Assuming that Eq.(\ref{eeqq}) 
is valid, we now analyse the behaviour of $L_{\nu_{\alpha}}$ as 
driven by $\nu_{\alpha} - \nu_{s}$ oscillations in isolation. 
Suppose that all initial asymmetries other than $L_{\nu_{\alpha}}$ 
can be neglected so that $L^{(\alpha)} \simeq 2 L_{\nu_{\alpha}}$.
Notice first of all that for $\delta m^2 > 0$ if follows from 
Eq.(\ref{ab3}) that $b$ is negative and $a$ has the opposite sign to 
$L_{\nu_{\alpha}}$.  Thus from Eq.(\ref{eeqq}) it is easy
to see that the point $L_{\nu_{\alpha}} = 0$ is always
a stable fixed point. That is,
when $L_{\nu_{\alpha}} > 0$ the rate of change $dL_{\nu_{\alpha}}
/dt$ is negative, while when $L_{\nu_{\alpha}} < 0$ the rate of change
$dL_{\nu_{\alpha}}/dt$ is positive, so $L_{\nu_{\alpha}}$ 
always tends to zero.
[In the realistic case, where the baryon and electron
asymmetries are not neglected, $L^{(\alpha)}$ is given by 
Eq.(\ref{Lsuper}). In this case $L^{(\alpha)} \sim 0$ is an approximate
fixed point. Note that even if all of the lepton numbers where
initially zero, lepton number would be generated such that 
$L^{(\alpha)} \simeq 0$, i.e. $2L_{\nu_{\alpha}} 
\simeq -\eta$ [see Eq.(\ref{Lsuper})].  Note that $2L_{\nu_{\alpha}}$ 
is only approximately $-\eta$ because of the $\Delta$ term proportional
to $L_{\nu_{\alpha}}$ in Eq.(\ref{eeqq})].

Now consider neutrino oscillations with $\delta m^2 < 0$. 
In this case $b$ is positive and $a$ has the same sign as 
$L_{\nu_{\alpha}}$.  From Eq.(\ref{eeqq}), $L_{\nu_{\alpha}} 
\simeq 0$ is a stable fixed point only 
when $b > \cos 2\theta_0$.  When $b < \cos2\theta_0$, 
the point $L_{\nu_{\alpha}} \simeq 0$ is unstable. 
[That is, if $L_{\nu_{\alpha}} > 0$, then 
$dL_{\nu_{\alpha}} /dt > 0$, while if $L_{\nu_{\alpha}} < 0$ then
$dL_{\nu_{\alpha}}/dt < 0$].
Since $b \sim T^6$, at some point during the evolution of
the Universe $b$ becomes less than $\cos2\theta_0$ and
$L_{\nu_{\alpha}} = 0$ becomes unstable. If $|\delta m^2|
\stackrel{>}{\sim} 10^{-4}\ eV^2$, then this point (where $b = 
\cos2\theta_0$) occurs for temperatures greater than about three
MeV (assuming $\cos 2\theta_0 \simeq 1$). In this region
the rate of change of lepton number is dominated by collisions and 
Eq.(\ref{eeqq}) is approximately valid.  When the critical point 
where $b = \cos2\theta_0$ is reached, the lepton asymmetries are small 
and hence $|a| \ll \cos2\theta_0 \simeq 1$. Equation (\ref{eeqq}) 
then implies that $dL_{\nu_{\alpha}}/dt$ is approximately 
proportional to $L_{\nu_{\alpha}}$, which leads to a brief but 
extremely rapid period of exponential growth of $L_{\nu_{\alpha}}$ 
\cite{ftv}. Furthermore note that the constant of proportionality 
is enhanced by resonances for both neutrinos and antineutrinos at 
this critical point ($a \simeq 0$, $b = \cos2\theta_0$). The exponent 
governing the exponential increase in $L_{\nu_{\alpha}}$ is thus 
a large number (unless $\sin^2 2\theta_0$ is very small). Note 
that the critical point $b = \cos2\theta_0 $ occurs when
\begin{equation}
T_c \simeq 13(16)\left({|\delta m^2|\cos2\theta_0 \over eV^2}\right)^{1/6}
\  {\rm MeV},
\label{critT}
\end{equation} 
for the $\nu_e - \nu_{s}$ ($\nu_{\mu, \tau} - \nu_{s}$)
oscillations we have been focusing on.

As the system passes through this critical temperature, lepton
number is rapidly created until $a\stackrel{>}{\sim} \cos2\theta_0
-b$. The resonance at $a = \cos2\theta_0 -b$ acts like
a barrier which keeps $a > \cos2\theta_0 - b$ as the temperature
falls below $T_c$. Since the parameter $a$ is proportional to 
$L_{\nu_{\alpha}}T^4$, it follows that the lepton number continues 
to grow approximately like $T^{-4}$ after the resonance as the 
temperature falls. 

As the temperature drops, eventually the
oscillations cannot keep up with the expansion of
the Universe. For temperatures well below the resonance, $a \simeq 
\cos 2\theta_0$ (assuming that $L_{\nu_{\alpha}} > 0 $ for
definiteness). In this region, the rate of change of
$a$ due to the oscillations is balanced by the rate of
change of $a$ due to the expansion of the Universe.  That is,
\begin{equation}
{da \over dt} = {\partial a \over \partial L_{\nu_{\alpha}}}
{\partial L_{\nu_{\alpha}} \over \partial t} + 
{\partial a \over \partial t} \simeq 0.
\label{sat1}
\end{equation}
Eventually, the rate of change of $a$ due to the expansion
of the Universe becomes larger in magnitude than the maximum rate
of change of $a$ due to oscillations. At this point, $a$ falls
below the resonance point (i.e. $a <  \cos2\theta_0 - b $) and 
the value of $L_{\nu_{\alpha}}$ will be approximately frozen.
The point in time when this occurs is thus governed by the
equation
\begin{equation}
{\partial a \over \partial L_{\nu_{\alpha}}}
{\partial L_{\nu_{\alpha}} \over \partial t}|_{{\rm max}} = 
-{\partial a \over \partial t}.
\label{sat2}
\end{equation}
The maximum rate of change of $L_{\nu_{\alpha}}$ occurs at
the resonance where $a = \cos2\theta_0 - b$.  Using Eq.(\ref{eeqq}), 
we can easily evaluate $dL_{\nu_{\alpha}}/dt$ at this
point. Assuming that $\cos 2\theta_0 \simeq 1$, we find at
the resonance,
\begin{equation}
{dL_{\nu_{\alpha}} \over dt} =
{3 \over 32} \Gamma_{\nu_{\alpha}} a,
\label{sat3}
\end{equation}
where we have assumed that $x \simeq \sin^2 2\theta_0$,
which should be valid since we are in the region of low
temperatures $T \sim 3$ MeV [recall that $x$ is
defined in Eq.(\ref{jjds})]. Also note that
\begin{equation}
{\partial a \over \partial t} = {\partial a\over \partial
T}{dT \over dt} \simeq -{4a \over T}
{5.5T^3 \over M_P},
\label{sat4}
\end{equation}
where we have used the result that the parameter $a$ is proportional to
$T^4$, and $dT/dt \simeq -5.5T^3/M_P$ (which
is approximately valid for $1 \ {\rm MeV}\stackrel{<}{\sim} T 
\stackrel{<}{\sim} 100 \ {\rm MeV}$, and $M_P \simeq 1.2 \times 
10^{22}\ {\rm MeV}$ is the Planck mass). 
Thus, using Eqs.(\ref{sat3},\ref{sat4}), the condition Eq.(\ref{sat2})
can be solved for $T$. Doing this exercise, and
denoting this value of $T$ by $T_f$, we find
\begin{equation}
T_f \simeq 
\left[ {50 \delta m^2 \over M_P y_{\alpha}G_F^3}\right]^{1 \over 7}
\simeq \left[ {\delta m^2 \over eV^2}\right]^{1 \over 7} \ {\rm MeV},
\label{sat5}
\end{equation}
where we have used Eq.(\ref{Gammas}),  Eq.(\ref{ab3}). 
Thus, we  expect $L_{\nu_{\alpha}}$ to evolve like $T^{-4}$ 
until quite low temperatures of order $1$ MeV. Note however 
that when the momentum distribution is taken into account, 
the situation is somewhat different. This is because only a small 
fraction of neutrinos (typically of order 1 percent or
less) will be at the resonance, so that the magnitude
of the maximum value of $\partial L_{\nu_{\alpha}}/\partial t$
will be reduced by a few orders of magnitude. 
Because of the $1/7$ power in 
Eq.(\ref{sat5}), the temperature where $L_{\nu_{\alpha}}$ is
approximately frozen, $T_f$, increases by only a relatively small
factor of 2 or 3.  Finally recall that for temperatures below the
initial resonance, the MSW effect can also contribute significantly.
This is because for low temperatures near $T_f$, there will be a 
significant number of neutrinos which will be passing through the 
resonance. For low temperatures, the adiabatic condition is expected 
to hold [for most of the parameter space of interest, see 
Eq.(\ref{zyx})]. Also, recall that the oscillations will not be damped by 
collisions for low temperatures [see Eq.(\ref{9999})] and thus
ordinary neutrinos can be converted into sterile neutrino states 
simply by passing through the resonance\cite{msw}.  This effect 
will help keep $a \simeq 1$ for even lower temperatures.

Clearly these factors (the momentum distribution and the MSW flavour
conversion of the neutrinos passing through the resonance) will be 
important if one wants to know the final magnitude of 
$L_{\nu_{\alpha}}$. For example, the final magnitude of 
$L_{\nu_{e}}$ is very important if one wants to calculate the region
of parameter space where the $L_{\nu_e}$ is large enough to
affect big bang nucleosynthesis through nuclear reaction rates. 
However, for the application in this paper, the precise value
of $L_{\nu_{\alpha}}$ at low temperatures is not required,
so we will leave a study of this issue to the future.

In order to illustrate the evolution of $L_{\nu_{\alpha}}$ 
we take some examples. It is illuminating to compare the
evolution expected from the simple Eq.(\ref{eeqq}) [based
on the assumptions (1)-(5) discussed above], with
the evolution governed by the more complicated density matrix equations.
[Eqns.(\ref{jk}), see next section for some discussion of the density
matrix formalism]. The evolution of $L_{\nu_{\alpha}}$ as governed by the 
density matrix equations hold more generally than Eq.(\ref{eeqq}).
This is because they do not require the assumptions (2),(3),(4) 
or (5) [discussed above] to hold.  They do still incorporate assumption 
(1), that is the thermal distribution of the neutrino momentum 
is neglected.

In Figure 1,2 we plot the evolution of $L_{\nu_{\alpha}}$ for 
some typical parameters. We consider for example $\nu_{\mu,\tau} - \nu_s$
oscillations. In Figure 1 we take $\delta m^2 =
-1 \ eV^2$, and $\sin^2 2\theta_0 = 10^{-4}, 10^{-8}$.
Figure 2 is the same as figure 1 except that $\delta m^2 
= -1000 \ eV^2$ and $\sin^2 2\theta_0 = 10^{-6}, 10^{-9}$.
The solid lines are the result of numerically 
integrating the density matrix equations, while the dashed
lines are the results of numerically integrating Eq.(\ref{eeqq}).
We stress that in both the density matrix equations and in
Eq.(\ref{eeqq}), the momentum distribution of the
neutrino has been neglected. The effect of the momentum distribution
will be considered in detail in sections IV, V.

In the examples in Figure 1,2 the initial lepton asymmetry was taken 
as zero.  The generation of lepton number is essentially independent
of the initial lepton number asymmetry provided that it is less
than about $10^{-5}$\cite{ekmpl,fv1}. This is because for temperatures 
greater than the resonance temperature, the oscillations destroy or 
create lepton number until $L^{(\alpha)} \approx 0$ independently of 
the initial value of $L_{\nu_{\alpha}}$ [which we denote as $L_{init}$], 
provided that $|L_{init}|$ is less than about $10^{-5}$.  For 
$|L_{init}| \stackrel{>}{\sim} 10^{-5}$, the oscillations at
temperatures above the resonance temperature are not strong
enough to destroy the initial asymmetry. Consequently, 
$L_{\nu_{\alpha}}$ remains large, and it will become larger due to the
oscillations which create lepton number at temperatures below
the resonance temperature. 

As the Figures show, the behaviour expected from Eq.(\ref{eeqq})
occurs. The main difference arises at the resonance where
the magnitude of the lepton number is somewhat larger than
expected from Eq.(\ref{eeqq}). This occurs because the
assumptions (3) and (5) [discussed above], which lead to
Eq.(\ref{eeqq}) are not valid at this resonance. Actually,
in Figure 1,2 we have plotted $|L_{\nu_{\mu}}|$. Integration
of the density matrix equation reveals that in example 1 (but
not in example 2, although $L_{\nu_{\alpha}}$ does change sign), 
the generated lepton number oscillates at the 
resonance and changes sign a few times (see Ref.\cite{shi}, for 
a figure illustrating this).  Note that this effect can be 
understood from Eq.(\ref{bob}).  To see this, observe that when 
$L_{\nu_{\alpha}}$ is initially created at the resonance, the 
parameter $\lambda^-$ grows very rapidly because it is proportional 
to $L_{\nu_{\alpha}}$.  The creation of $L_{\nu_{\alpha}}$ may be
so rapid that $\int^t_{t-\tau} \lambda^- dt'$ is approximately 
independent of $\tau$ when the initial rapid growth of 
$L_{\nu_{\alpha}}$ occurs.  If this happens then at this 
instant Eq.(\ref{bob}) can be simplified to the approximate 
form
\begin{equation}
{dL_{\nu_{\alpha}} \over dt} \sim
{3\beta^2 \over 8} \sin\left[ \int^t_{t-\omega_0} \lambda^- dt'
\right] \int^t_0 e^{-\tau/\omega_0} \sin\left[\int^t_{t-\tau} \lambda^+ dt''
\right] d\tau.
\end{equation}
The oscillations
occur because of the factor $\sin\int^t_{t-\omega_0} \lambda^- dt'$
which oscillates between $\pm 1$.
Note however, that this oscillation of lepton number would not
be expected to occur in the realistic case where the
thermal spread of neutrino momenta is considered.

Note that it may be possible to predict the sign of the asymmetry 
in principle. Assuming that the resonance is smooth enough so that 
Eq.(\ref{eeqq}) is valid, the equation governing the evolution of 
$L_{\nu_{\alpha}}$ has the approximate form
\begin{equation}
{dL_{\nu_{\alpha}} \over dt} = 
A(2L_{\nu_{\alpha}} + \stackrel{\sim}{\eta}) - BL_{\nu_{\alpha}} 
= (2A - B)L_{\nu_{\alpha}} + A\stackrel{\sim}{\eta},
\label{frid}
\end{equation}
where $\stackrel{\sim}{\eta} \equiv \eta + L_{\nu_e} + L_{\nu_{\mu}}
+ L_{\nu_{\tau}} - L_{\nu_{\alpha}}$ (we have
defined $\stackrel{\sim}{\eta}$ such that  it is 
independent of $L_{\nu_{\alpha}}$). Note that $A$ and $B$ [which
can be obtained from Eq.(\ref{eeqq})] are complicated 
functions of time. Observe however that
$B > 0$ and $A$ is initially less than zero, and at
the resonance $A$ changes sign and becomes positive after
that.  In the region where $2A < B$,
the lepton number evolves such that
\begin{equation}
(2A - B)L_{\nu_{\alpha}} + A\stackrel{\sim}{\eta} \to 0.
\end{equation}
Thus $L_{\nu_{\alpha}}$ will evolve such that it has a sign
opposite to $\stackrel{\sim}{\eta}$ just before
the resonance.  When $2A > B$, $L_{\nu_{\alpha}}$ will become 
unstable and grow rapidly.  Note that at the point $A = B/2$,
\begin{equation}
{dL_{\nu_{\alpha}} \over dt}  
= A\stackrel{\sim}{\eta}.
\label{frid9}
\end{equation}
Hence, at the point where the initial rapid creation of
$L_{\nu_{\alpha}}$ occurs, the rate of change of 
$L_{\nu_{\alpha}}$ will be proportional to $\stackrel{\sim}{\eta}$.
Thus, we might expect that the sign of $L_{\nu_{\alpha}}$ will
be the same as the sign of the asymmetry $\stackrel{\sim}{\eta}$ after
the resonance.  This means that $L_{\nu_{\alpha}}$ should change 
sign at the resonance.  Note however that because $\stackrel{\sim}{\eta}$ 
depends on the initial values of the lepton asymmetries which are 
unknown at the moment, it seems that the sign of $L_{\nu_{\alpha}}$ 
cannot yet be predicted.  However the above calculation shows that the 
sign of $L_{\nu_{\alpha}}$ should not depend on statistical fluctuations, 
as we initially thought likely\cite{ftv}.  

Finally we would like to comment on the region of parameter space
where significant generation of lepton number occurs. Firstly, we require
that $\delta m^2 < 0$ and that
$|\delta m^2| \stackrel{>}{\sim} 10^{-4} \ eV^2$, so that 
$T_c^{\alpha s} \stackrel{>}{\sim} 3$ MeV. For
$|\delta m^2| \stackrel{<}{\sim} 10^{-4}\ eV^2$, lepton number
can still be generated but it is dominated by the oscillations
between collisions and is oscillatory\cite{shi, ekm}. Note that 
in the realistic case where the spread of momenta is taken
into account, oscillations of lepton number would be
smoothed out and may not occur.
A numerical study in Ref.\cite{shi} shows that $\sin^2 2\theta_0 
\stackrel{>}{\sim} 10^{-11}(eV^2/|\delta m^2|)^{1/6}$ 
is also necessary (see also Ref.\cite{ftv} for an approximate analytical 
derivation). Finally, we must require that $\sin^2 2\theta_0$
be small enough so that the sterile neutrinos do not come
into equilibrium. [For example, if there are equal numbers of $\nu_{\mu}$
and $\nu'_{e}$ then the rate $n_{\nu_{\mu}}\Gamma (\nu_{\mu} \to 
\nu'_{e}) = n_{\nu'_e}\Gamma (\nu_{e}'\to \nu_{\mu})$ and $L_{\mu}$
cannot be generated].  We will re-examine the region of parameter 
space where significant generation of lepton number occurs in 
section V (where the effects of the Fermi-Dirac momentum distribution 
of the neutrino will be taken into account).

Note that in Ref.\cite{shi}, it is argued that lepton number
generation only occurs provided that $|\delta m^2| \stackrel{<}{\sim}
100\ eV^2$. We have not been able to verify this result, either
analytically or numerically. 
In fact, we have been able to obtain any significant
upper bound on $|\delta m^2|$.
\vskip 0.6cm
\noindent
{\bf III. Lepton number generation due to neutrino oscillations - A more
exact treatment}
\vskip 0.5cm
\noindent
In this section we derive a more general equation describing
lepton number generation in the early Universe
which can be applied when the system is 
changing rapidly, as occurs, for instance, at the resonance.
The only assumptions that we will make are the assumptions (1),
(2) and (4) (discussed in the previous section). That is we will 
neglect the spread of neutrino momenta and set $\langle p \rangle 
\simeq 3.15 T$, and we will also assume that there are negligible 
numbers of sterile neutrinos generated.
In the appendix an alternative derivation (with the same
end result) based on the Hamiltonian formalism is presented.
Although not yet realistic because of assumptions (1) and (2),
this derivation turns out to be particularly useful because it
allows us to work out the region of parameter space where the
simple Eq.(\ref{eeqq}) is approximately valid. As we will show,
it turns out that Eq.(\ref{eeqq}) has a wider applicability than
might be expected from the adiabatic condition Eqs.(\ref{zyx2}).

The system of an active neutrino oscillating with a sterile
neutrino can be described by a density
matrix. See, for example, Ref.\cite{bm} for details.
Below we very briefly outline this formalism and show how
it leads to an integro-differential equation
which reduces to Eq.(\ref{eeqq}) in the static limit.

The density matrices for the neutrino system are given by
\begin{equation}
\rho_{\nu} = {P_0 + {\bf P}.\sigma \over 2},
\  \rho_{\bar \nu} = {\bar P_0 + \bar{{\bf P}}.\sigma \over 2},
\end{equation}
where $P_0$ and $\bar P_0$ are the relative number densities
of the mixed neutrino and anti-neutrino species, and ${\bf P}$
and $\bar{{\bf P}}$ are the polarization vectors that describe
the internal quantum state of the mixed neutrinos in terms of an
expansion in the Pauli matrices $\sigma$.  The number densities of
$\nu_{\alpha}$ and $\nu_s$ are given by
\begin{equation}
{n_{\nu_{\alpha}}\over P_0} = {1 + P_Z\over 2},\
{n_{\nu_{s}} \over P_0} = {1 - P_Z\over 2}, 
\label{anot}
\end{equation}
with analogous equations for the anti-neutrinos.
The evolution of $P_0, {\bf P}$ are governed by the equations
\cite{bm}:
\begin{eqnarray}
&{d\over dt}{\bf P} = {\bf V} \times {\bf P} + 
(1 - P_Z)({d \over dt}lnP_0) {\bf \hat{z}}
- (D^E + D^I + {d \over dt}lnP_0)(P_x {\bf \hat{x}} + P_y {\bf\hat{y}}) 
\nonumber \\
&{d \over dt}P_0 = \sum_{i =e, \nu_{\beta}; \beta \neq \alpha}
\langle \Gamma(\nu_{\alpha} \bar \nu_{\alpha} 
\to i \bar i)\rangle(\lambda_{\alpha}n_{\alpha}n_{\bar \alpha}
- n_{\nu_e}n_{\bar \nu_e}),
\label{jk}
\end{eqnarray}
where $\lambda_{\nu} = 1$ and $\lambda_e = 1/4$, and $\langle ...
\rangle$ indicates the average over the momentum distributions.
The quantity ${\bf V}$ is given by
\begin{equation}
{\bf V} = \beta {\bf \hat{x}} + \lambda {\bf \hat{z}},
\end{equation}
where $\beta, \lambda$ are defined by
\begin{equation}
\beta = {\delta m^2 \over 2p} \sin2\theta_0, \
\lambda = {\delta m^2 \over 2p }(\cos2\theta_0 - b \pm a),
\label{99}
\end{equation}
where the $+(-)$ sign corresponds to neutrino (anti-neutrino) 
oscillations.  The quantities $D^E$ and $D^I$ are quantum damping 
parameters resulting from elastic and inelastic processes respectively.
According to ref.\cite{bm}, $D^E + D^I = \Gamma_{\nu_{\alpha}}/2$.
The function
$\langle \Gamma (\phi \psi \to \phi'\psi')\rangle$ is the
collision rate for the process $\phi \psi \to \phi'\psi'$ 
averaged over the distribution of collision parameters at
the temperature $T$ assuming that all species are in equilibrium.

Expanding out Eq.(\ref{jk}), we have: 
\begin{eqnarray}
& {dP_z \over dt} = \beta P_y + (1 - P_z)({d \over dt}
log P_0),  \nonumber \\
& {dP_y \over dt} =  \lambda P_x - \beta P_z - P_y/\omega_0, 
\nonumber \\
& {dP_x \over dt} = - \lambda P_y - P_x/\omega_0,
\label{kl}
\end{eqnarray}
where $\omega_0 \equiv 1/(D^E + D^I + {d \over dt}log P_0) \simeq
1/D$ (where $D \equiv D^E + D^I$).
If we make the approximation of setting all of the number densities to
their equilibrium values, and also assume that the number of
sterile species is small, then $P_z \simeq 1$ 
and Eq.(\ref{kl}) simplifies to
\begin{eqnarray}
& {dP_z \over dt} \simeq \beta P_y,  \nonumber \\
& {dP_y \over dt} \simeq  \lambda P_x - \beta  - P_y/\omega_0, 
\nonumber \\
& {dP_x \over dt} \simeq - \lambda P_y - P_x/\omega_0.
\label{cde}
\end{eqnarray}
Strictly speaking, the approximation of setting $P_z  = 1 = 
constant$ can only be valid when $\beta P_z$ is small enough, so that
MSW flavour conversion cannot occur, i.e. when
\begin{equation}
|\beta| \stackrel{<}{\sim} |\lambda| \ {\rm or}\ {1 \over \omega_0}.
\label{smallen}
\end{equation}
It is useful to introduce the complex variable $\stackrel{\sim}{P}(t)$ 
defined by $\stackrel{\sim}{P} \equiv P_x + iP_y$.
It is easy to see that the resulting equation describing the
evolution of $\stackrel{\sim}{P}(t)$ is given by
\begin{equation}
i{d\stackrel{\sim}{P} \over dt} =
-\lambda \stackrel{\sim}{P} -i{\stackrel{\sim}{P} \over
\omega_0} + \beta.
\label{wwee}
\end{equation}
The solution to this equation with initial condition $\stackrel{\sim}
{P}(0) = 0$ is:
\begin{equation}
\stackrel{\sim}{P}(t) = 
-i\int^t_0 \beta (t') e^{(t'-t)/\omega_0} e^{i\int^{t}_{t'} 
\lambda dt''} dt',
\label{wwew}
\end{equation}
where $\omega_0$ has been assumed to be independent of time
which is approximately valid for temperatures above a few MeV where the 
expansion rate is less than the collision rate.
One can easily verify that Eq.(\ref{wwew}) is indeed the solution
of Eq.(\ref{wwee}) by direct substitution.
Thus, taking the imaginary part of both sides of Eq.(\ref{wwew}),
we find that
\begin{equation}
P_y = 
-\int^t_0 \beta e^{(t'-t)/\omega_0} \cos\left[\int^{t'}_t 
\lambda dt''\right] dt'.
\label{abc}
\end{equation}
From Eq.(\ref{anot}), it follows that:
\begin{equation}
{dL_{\nu_{\alpha}} \over dt} = {3 \over 16} {d \over dt}(P_z - \bar P_z),
\end{equation}
where $\bar P_z$ denotes the z-component of the polarization
vector for anti-neutrinos.  Thus using Eq.(\ref{cde}) the above 
equation becomes
\begin{equation}
{dL_{\nu_{\alpha}} \over dt} = {3 \beta \over 16}(P_y - \bar P_y).
\end{equation}
Note that $P_y$ is given by Eq.(\ref{abc}) and $\bar P_y$
is defined similarly to $P_y$ except that we must replace 
$a \to -a$. Thus, we obtain
\begin{equation}
{dL_{\nu_{\alpha}} \over dt}  \simeq 
{-3 \beta^2\over 16}\int^t_0  e^{(t'-t)/\omega_0} 
\left(\cos\left[\int^{t'}_t  \lambda dt'' \right] - 
\cos\left[\int^{t'}_t \bar \lambda dt''  
\right] \right)dt' ,
\end{equation}
where $\lambda = \delta m^2 (c - b + a)/2p,\ \bar \lambda = 
\delta m^2 (c - b - a)/2p$. Note that
we have taken $\beta$ outside the integral, which is
valid for $T \stackrel{>}{\sim} 2\ {\rm MeV}$, because $\beta$ is 
approximately constant over the interaction time scale 
$t - t'$\cite{last}.  Changing variables from $t'$ to the variable 
$\tau$ where $ \tau \equiv t - t'$, this equation reduces
to 
\begin{equation}
{dL_{\nu_{\alpha}} \over dt} =
{-3\beta^2 \over 16}\int^t_0 e^{-\tau/\omega_0} 
\left(\cos\left[\int^t_{t-\tau} \lambda 
dt'\right]
- \cos\left[\int^t_{t-\tau} \bar \lambda
dt'\right] \right) d\tau,
\label{whhhh1}
\end{equation}
or equivalently,
\begin{equation}
{dL_{\nu_{\alpha}} \over dt} = {-3\beta^2 \omega_0\over 16}
\left[ \langle \cos \int^t_{t-\tau}\lambda dt'\rangle 
- \langle \cos \int^t_{t-\tau}\bar \lambda dt'\rangle 
\right].
\end{equation}
Note that the above equation can be re-written using
a trigonometric identity, so that
\begin{equation}
{dL_{\nu_{\alpha}} \over dt} =
{3\beta^2 \over 8}\int^t_0 e^{-\tau/\omega_0} 
\sin\left[\int^t_{t-\tau} \lambda^+ 
dt'\right]
\sin\left[\int^t_{t-\tau} \lambda^-
dt''\right] d\tau,
\label{wh1}
\end{equation}
where $\lambda^{\pm} = (\lambda \pm \bar \lambda)/2$.

The phenomenon of neutrino oscillations can also
be described by the Hamiltonian formalism.
We show in the appendix that this formalism also leads to 
Eq.(\ref{wh1}) under the same assumptions.
The density matrix formalism is particularly useful if one
wants to keep track of the various number densities. 
In this more general case, it is very difficult to solve 
the system analytically and so far this more general case has only
been studied numerically.

In the static limit where $\lambda, \bar \lambda$ are approximately 
constant, it is straightforward to show that
Eq.(\ref{whhhh1}) reduces to Eq.(\ref{eeqq}) with $x$ given by 
\begin{equation}
x = {1 \over 4}\Gamma_{\nu_{\alpha}}^2\left(
{2p \over \delta m^2}\right)^2,
\label{xxz}
\end{equation}
rather than by Eq.(\ref{jjds}) [note that Eq.(\ref{xxz}) reduces to 
Eq.(\ref{jjds}) for most of the parameter space of interest except 
for quite low temperatures].  This difference is due to the 
fact that in deriving Eq.(\ref{whhhh1}) we have made the assumption 
Eq.(\ref{smallen}). Because Eq.(\ref{eeqq}) is much simpler than 
Eq.(\ref{wh1}), it is particularly useful to determine the region 
of parameter space where the static limit [Eq.(\ref{eeqq})] is an 
acceptable approximation.  We now study this issue.

Expand $\lambda_{t'}$ (note that we are using
the notation that $\lambda_x$ denotes $\lambda$ 
evaluated at the point $x$) in a Taylor series around the 
point $t'= t$, that is
\begin{equation}
\lambda_{t'} = \lambda_t + [t' - t]\left({d 
\over dt'}\lambda\right)_{t} + ...
\label{tay}
\end{equation}
Using this Taylor series, the integrals $\int^t_{t-\tau}
\lambda dt'$ can be expanded as follows
(with a similar expansion for $\int^t_{t - \tau}
\bar \lambda dt'$),
\begin{equation}
\int^t_{t-\tau} \lambda dt' =
\lambda_t \tau - {\tau^2 \over 2}\left({d  \over
dt'}\lambda\right)_t  + ....
\label{tay1}
\end{equation}
The static approximation will be valid provided that
\begin{equation}
\langle \cos \int^t_{t - \tau} \lambda dt'\rangle - 
\langle \cos \int^t_{t - \tau} \bar\lambda dt'\rangle \simeq 
\langle \cos \tau \lambda_t \rangle
- \langle \cos \tau \bar \lambda_t \rangle.
\label{hap1}
\end{equation}
Using the expansion Eq.(\ref{tay1}), observe that
\begin{equation}
\langle \cos \int^t_{t - \tau} \lambda dt'\rangle = 
\langle \cos \tau \left[ \lambda_t - {\tau \over 2}
\left({d \lambda \over dt'}\right)_t + ...\right]\rangle.
\label{xx}
\end{equation}
The above equation can be used to determine the
region of validity of the static approximation Eq.(\ref{hap1}).
The region of validity of Eq.(\ref{hap1}) depends on the values 
of the parameters $\lambda, \bar \lambda$. There are 
essentially four regions to consider.
\vskip 0.3cm
\noindent
(a) $\omega_0 |\lambda_t|, \ \omega_0 |\bar \lambda_t| 
\stackrel{>}{\sim} 1$. In this region, Eq.(\ref{hap1}) 
is approximately valid provided that
\begin{equation}
|{\omega_0 \over 2}\left({d\lambda \over dt}\right)_t| 
\stackrel{<}{\sim} |\lambda_t|,\
|{\omega_0 \over 2}\left({d\bar \lambda \over dt}\right)_t| 
\stackrel{<}{\sim} |\bar \lambda_t|.
\label{ess1}
\end{equation}
\vskip 0.3cm
\noindent
(b) $\omega_0 |\lambda_t| \simeq 0, \ \omega_0 |\bar \lambda_t| 
\stackrel{>}{\sim} 1$. In this region, Eq.(\ref{hap1}) 
is approximately valid provided that Eq.(\ref{ess1}) holds and
\begin{equation}
\langle \cos \int^t_{t - \tau} \lambda dt' \rangle \simeq 0.
\end{equation}
From Eq.(\ref{xx}) this equation implies that
\begin{equation}
|{\omega_0^2 \over 2}\left({d\lambda \over dt}\right)_t| 
\stackrel{<}{\sim} 1.
\label{ess2}
\end{equation}
\vskip 0.3cm
\noindent
(c) $\omega_0 |\bar \lambda_t| \simeq 0, \ \omega_0 |\lambda_t| 
\stackrel{>}{\sim} 1$. In this region, Eq.(\ref{hap1}) is 
approximately valid provided that Eq.(\ref{ess1}) holds and
\begin{equation}
|{\omega_0^2 \over 2}\left({d\bar \lambda \over dt}\right)_t| 
\stackrel{<}{\sim} 1.
\label{ess3}
\end{equation}
\vskip 0.3cm
\noindent
(d) $\omega_0 |\lambda_t| \simeq 0, \ \omega_0 |\bar \lambda_t| 
\simeq 0$. In this region, Eq.(\ref{hap1}) 
can never be a strictly valid approximation because
the right-hand side of Eq.(\ref{hap1}) is zero at this point.
Note however that the static approximation will be
acceptable provided that the left-hand side of Eq.(\ref{hap1})
is small at this point, which is true if
Eq.(\ref{ess2}) and Eq.(\ref{ess3}) are valid.

Observe that Eqs.(\ref{ess2},\ref{ess3}) are more stringent
than Eq.(\ref{ess1}). Evaluating Eq.(\ref{ess2}) at the resonance, 
we find
\begin{equation}
|{\omega_0^2 \over 2}{d \over dt'}\left[ {\delta m^2 \over 2p}(
\cos2\theta_0 - b + a)\right]| \stackrel{<}{\sim} 1.
\label{above}
\end{equation}
For Eq.(\ref{ess3}) we only need to replace $a \to -a$ in
the above equation.  Assuming that there is no
accidental cancellation between the various independent
terms, Eq.(\ref{above}) implies,
\begin{equation}
|{\delta m^2 \over 2p}{6b \over T}{dT
\over dt}| \stackrel{<}{\sim} {\Gamma^2_{\nu_{\alpha}} \over 2},\
|{da \over dT}| \stackrel{<}{\sim}
|{\Gamma^2_{\nu_{\alpha}} \over 2} {2p \over \delta m^2}
{dt \over dT}|,
\label{hohum}
\end{equation}
where we have used $\partial b/\partial T = 6b/T$ and recall that
$\omega_0 = 2/\Gamma_{\nu_{\alpha}}$. In deriving Eq.(\ref{hohum})
we have also neglected a term proportional to $(\cos2\theta_0 - b + a)$
which is less stringent than Eq.(\ref{hohum}) because $(\cos 2\theta_0 - 
b + a) \approx 0$ is just the resonance condition.  The first 
condition in Eq.(\ref{hohum}) is satisfied provided that 
\begin{equation}
T \stackrel{>}{\sim} \left({11\delta m^2 \cos 2\theta_0\over 
M_P y_{\alpha}^2G_F^4}
\right)^{1 \over 9} \simeq 11 \left({\delta m^2 \over eV^2}
\right)^{1 \over 9}\ {\rm MeV},
\label{abgh}
\end{equation}
where we have set $b = \cos2\theta_0$ (which leads to the
most stringent condition) and we have used $dT/dt \simeq -5.5T^3/M_P$. 
In order to evaluate the second condition in Eq.(\ref{hohum}), 
observe that
\begin{equation}
{da \over dT} = {\partial a \over \partial L_{\nu_{\alpha}}}
{\partial L_{\nu_{\alpha}} \over \partial T} + {\partial a \over
\partial T}.
\end{equation}
Assuming that there is no accidental cancellation between the
two terms on the right-hand-side of the above equation, the 
second term in Eq.(\ref{hohum}) implies the following conditions
at the resonance,
\begin{equation}
|{\partial a \over \partial T}| \stackrel{<}{\sim}
|{\Gamma^2_{\nu_{\alpha}} \over 2}{2p \over \delta m^2}
{dt \over dT}|,\
|{\partial a \over \partial L_{\nu_{\alpha}}} {\partial L_{\nu_{\alpha}}
\over \partial T}| \stackrel{<}{\sim}
|{\Gamma^2_{\nu_{\alpha}} \over 2}{2p \over \delta m^2}
{dt \over dT}|.
\label{uio}
\end{equation}
Using $\partial a/\partial T \simeq 4a/T$, and $a \simeq 1$, then the 
first equation above gives approximately the same condition as the 
first equation in Eq.(\ref{hohum}). The second condition in 
Eq.(\ref{uio}) gives a condition on the rate of change of lepton number 
at the resonance.  Expanding this equation out we find that
\begin{equation}
| {\partial L_{\nu_{\alpha}} \over \partial T}| \stackrel{<}{\sim}
|{\Gamma^2_{\nu_{\alpha}} \over 2}
{dt \over dT}{1 \over 2\sqrt{2}
G_Fn_{\gamma}}| \simeq 
{y_{\alpha}^2 M_P G_F^3 4.1 T^4 \over 22\sqrt{2}} \simeq
4\times 10^{-11} \left({T \over {\rm MeV}} \right)^4{1 \over {\rm MeV,}} 
\label{hohum2}
\end{equation}
where we have used $n_{\gamma} = 2\zeta (3) T^3/\pi^2 \simeq T^3/4.1$.
Note that we have also used Eq.(\ref{Gammas}) for the collision frequency.
Thus, for example, if we are interested in studying the region
where the lepton number is initially created, then a necessary
condition for Eq.(\ref{eeqq}) to be approximately valid is that 
the resonance must occur for temperatures satisfying Eq.(\ref{abgh}).
From Eq.(\ref{critT}) (with $\cos 2\theta_0 \simeq 1$),
this implies that
\begin{equation}
\delta m^2 \stackrel{>}{\sim} 9 \times 10^{-2} \ (5 \times 10^{-3})\ 
eV^2,
\label{76}
\end{equation}
for $\nu_{e} - \nu_s$ ($\nu_{\mu, \tau} - \nu_s$) oscillations.
The creation of $L_{\nu_{\alpha}}$ must also satisfy Eq.(\ref{hohum2})
at the resonance. This condition should be checked
when using Eq.(\ref{eeqq}) for self consistency.

Perhaps surprisingly, there is a significant region of
parameter space where the oscillations are not
adiabatic at the resonance [i.e. $\gamma \stackrel{>}{\sim} 1$
in Eqs.(\ref{zyx},\ref{zyx2})] but Eqs.(\ref{hohum}) are 
nevertheless satisfied.
This is possible because Eqs.(\ref{hohum}) are not equivalent
to the adiabatic conditions Eqs.(\ref{zyx},\ref{zyx2}).
This is because Eqs.(\ref{hohum}) arise from demanding that the
total contribution to $dL_{\nu_{\alpha}}/dt$ reduce approximately to the 
simple Eq.(\ref{eeqq}). Recall that the total contribution to
$dL_{\nu_{\alpha}}/dt$
can be separated into two distinct contributions: 
from oscillations due to collisions and 
from oscillations between collisions.
The adiabatic condition, on the other hand, is a necessary condition for
the contribution of $dL_{\nu_{\alpha}}/dt$ from collisions to
reduce to Eq.(\ref{eeqq}). Thus it turns out that in the region when
the system is both non-adiabatic and Eqs.(\ref{hohum}) are
satisfied, the modification to the equation for $dL_{\nu_{\alpha}}/dt$ 
from collisions which arises from the non-adiabaticity 
cancels with the extra contribution to $dL_{\nu_{\alpha}}/dt$
from oscillations between collisions.
This type of cancellation is more transparent in the Hamiltonian 
formalism (see the appendix).  

Finally, to illustrate the analysis of this section, consider
the examples given in Figures 1 and 2.
Recall that the solid and dashed lines correspond
to the density matrix Eqs.(\ref{jk}) and Eq.(\ref{eeqq}) respectively.
Observe that for the example in Figure 1 (which has $\delta m^2 = 
-1 \ eV^2$), Eq.(\ref{eeqq}) is not a very good approximation
at the resonance where the lepton number is initially created (although
it is a reasonable approximation for small $\sin^2 2\theta_0$). This is
because the lepton number is created so rapidly that Eq.(\ref{hohum2})
is not valid. However, for the example shown in Figure 2, where
$\delta m^2 = -1000 \ eV^2$, the temperature where
the lepton number is created is much higher. Observe that
Eq.(\ref{hohum2}) is not as stringent for high temperatures and
it is therefore not surprising that the static approximation is 
approximately valid for this case.  [Note that the result that the 
static approximation tends to be a good approximation at high 
temperatures can also be seen by observing that for high temperatures, 
$\omega_0 \to 0$, and in this limit, Eq.(\ref{hap1}) will be satisfied].
\vskip 0.5cm
\noindent
{\bf IV The Thermal Momentum distribution of the neutrino}
\vskip 0.5cm
\noindent
Hitherto we have made the assumption that the neutrinos 
are monochromatic. This assumption is not expected to hold
for the neutrinos in the early Universe. The momentum distribution
of these neutrinos will be the usual Fermi-Dirac distribution.
Note that the width of the initial resonance
in momentum space is much smaller than the spread of neutrino momenta.
This means that only a few of the neutrinos will be at
resonance at a given time. Also, not all of the neutrinos
will be creating lepton number.  Neutrinos in the region defined 
by $b^p > \cos2\theta_0$ (which includes part of the resonance)
destroy lepton number, and those in the region 
$b^p < \cos2\theta_0$ create lepton number.
The point where net lepton number is created only occurs when
the lepton number creating neutrino oscillations dominate
over the lepton number destroying oscillations.
Recall that in the unphysical case where all of the
neutrinos are assumed to be monochromatic,
all of the neutrinos enter the resonance at the same time,
where they all destroy lepton number if $b > \cos2\theta_0$,
or all create lepton number if $b < \cos2\theta_0$.
Clearly, the effect of the thermal spread of momentum will make the
creation of lepton number much smoother.
An important consequence of this is that there will be even larger 
regions of parameter space where the system is smooth 
enough so that the static approximation is valid and hence
Eq.(\ref{eeqq}) will be a good approximation (modified to
incorporate the momentum dependence).  

In the static limit, we can simply re-derive Eq.(\ref{eeqq}), 
assuming that the neutrino momenta form the usual Fermi-Dirac 
distribution.  In this case, 
Eq.(\ref{332}) becomes
\begin{eqnarray}
&n_{\gamma}\frac{dL_{\nu_{\alpha}}}{dt} \simeq 
{1 \over 2}\int \left[- \Gamma(\nu_{\alpha}
\to \nu_{s}) + \Gamma(\bar \nu_{\alpha} \to 
\bar \nu_{s})\right] 
(dn_{\nu_{\alpha}} - dn_{\nu_s} + dn_{\bar\nu_{\alpha}} -  
dn_{\bar \nu_s}) \nonumber \\
& - {1 \over 2}\int \left[ \Gamma(\nu_{\alpha}
\to \nu_{s}) + \Gamma(\bar \nu_{\alpha} \to 
\bar \nu_{s})\right] (dn_{\nu_{\alpha}} -  dn_{\nu_s} - dn_{\bar \nu_{\alpha}}
+ dn_{\bar \nu_s}), 
\label{io1}
\end{eqnarray}
where   
\begin{equation}
dn_{\nu_{\alpha}} = {1\over 2\pi^2}{p^2dp \over 1 + e^{(p-\mu)/T}},\
dn_{\bar \nu_{\alpha}} = {1 \over 2\pi^2}{p^2dp \over 1 + e^{(p+\mu)/T}},
\label{tues}
\end{equation}
and $dn_{\nu_s}, dn_{\bar \nu_s}$ are the differential number
densities for the sterile and anti-sterile neutrinos respectively.
In Eq.(\ref{tues})
$\mu$ is the chemical potential.  

The reaction rates can easily be obtained 
following a similar derivation as before [Eqns.(\ref{Gamma}-\ref{ab})],
but this time we keep the momentum dependence (rather
than setting $p = \langle p \rangle$). 
Doing this, we find the following equation 
for the rate of change of lepton number in the static limit:
\begin{equation}.
{dL_{\nu_{\alpha}} \over dt} =
{\pi^2 \over 4\zeta(3)T^3} \int {s^2 \Gamma^p_{\nu_{\alpha}} 
a^p (c - b^p)
(dn_{\nu_{\alpha}}^{+} - dn_{\nu_s}^+)
\over [x^p + (c - b^p + a^p)^2][x^p + (c - b^p -a^p)^2]
} + \Delta,
\label{eeqq1}
\end{equation}
where $\Delta$ is a small correction term
\begin{equation}
\Delta \simeq {- \pi^2 \over 8\zeta(3) T^3} \int
{s^2 \Gamma^p_{\nu_{\alpha}} [x^p + (a^p)^2 + (b^p - c)^2]
(dn_{\nu_{\alpha}}^{-} - dn_{\nu_s}^-) \over 
[x^p + (c - b^p + a^p)^2][x^p + (c - b^p -a^p)^2]}, 
\end{equation}
and $dn_{\nu_{\alpha}}^{\pm} \equiv dn_{\nu_{\alpha}} \pm dn_{\bar
\nu_{\alpha}}$.  Recall that $c \equiv \cos 2\theta_0,\ 
s \equiv \sin 2\theta_0$.
In these equations note that the quantities, $b^p, a^p, x^p, 
\Gamma^p_{\nu_{\alpha}}$ are all functions of momentum of the form:
\begin{equation}
b^p =  b {p^2 \over \langle p \rangle^2},\
a^p =  a  {p \over \langle p \rangle},\
x^p = s^2 + {\Gamma^2_{\nu_{\alpha}} \over 4}
\left({p \over \langle p \rangle}\right)^2
\left({2p\over \delta m^2}\right)^2,
\ \Gamma^p_{\nu_{\alpha}} = \Gamma_{\nu_{\alpha}} 
{p \over \langle p \rangle},
\label{rand2}
\end{equation}
where $a, b, \Gamma_{\nu_{\alpha}}$ are defined in 
Eqs.(\ref{ab3}, \ref{Gammas}).  Eq.(\ref{eeqq1}) reduces to 
Eq.(\ref{eeqq}), in the limit where all of the neutrino momenta 
are fixed to $p = \langle p \rangle$.
[Note that $a^p = a$ when $p = \langle p \rangle \simeq 3.15T$
and similarly for $b^p, x^p$ and $\Gamma^p_{\nu_{\alpha}}$].

Note that the chemical potential 
is related to the lepton number by the equation
\begin{equation}
n_{\gamma} L_{\nu_{\alpha}} \equiv 
n_{\nu_{\alpha}} - n_{\bar \nu_{\alpha}} =
{T^3 \over 6}\left({\mu \over T}\right) + {\cal O}(\mu^3).
\label{tues2}
\end{equation}
Using Eqs.(\ref{tues},\ref{tues2}) we find
\begin{eqnarray}
&dn_{\nu_{\alpha}}^+ =
{1 \over \pi^2} {p^2 dp \over 1 + e^{p/T}}
+ {\cal O}(L_{\nu_{\alpha}}^2),
\nonumber \\
&dn_{\nu_{\alpha}}^{-} =
n_{\gamma}L_{\nu_{\alpha}}{6 \over \pi^2 T^3} {p^2 e^{p/T}dp 
\over (1 + e^{p/T})^2}
+ {\cal O}(L^3_{\nu_{\alpha}}).
\end{eqnarray}
Thus substituting the above relations into Eq.(\ref{eeqq1}),
we find that
\begin{equation}
{dL_{\nu_{\alpha}} \over dt} =
{\pi^2 \over 4\zeta(3)T^3} \int^{\infty}_0 {s^2 \Gamma^p_{\nu_{\alpha}} 
a^p (c - b^p)\over
[x^p + (c - b^p + a^p)^2][x^p + (c - b^p -a^p)^2]
} \left[ {p^2  \over \pi^2(1 + e^{p/T})} - 
{dn^+_{\nu_s} \over dp}\right] dp + \Delta,
\label{eeqq2}
\end{equation}
where $\Delta$ is a small correction term
\begin{equation}
\Delta \simeq {- \pi^2 \over 8\zeta(3) T^3} \int^{\infty}_0
{s^2 \Gamma^p_{\nu_{\alpha}} [x^p + (a^p)^2 + (b^p - c)^2]\over 
[x^p + (c - b^p + a^p)^2][x^p + (c - b^p -a^p)^2]
} \left[ {12\zeta(3) L_{\nu_{\alpha}} p^2 e^{p/T} \over  
\pi^4 (1 + e^{p/T})^2} - {dn^-_{\nu_s} \over dp}\right] dp. 
\end{equation}
Eq.(\ref{eeqq2}) can be integrated numerically to obtain
$L_{\nu_{\alpha}}$ as a function of time (or temperature).
We will give some examples in the next section after we
discuss how to calculate the sterile neutrino number distributions.

The main effect of the thermal spread of neutrino momenta is
to make the generation of lepton number much smoother.
From a computational point of view, this is very fortunate.
This is because Eq.(\ref{eeqq2}), like Eq.(\ref{eeqq}), is
only valid provided the lepton number generation is sufficiently
smooth (see the previous section for a detailed discussion of
this point).  In particular, Eq.(\ref{eeqq2}) should be
a much better approximation to reality at the resonance
where significant lepton number is initially generated. 

To complete this section, we comment on the rate of change of 
lepton number due to ordinary-ordinary neutrino oscillations.
For definiteness consider $\nu_e - \nu_{\mu}$ oscillations.
The rate of change of $L_{\nu_{\mu}} - L_{\nu_e}$ is given by
\begin{eqnarray}
&{n_{\gamma} \over 2}{d(L_{\nu_{\mu}} - L_{\nu_e}) \over dt} =
-\int \Gamma(\nu_{\mu} \to \nu_e) dn_{\nu_{\mu}}
+ \int \Gamma(\bar \nu_{\mu} \to \bar \nu_e)
dn_{\bar \nu_{\mu}}
\nonumber \\
& + \int \Gamma(\nu_{e} \to \nu_{\mu}) dn_{\nu_{e}}
- \int \Gamma(\bar \nu_{e} \to \bar \nu_{\mu})
dn_{\bar \nu_{e}}.
\label{hum}
\end{eqnarray}
Using $\Gamma (\nu_{\mu} \to \nu_e) = \Gamma (\nu_e \to \nu_{\mu})$
(and similarly for the anti-neutrino rates), Eq.(\ref{hum}) becomes
\begin{eqnarray}
&{n_{\gamma} \over 2}{d(L_{\nu_{\mu}} - L_{\nu_e})\over dt}
= -\int^{\infty}_0 \Gamma(\nu_{\mu} \to \nu_e) 
\left({dn_{\nu_{\mu}}\over dp} - {dn_{\nu_{e}} \over dp}\right) dp 
\nonumber \\
& + \int^{\infty}_0 \Gamma(\bar \nu_{\mu} \to \bar \nu_e) 
\left({dn_{\bar \nu_{\mu}}\over dp} - 
{dn_{\bar \nu_{e}} \over dp}\right) dp,
\label{55}
\end{eqnarray}
where
\begin{eqnarray}
&{dn_{\nu_{e}} \over dp} =
{1 \over 2\pi^2} {p^2 \over 1 + e^{(p-\mu_1)/T}},\
{dn_{\bar \nu_{e}} \over dp} =
{1 \over 2\pi^2}{p^2 \over 1 + e^{(p+\mu_1)/T}}, \nonumber \\
&{dn_{\nu_{\mu}} \over dp} =
{1 \over 2\pi^2} {p^2 \over 1 + e^{(p-\mu_2)/T}},\
{dn_{\bar \nu_{\mu}} \over dp} =
{1 \over 2\pi^2}{p^2 \over 1 + e^{(p+\mu_2)/T}}.
\end{eqnarray}
The chemical potentials $\mu_{1,2}$ are related to the lepton
numbers $L_{\nu_{e,\mu}}$ by the equations
\begin{equation}
{\mu_1 \over T} =  {6 n_{\gamma} \over T^3}L_{\nu_e},\
{\mu_2 \over T} = {6 n_{\gamma} \over T^3} L_{\nu_{\mu}},
\end{equation}
where we have assumed that $\mu_i/T \ll 1$.  Using these relations  
and expanding out Eq.(\ref{55}) (again assuming that 
$\mu_i/T \ll 1$) we find to leading order that
\begin{equation}
{d(L_{\nu_{\mu}} - L_{\nu_e}) \over dt} 
\simeq -{6(L_{\nu_{\mu}} - L_{\nu_{e}}) 
\over \pi^2 T^3} \int^{\infty}_0 {p^2 e^{p/T} \over (1 + e^{p/T})^2}
\left[\Gamma(\nu_{\mu} \to \nu_e) + \Gamma(\bar \nu_{\mu} 
\to \bar \nu_e)\right]dp.
\label{rand}
\end{equation}
From the above equation we see that $L_{\nu_{\mu}} - L_{\nu_e}$ 
always evolves such that $(L_{\nu_{\mu}} - L_{\nu_e}) \to 0$.
Also note that Eq.(\ref{rand}) shows that the rate of change of
lepton number due to ordinary-ordinary neutrino oscillations
is generally smaller than the rate of change of lepton number
due to ordinary - sterile neutrino oscillations (assuming
$L_{\nu_{\alpha}} \ll 1$). [Actually Eq.(\ref{rand}) 
has a strength comparable to the correction term
$\Delta$ for ordinary-sterile neutrino oscillations Eq.(\ref{delta}), 
although note that the mixing angle between ordinary neutrinos
can be significantly larger than the mixing angle between ordinary
and sterile neutrinos].

For ordinary-ordinary neutrino oscillations,
neutral current interactions do not collapse the wavefunction
because they cannot distinguish different neutrino species.
Only the charged current interactions can distinguish
the neutrino species. For example, for temperatures $1 \ {\rm MeV}
\stackrel{<}{\sim} T \stackrel {<}{\sim} 30 \ {\rm MeV}$, there are near
equilibrium number densities of electrons and positrons.  The number of
muons and anti-muons will be much less than the number of electrons
and positrons, and we will neglect them (actually
these are important for $\nu_{\tau} -\nu_{\mu}$ oscillations).
The rate at which charged current interactions occur
is given approximately by $\Gamma_{\nu} \sim |\Gamma_{\nu_e} - 
\Gamma_{\nu_{\mu}}|\simeq y_e' G_F^2 T^5$, where
$y_e'\sim y_e - y_{\mu}$ [$\simeq 1.1$ see Eq.(\ref{Gammas})].  
Also note that anti-neutrino -neutrino\cite{markbruce}
and neutrino - neutrino\cite{Pant} forward scattering amplitudes 
induce off diagonal contributions to the effective potential.
[Note that these contributions do not occur for the effective
potential governing ordinary - sterile neutrino oscillations].
It would be necessary to include these effects in-order
to evaluate the reaction rates. We leave this as a
take home exercise for the reader.
\vskip 0.5cm
\noindent
{\bf V. The effects of non-negligible sterile neutrino number 
densities and the parameter space for large lepton number asymmetry 
generation}
\vskip 0.5cm
\noindent
In this section we will do three things.
We will study the effects of non-negligible sterile
neutrino number densities, which can arise for the case of
relatively large, or even moderate values of $\sin^2 2\theta_0$.  
We will examine the parameter space where significant 
generation of lepton number occurs. Finally, we will obtain the 
BBN bound on the parameter space for $\nu_{\alpha}-\nu_s$ 
oscillations with $\delta m^2 <0,$ and with $|\delta m^2| 
\stackrel{>}{\sim} 10^{-4} \ eV^2, \ \sin^2 2\theta_0 
\stackrel{<}{\sim} 10^{-2}$. 

There are several ways in which the creation of lepton number(s)
can prevent the sterile neutrinos from coming into
equilibrium. One way is that one set of oscillations $\nu_{\alpha}-
\nu_{s}$ creates $L_{\nu_{\alpha}}$. The lepton number
$L_{\nu_{\alpha}}$ can then suppress other, independent
oscillations such as $\nu_{\beta} - \nu_{s}$ oscillations (with
$\beta \neq \alpha$) for example.  A more direct, but less dramatic 
way in which the creation of lepton number can help
prevent the sterile neutrinos from coming into
equilibrium, is that the lepton number generated from
say $\nu_{\alpha}-\nu_{s}$ oscillations  itself suppresses the 
$\nu_{\alpha}-\nu_{s}$ oscillations\cite{kc}. 
We will examine the latter effect here (some examples of the former
effect will be studied in the next section). Previous work\cite{B, B2,
B3, B4} obtained the BBN bound for large $|\delta m^2| \stackrel{>}{\sim} 
10^{-4} \ eV^2$ (with $\delta m^2 < 0$) and small 
$\sin^2 2\theta_0 \stackrel{<}{\sim} 10^{-2}$ which can be
approximately parametrized as follows\cite{B4}
\begin{equation}
|\delta m^2| \sin^4 2\theta_0 \stackrel{<}{\sim} 10^{-9} \ eV^2.
\label{blob}
\end{equation} 
This bound arises by assuming that the 
$\nu_{\alpha} - \nu_{s}$ oscillations do not bring the sterile
$\nu_s$ state into equilibrium. Note that this bound neglected
the creation of lepton number and it also did not include the effects
of the distribution of neutrino momentum. However, in the realistic 
case, the creation of $L_{\nu_{\alpha}}$ (after it occurs) will 
suppress the $\nu_{\alpha} - \nu_{s}$ oscillations and the actual 
bound would be expected to be somewhat less stringent than Eq.(\ref{blob}).

To proceed we will need to examine the effects of non-negligible 
sterile neutrino number densities.  The evolution of the number 
distribution of sterile neutrinos is governed by the rate equation
\begin{equation}
{d \over dt}\left[ {dn_{\nu_s}/dp \over dn_{\nu_{\alpha}}/dp} \right] =
\left[ {dn_{\nu_{\alpha}}/dp  - dn_{\nu_s}/dp \over
dn_{\nu_{\alpha}}/dp} \right]
\Gamma (\nu_{\alpha} \to \nu_s).
\label{rate}
\end{equation}
A similar equation holds for the number distribution of
sterile anti-neutrinos.  Introducing the notation, 
$z \equiv {dn_{\nu_s}/dp \over
dn_{\nu_{\alpha}}/dp}\ $ 
(for anti-neutrinos we use the corresponding notation,
$\bar z \equiv {dn_{\bar \nu_s}/dp \over
dn_{\bar \nu_{\alpha}}/dp}$), Eq.(\ref{rate}) becomes
\begin{equation}
{dz \over dt} = (1 - z)\Gamma (\nu_{\alpha} \to \nu_s)
 = {(1 - z)\over 4}{\Gamma^p_{\nu_{\alpha}} s^2 \over x^p + 
(c - b^p + a^p)^2}.
\label{rate2}
\end{equation}
The corresponding equation for anti-neutrinos can
be obtained by replacing $z \to \bar z$ and $a^p \to -a^p$ 
in the above equation.  In solving the above differential 
equation, we will assume the initial condition $z = 0$. We will
also assume that the number densities of the ordinary neutrinos are
given by their equilibrium values.  Note that the quantity $z$ depends
only on the reaction rates and is otherwise independent of the 
expansion. 

From the definition of $z$, it follows that 
$dn_s = z dn_{\nu_{\alpha}},
\ d\bar n_s = \bar z dn_{\bar \nu_{\alpha}}$. Thus, from
Eq.(\ref{eeqq2}),
\begin{equation}
{dL_{\nu_{\alpha}} \over dt} \simeq
{1 \over 4\zeta(3)T^3} \int^{\infty}_0 {s^2 \Gamma^p_{\nu_{\alpha}} 
a^p (c - b^p)\over
[x^p + (c - b^p + a^p)^2][x^p + (c - b^p -a^p)^2]
} {(1 - z^+)p^2 dp \over 1 + e^{p/T}} + \Delta,
\label{eeqq3}
\end{equation}
where $\Delta$ is a small correction term
\begin{equation}
\Delta \simeq { 1 \over 8\zeta(3) T^3} \int^{\infty}_0
{s^2 \Gamma^p_{\nu_{\alpha}} [x^p + (a^p)^2 + (b^p - c)^2]\over 
[x^p + (c - b^p + a^p)^2][x^p + (c - b^p -a^p)^2]
} {z^- p^2 dp \over  1 + e^{p/T}},
\end{equation}
with $z^{\pm} \equiv (z \pm \bar z)/2$ and we have neglected 
a small correction term which is proportional to $L_{\nu_{\alpha}}$.
Note that Eq.(\ref{rate2}) and Eq.(\ref{eeqq3}) must be solved
simultaneously.
 
For the numerical work, the continuous variable $p/T$ is replaced
by a finite set of momenta $p_n/T$ (with $n = 0, 1, ..., N$) and
the integral over momentum in Eq.(\ref{eeqq3})
is replaced by the sum of a finite number of terms.
Correspondingly, the variable $z(t, p/T)$ is replaced by the 
set of variables, $z_n (t)$, where the evolution of
each variable $z_n (t)$ is governed by the
differential equation, Eq.(\ref{rate2}) [with $p/T = p_n/T$ for
$z = z_n(t)$, $n = 0,1,...,N$].
Thus, the single differential equation, Eq.(\ref{rate2}) is
replaced by a set of $N$ differential equations, one for each
momentum step.  These differential equations, together with 
Eq.(\ref{eeqq3}), are coupled differential equations which 
must be integrated simultaneously. 

We now illustrate the creation of lepton number as governed by
Eqs.(\ref{eeqq3},\ref{rate2}) with some examples.
We have numerically integrated Eqs.(\ref{eeqq3},\ref{rate2})
for the following parameter choices. 
In Figure 3 we have considered $\nu_{\mu, \tau} - \nu_s$
oscillations with the parameter choice
$\delta m^2 = -1 \ eV^2$, $\sin^2 2\theta_0 = 10^{-4}\
$ (dashed line), $\sin^2 2\theta_0 = 10^{-6}\ $ (dashed-dotted
line) and $\sin^2 2\theta_0 = 10^{-8} \ $ (solid line). 
Figure 4 is the same as Figure 3, except that
$\delta m^2 = -1000 \ eV^2$, $\sin^2 2\theta_0 = 10^{-6}\
$ (dashed line), $\sin^2 2\theta_0 = 10^{-7}\ $ (dashed-dotted
line) and $\sin^2 2\theta_0 = 10^{-9}$ (solid line).
In both examples we have assumed that the initial lepton asymmetry
is zero. Recall that the generation of lepton number is essentially 
independent of the initial lepton number asymmetry provided that it 
is less than about $10^{-5}$ (for more discussion about this
point see section 2).  Note that for convenience we have 
plotted $|L_{\nu_{\alpha}}|$.  The lepton asymmetry $L_{\nu_{\alpha}}$
changes sign at the point where it is created. Before this
point $L_{\nu_{\alpha}}$ evolves such that it has the
opposite sign to $\eta$ while for evolution subsequent to the
point where $L_{\nu_{\alpha}}$ is initially created,
$L_{\nu_{\alpha}}$ has the same sign as $\eta$. Recall
that this behaviour is expected (see the earlier discussion
in section II).

In these examples, the generation of lepton number is considerably
smoother than in the earlier case where the momentum
distribution was neglected (see Figures 1,2).
For this reason, it turns out that throughout most of the evolution
of $L_{\nu_{\alpha}}$, the rate of change of $L_{\nu_{\alpha}}$ 
satisfies the condition Eq.(\ref{hohum2}) and thus Eq.(\ref{eeqq2})
should be approximately valid (except at quite low temperatures
where the MSW effect will be important).

In order to gain insight into the effects of the neutrino
momentum distribution, it is useful to compare Figures 3,4 
(which incorporate the neutrino momentum distribution) with
the Figures 1,2 (where all of the momentum of all of the
neutrinos were set equal to the mean momentum).
Qualitatively, there is not a great deal of difference.
However there are several very important effects, which
we summarize below.
\vskip 0.3cm
\noindent
(1)  For the examples with relatively small $\sin^2 2\theta_0$,
lepton number creation generally begins somewhat earlier 
(i.e. at a higher temperature) than in the case where momentum 
distribution is neglected.  For the examples shown in Figures 3,4
with $\delta m^2 = -1\ eV^2, \ \sin^2 2\theta_0 = 10^{-8}$ 
($-1000 \ eV^2, \ \sin^2 2\theta_0 = 10^{-9}$), lepton number 
is created when $T \simeq 20$ MeV ($T \simeq 65$ MeV).   This can 
be compared with the simplistic case where the neutrino momentum 
distribution was neglected.  In this case, we see from Figures 1, 2 
that lepton number creation begins at $T \simeq 16.0$ MeV 
($T \simeq 50$ MeV) for $\delta m^2 = -1\ eV^2$ ($-1000\ eV^2$).
The fact that the critical temperature
increases can be explained rather simply. Note that the neutrino
number density distribution peaks at about $p \simeq 2.2T$, 
which should be compared with the average momentum of about 
$3.15T$ used in Figures 1 and 2. Using the former
approximation instead of the latter leads the critical temperature 
to increase by about $12\%$. This explains qualitatively why the critical
temperature increases. Note, however, that 
the accurate numerical calculations
displayed in Figures 3 and 4 actually show that the temperature increases
by more than this, and also that the temperature increase depends
on the mixing angle.
\vskip 0.3cm
\noindent
(2)  For the examples with large $\sin^2 2\theta_0$,
the point where significant generation of lepton number is created
occurs much later than in the examples with small $\sin^2 2\theta_0$.
The reason for this is that for large $\sin^2 2\theta_0$, the
number density of sterile neutrinos is larger. 
In the region before significant lepton
number is generated, $a \simeq 0$ and all of the neutrino
oscillations with $b^p < \cos 2\theta_0$ have already passed 
through the resonance while the neutrino oscillations with 
$b^p > \cos 2\theta_0$ have yet to pass through the resonance. 
Since the creation of sterile neutrinos is dominated by the 
oscillations at the resonance, it follows that the sterile neutrino 
number distribution with momenta in the region where 
$b^p < \cos 2\theta_0$ will be much greater than for sterile neutrinos 
with momenta in the region where $b^p > \cos2\theta_0$.
Thus, from Eq.(\ref{eeqq3}), the lepton number
creating oscillations (with $b^p < \cos 2\theta_0$) are
suppressed if the number density of sterile neutrnos is non-negligible,
as occurs for large $\sin^2 2\theta_0$. The lepton number destroying
oscillations (with $b^p > \cos 2\theta_0$), on the other hand, 
are not suppressed because the number
of sterile neutrinos with $b^p > \cos 2\theta_0$ are negligible.

\vskip 0.3cm
\noindent
(3) The creation of lepton number is considerably smoother in
the realistic case. For instance, in the 
example where $\delta m^2 = -1\ eV^2$, $\sin^2 2\theta_0 = 
10^{-8}$, in the realistic case (shown in Figure 3), $L_{\nu_{\mu}}$ 
goes from $10^{-10}$ to $10^{-6}$ in about $\Delta T \approx 1$ MeV, 
whereas in the unrealistic case where the neutrino momentum distribution 
was neglected (shown in Figure 1), $|L_{\nu_{\mu}}|$ goes from 
$10^{-10}$ to $10^{-6}$ in about $\Delta T \approx 0.005$ MeV. 
\vskip 0.3cm
\noindent
(4) At low temperatures, the lepton number gets ``frozen'' 
at an earlier time. For example, in the 
case where $\delta m^2 = -1\ eV^2$ and $\sin^2 2\theta_0 =
10^{-8}$, with momentum dependence (Figure 3),
the final value for the lepton number
is $\sim 4 \times 10^{-4}$, whereas in the unrealistic case without
the neutrino momentum distribution,
the final value for the lepton number for this example (Figure 1)
is $\sim 10^{-1}$.
As discussed briefly in section II, this effect is
expected because the temperature where the lepton number
gets frozen occurs when the rate of change of the
variable $a$ due to the expansion of the Universe 
dominates over the rate of change of $a$ due to
neutrino oscillations. In the realistic case where 
the momentum distribution is taken into account, the
maximum value of the rate of change of $a$ due
to neutrino oscillations is suppressed because only
a small fraction of the neutrinos will be at the resonance.

\vskip 0.5cm
This last point suggests that the momentum distribution cannot
be ignored if one is interested in finding out the 
precise final value of the lepton number generated.
However, note that Eq.(\ref{eeqq2}) does not incorporate
flavour conversion due to the MSW effect [see assumption (4)
in section II for some discussion about this point].
The effect of the MSW flavour conversion should be
to be to keep $a \simeq 1$ for lower temperatures.
This means that the final value of $L_{\nu_{\alpha}}$ should
be significantly larger than suggested by Figures 3,4.
This effect will need to be incorporated if one wants to 
calculate the precise value of the final lepton number generated.
[The precise value of the final lepton number can be
obtained by numerically integrating the density matrix
equations Eqs.(\ref{jk}) suitably modified to
incorporate the neutrino momentum distribution]. 
In particular, if one is interested in working out
the region of parameter space where the electron lepton
number is large enough to affect BBN through nuclear reaction
rates, then the final value of the electron lepton number
is very important\cite{idono,fn444}.

Note that we can check Eq.(\ref{blob}) by numerically integrating 
Eq.(\ref{rate2}) and Eq.(\ref{eeqq2}) assuming for definiteness that 
$\rho_{\nu_{s}}/\rho_{\nu_{\alpha}} \stackrel{<}{\sim} 0.6$ (where 
the $\rho$'s are the energy densities). This leads to the following 
constraint on $\delta m^2, \sin^2 2\theta_0,$
\begin{equation}
\sin^2 2\theta_0 
\stackrel{<}{\sim} 2 (4) \times 10^{-5}\left[ {
eV^2 \over |\delta m^2|}\right]^{1 \over 2}, 
\label{gh}
\end{equation}
for $\nu_{e} -\nu_s$ ($\nu_{\mu,\tau} -\nu_s$) oscillations.
Thus, we see that Eq.(\ref{blob}) turns out to be a good 
approximation after all. This is basically due to the 
result that the creation of a non-negligible number of sterile
neutrinos has the effect of delaying the point where
significant lepton number is created [see point (2) above].

Finally, the region of parameter space where significant neutrino
asymmetries are generated by ordinary sterile neutrino
oscillations can be obtained by
integrating Eq.(\ref{eeqq3}) and Eq.(\ref{rate2}).
The result of 
this numerical work is that significant neutrino asymmetry 
($|L_{\nu_{\alpha}}| \stackrel{>}{\sim} 10^{-5}$) is
generated by ordinary - sterile neutrino oscillations for
the following region of parameter space
\begin{equation}
6 (5) \times 10^{-10} 
\left[ {eV^2 \over |\delta m^2|}\right]^{1 \over 6}
\stackrel{<}{\sim}\ \sin^2 2\theta_0 
\stackrel{<}{\sim} 2 (4) \times 10^{-5}\left[ {
eV^2 \over |\delta m^2|}\right]^{1 \over 2}, 
\ {\rm \ and\ } |\delta m^2| \stackrel{>}{\sim} 10^{-4} \ eV^2,
\label{ghh}
\end{equation}
for $\nu_e - \nu_s$ ($\nu_{\mu,\tau}-\nu_s$) oscillations.
Note that we have assumed that $\rho_{\nu_s}/\rho_{\nu_{\alpha}}
\stackrel{<}{\sim} 0.6$ [Eq.(\ref{gh})].
In the general case where no bound on $\rho_{\nu_s}/\rho_{\nu_{\alpha}}$
is assumed, the upper bound on $\sin^2 2\theta_0$
is considerably weaker.
For example, $\nu_{\mu, \tau} - \nu_s$ oscillations
with $\delta m^2 = -1\ eV^2, \ \sin^2 2\theta_0 = 10^{-4}$
violate the bound, Eq.(\ref{gh}) but still generate
significant neutrino asymmetry, as illustrated in Figure 3.
[For this particular example, we found that
$\rho_{\nu_s}/\rho_{\nu_{\alpha}} \simeq 0.86$].

The parameter space in Eq.(\ref{ghh}) can be compared with
previous work where the momentum dependence was neglected\cite{shi,ftv}.
As we have mentioned above, the upper bound on $\sin^2 2\theta_0$
which assumes a BBN bound of $\rho_{\nu_s}/\rho_{\nu_{\alpha}}
\stackrel{<}{\sim} 0.6$, is not modified much when the momentum
distribution of the neutrino is incorporated.
For the lower limit of $\sin^2 2\theta_0$, the effect of the
momentum dependence is to reduce the region of parameter
space by nearly two orders of magnitude.

Finally, it may be possible for significant neutrino asymmetries 
to be generated for $|\delta m^2| \stackrel{<}{\sim} 10^{-4} \ eV^2$, 
however the mechanism of production of these asymmetries is dominated by
oscillations between collisions (rather than the mechanism of
collisions) and tend to be oscillatory\cite{shi, ekm, kc}.

\vskip 0.5cm
\noindent
{\bf VI. Consistency of the maximal vacuum oscillation solutions 
of the solar and atmospheric neutrino problems with BBN}
\vskip 0.5cm
\noindent
We now turn to another application of the phenomenon of 
lepton number creation due to ordinary-sterile neutrino
oscillations. First, in the context of a simple explanation of the
solar neutrino problem which involves large angle $\nu_e - \nu_s$
oscillations, we will determine the conditions under which 
the lepton number produced from $\nu_{\alpha} - \nu_s$
oscillations can suppress the oscillations $\nu_{\beta}-\nu_s$
(where $\beta \neq \alpha$).  This allows the BBN bounds on 
ordinary-sterile neutrino oscillations to be evaded by many orders 
of magnitude, as we will show.  We begin by briefly reviewing the 
maximal vacuum oscillation solution to the solar neutrino problem
\cite{fv}.

One possible explanation of the solar neutrino problem is that
the electron neutrino oscillates maximally (or near maximally) with
a sterile neutrino (which we here denote as $\nu_e'$ rather than
as $\nu_s$ in order to remind the reader that this sterile neutrino
is approximately maximally mixed with $\nu_e$)\cite{fv,gp}. We will 
denote the $\delta m^2$ for $\nu_e - \nu_e'$ oscillations by $\delta
m^2_{ee'}$.  As is well known, for a large range of parameters \cite{fn1}
\begin{equation}
3 \times 10^{-10}\ eV^2 \stackrel{<}{\sim}\ |\delta m^2_{ee'}| 
\stackrel{<}{\sim}\ 10^{-3} \ eV^2,
\label{solar}
\end{equation}
maximal vacuum oscillations imply that the flux of electron
neutrinos from the sun will be reduced by a factor of two for
all neutrino energies relevant to the solar neutrino experiments.
We will call this scenario the ``maximal vacuum oscillation
solution'' to the solar neutrino problem.
It is a very simple and predictive scheme which can either
be ruled out or tested more stringently with the 
{\it existing experiments}. Importantly, it also makes definite
predictions for the new experiments, SNO, Superkamiokande and 
Borexino. Our interest in this scheme is also motivated by the 
exact parity symmetric model (see Ref.\cite{flv} for a review of
this model). This model predicts that ordinary neutrinos will be
maximally mixed with mirror neutrinos (which are approximately 
sterile as far as ordinary matter is concerned)
if neutrinos have mass\cite{flv}. If we make the assumption that 
the mixing between the generations is small (as it is in the
quark sector) then the exact parity symmetric model predicts that the
three known neutrinos will each be (to a good approximation) 
maximal mixtures of two mass eigenstates.
There are also other interesting models which
predict that the electron neutrino is approximately maximally
mixed with a sterile neutrino\cite{oth}.
The maximal mixing of the electron neutrino ($\nu_e$) and the
sterile neutrino will reduce the
solar neutrino flux by an energy independent factor of two for the
large range of parameters given in Eq.(\ref{solar}).
This leads to definite  {\it predictions} for the expected solar
neutrino fluxes for the existing experiments. In Ref.\cite{fv}, we
compared these predictions with the existing experiments.
We summarize the results of that exercise in Table 1 which
we have updated to include the most recent data\cite{fn34}.

Note that in Table 1, the Kamiokande experiment has been used
as a measurement of the Boron flux\cite{bb, kp, kam}.
This is a sensible way to analyse the data (but not the only
way of course) because the flux of neutrinos coming from
this reaction chain is difficult to reliably calculate\cite{fn66}.
Clearly, the simple energy independent flux reduction by a factor
of two leads to predictions which are in quite reasonable agreement
with the data. If the minimal standard model had given such good 
predictions, few would have claimed that there is a solar neutrino problem.

Note that the maximal vacuum oscillation solution is distinct from 
the ``just so'' large angle vacuum oscillation solution\cite{js}. In 
the ``just so''  solution, the electron neutrino oscillation 
length is assumed to be about equal to the distance between the 
earth and the sun (which corresponds to $|\delta m^2| \simeq 10^{-10} \ 
eV^2$).  In this case the flux of neutrinos depends sensitively on
$\delta m^2$ and it is possible to fit the data to the free
parameters $\delta m^2, \ \sin^2 2\theta_0$\cite{js}.
The advantage of doing this is that a good fit to the data
can be obtained (however this is not so surprising since
there are two free parameters to adjust).
The disadvantage is that fine tuning is required and predictivity
is lost because of the two free parameters.  The maximal mixing 
solution on the other hand assumes maximal mixing and that 
$\delta m^2$ is in the range Eq.(\ref{solar}).  For this parameter 
range there is an energy independent flux reduction by a factor 
of two.  The advantage of this possibility is that it does not 
require fine tuning and it is predictive.  A consequence of this 
is that it is testable with the existing experiments.
The disadvantage of this scenario is that it does not give 
a perfect fit to the data.  However, in our opinion the predictions 
are in remarkably good agreement with the data given the simplicity 
and predictivity of the model.

With the range of parameters in Eq.(\ref{solar}) there is a 
potential conflict with BBN\cite{bbnrev,fn35}.  
For maximally mixed $\nu_{e}$ and $\nu'_{e}$ neutrinos, 
the following rather stringent BBN bound 
has been obtained {\it assuming that the lepton number
asymmetry could be neglected}\cite{B,B2,B3,B4}:
\begin{equation}
|\delta m^2_{ee'}| \stackrel{<}{\sim} 10^{-8} \ eV^2.
\label{naive}
\end{equation}
This bound arises by requiring that the sterile neutrinos do not
significantly modify the successful BBN calculations.
For temperatures above the kinetic decoupling temperature 
the requirement that the sterile neutrinos do not come into
equilibrium implies the bound $|\delta m^2_{ee'}| 
\stackrel{<}{\sim} 10^{-6}
\ eV^2$. Smaller values of $\delta m_{ee'}^2$ in the range
$10^{-8} \stackrel{<}{\sim} |\delta m_{ee'}^2|/eV^2 \stackrel{<}{\sim}
10^{-6}$ can be excluded because the oscillations deplete
the number of electron neutrinos (and anti-neutrinos) after
kinetic decoupling (so that they cannot be replenished).
The depletion of electron neutrinos increases the He/H primordial
abundance ratio.  This is because the temperature where the ratio of 
neutrons to protons freezes out is increased if there are less electron
neutrinos around.  For $|\delta m^2_{ee'}| 
\stackrel{<}{\sim} 10^{-8} \ eV^2$, 
the oscillation lengths are too long to have any significant effect
on the number densities of electron neutrinos during
the nucleosynthesis era. 
If the bound in Eq.(\ref{naive}) were valid then it would 
restrict much of the parameter space for the maximal vacuum oscillation
solution of the solar neutrino problem. However, this bound does not
hold if there is an appreciable lepton number asymmetry in the early 
Universe for temperatures between $1-100$ MeV\cite{fv1}.
This is because the generation of significant lepton number $L^{(e)}$
implies that the quantity $a_{ee'}$ [which is the $a$ parameter
defined in Eq.(\ref{ab3}) with $\delta m^2 = \delta m^2_{ee'}$]
is very large thereby suppressing the oscillations [note that
for $a_{ee'} \gg 1$, $\sin^2 2\theta_m \ll \sin^2 2\theta_0$, see
Eq.(\ref{bel})]. We will now show in detail how the creation of lepton 
number can relax the BBN bound Eq.(\ref{naive}) by many orders of
magnitude.

We will assume that the various oscillations can be approximately
broken up into the pairwise oscillations $\nu_e - \nu'_e$,
$\nu_{\mu} - \nu_e'$ and $\nu_{\tau} - \nu'_e$.
We will denote the various oscillation parameters in a self-evident
notation,
\begin{equation}
b_{\alpha e'}, a_{\alpha e'} \ {\rm for \ } \nu_{\alpha} 
- \nu'_e \ {\rm oscillations,\ }
\end{equation}
where $\alpha = e, \mu, \tau$.
We will denote the mixing parameters, $\delta m^2, \ \sin^2 2\theta_0$ 
appropriate for $\nu_{\alpha} - \nu'_e$ oscillations by
$\delta m^2_{\alpha e'}, \ \sin^2 2\theta_0^{\alpha e'}$.
Note that lepton number cannot be created by $\nu_{\alpha} 
- \nu'_e$ oscillations until $b_{\alpha e'} < \cos2\theta_0^{\alpha
e'}$.  Recall that the $b$ parameter is inversely
proportional to $\delta m^2$ [see Eq.(\ref{ab3})].
Thus, the earliest point during the evolution of the Universe
where lepton number can be created due to ordinary - sterile
neutrino oscillations occurs for oscillations which have the largest
$|\delta m^2|$. Note that these oscillations must satisfy the
bound in Eq.(\ref{ghh}) if they are to produce lepton number,
and they should also satisfy the BBN bound Eq.(\ref{gh}) if we
demand that the sterile neutrino energy density be small
enough so that BBN is not significantly modified.
Note that the $\nu_e - \nu'_e$ oscillations have very small
$|\delta m^2_{ee'}| \stackrel{<}{\sim} 10^{-3}\ eV^2$\cite{fn1}, and
$\cos 2\theta_0^{ee'} \sim 0$ (assuming maximal or
near maximal mixing), and thus these oscillations 
themselves cannot produce significant lepton number.
However, the $\delta m^2$ for $\nu_{\tau} - \nu'_e$
or $\nu_{\mu} - \nu'_e$ oscillations can
have much larger $\delta m^2$ (and they should also have $\delta m^2 < 0$
if $m_{\nu_{\mu}}, m_{\nu_{\tau}} > m_{\nu'_e}$)\cite{fn345}.
We will assume for definiteness that $m_{\nu_{\tau}} > m_{\nu_{\mu}}
> m_{\nu'_e}$ so that $|\delta m^2_{\tau e'}| > |\delta m^2_{\mu e'}|$ 
and the $\nu_{\tau} - \nu'_e$ oscillations create $L_{\nu_{\tau}}$
first (with $L_{\nu_{\mu}}, L_{\nu_e}$ assumed to be initially
negligible). If $m_{\nu_{\mu}} > m_{\nu_{\tau}}$ then we only need to
replace $\nu_{\tau} $ by $\nu_{\mu}$ in the following analysis.

Thus, we will consider the system comprising  
$\nu_{\tau}, \nu_e$ and $\nu'_e$ (and their anti-particles).
Our analysis will be divided into two parts. First, we will
calculate the condition that the $L^{(e)}$ created by 
$\nu_{\tau} - \nu_e'$ oscillations survives without being
subsequently destroyed by $\nu_e - \nu'_e$ oscillations.
We will then establish the conditions under which $L^{(e)}$ is created
early enough and is large enough to suppress the 
$\nu_e - \nu'_e$ oscillations so that only a negligible 
number of $\nu_e'$ is produced.  

For simplicity we will first analyse the system neglecting the
momentum distribution of the neutrino. This is useful because 
under this assumption it turns out that this system can be 
approximately solved analytically
as we will show. We will then consider the realistic case where
the spread of momenta is taken into consideration.

It is important to observe that the generation of $L_{\nu_{\tau}}$ 
also leads to the generation of $L^{(e)}$[through Eq.(\ref{Lsuper})]. 
If we assume that negligible $L_{\nu_e}$ is generated, 
then $L^{(e)} \simeq L^{(\tau)}/2$.  However
$\nu_{e}-\nu_{e}' $ oscillations can potentially generate 
$L_{\nu_e}$ such that $L^{(e)} \to 0$. (Recall that $L^{(e)} 
\simeq 0$ is an approximately stable fixed point for the 
$\nu_e-\nu'_e$ system for temperatures greater than a few MeV).  
The effect of the $\nu_e - \nu'_e$ oscillations will be greatest when
the $\nu_e - \nu'_e $ oscillations are at resonance. 
If negligible $L_{\nu_e}$ is generated, then
$|a_{ee'}| \simeq R|a_{\tau e'}|/2$ and
$|b_{e e'}| \simeq R(A_e/A_{\tau})|b_{\tau e'}|$ (where
$R \equiv |\delta m^2_{\tau e'}/\delta m^2_{e e'}|$). 
Hence the $\nu_e - \nu'_e$ resonance condition
($a_{ee'} = b_{ee'}$), will be satisfied when 
\begin{equation}
|a_{\tau e'}| = 2(A_e/A_{\tau})|b_{\tau e'}|.  
\label{kls}
\end{equation}
Recall that the $\nu_{\tau} - \nu'_e$ oscillations generate 
$L_{\nu_{\tau}}$ such that 
\begin{equation}
a_{\tau e'} \simeq 1 - b_{\tau e'},
\label{kls2}
\end{equation}
where we have assumed that $\cos2\theta_0 ^{\tau e'} \sim 1$ and
that $L_{\nu_{\tau}} > 0 $ for definiteness.
Observe that Eqs.(\ref{kls},\ref{kls2}) imply that the system
inevitably passes through the $\nu_e - \nu_e'$ resonance.
This event will occur when
\begin{equation}
|b_{\tau e'}| \simeq {A_{\tau} \over A_{\tau} + 2A_{e}}.
\end{equation}
Using the definition of $b_{\tau e'}$ which can be 
obtained from Eq.(\ref{ab3}), the above equation can be
solved for the $\nu_e - \nu'_e$ resonance temperature
\begin{equation}
T_{res}^{ee'} 
\simeq \left[ {\delta m^2 M_W^2 4.1 \over 6.3\sqrt{2}
G_F A_{\tau}}{A_{\tau} \over A_{\tau} + 2 A_e}\right]^{1 \over 6}
\simeq 11\left(
{\delta m^2_{\tau e'} \over eV^2} \right)^{1 \over 6}\ {\rm MeV}.
\label{w}
\end{equation}
Thus, when $T = T^{ee'}_{res}$, the $\nu_{\tau} - \nu'_e$ oscillations
have created enough $L^{(e)}$ so that the $\nu_e - \nu'_e$ 
oscillations will be at the resonance,
assuming that negligible $L_{\nu_e}$ has been generated.
In general the $\nu_e - \nu'_e$ resonance 
temperature depends on both $L_{\nu_e}$ and $L_{\nu_{\tau}}$.
The $\nu_e - \nu'_e$ resonance condition $a_{ee'} = b_{ee'}$ implies
that the resonance temperature for $\nu_e - \nu'_e$ oscillations
is related to $L_{\nu_e}$ and $L_{\nu_{\tau}}$ by
the equation
\begin{equation}
T_{res}^{ee'} = \sqrt{{M_W^2 \over A_e} L^{(e)}}
\simeq \sqrt{{M_W^2 \over A_e}(2L_{\nu_e} + L_{\tau})},
\label{mm}
\end{equation} 
where we have neglected the small baryon and electron asymmetries 
and a possible mu neutrino asymmetry (we will discuss the effects 
of the mu neutrino later).
Thus the resonance temperature will change when $L_{\nu_e}$ and 
$L_{\nu_{\tau}}$ change due to oscillations.  

Let us consider the rate of change of the quantity $(T_{res}^{ee'} - T)$,  
\begin{equation} 
{d(T_{res}^{ee'} - T) \over dt} =
{\partial T^{ee'}_{res} \over \partial L_{\nu_e}}{\partial L_{\nu_e} \over
\partial t}
+ {\partial T^{ee'}_{res} \over \partial L_{\nu_{\tau}}}{\partial 
L_{\nu_{\tau}} \over \partial t} - {dT \over dt},
\label{y}
\end{equation}
evaluated at the temperature $T = T_{res}^{ee'}$.
Note that the first term on the right-hand side of Eq.(\ref{y}) 
represents the rate of change of $T_{res}^{ee'}$ due 
to $\nu_{e} - \nu'_e$ oscillations
(and its sign is negative), while the second term is the rate
of change of $T_{res}^{ee'}$ due to $\nu_{\tau}-\nu'_e$ oscillations
(and the sign of this term is positive).
The third term in Eq.(\ref{y}) is the rate of change of 
$(T_{res}^{ee'} - T)$ due to the expansion of the Universe 
(${-dT \over dt} \simeq 5.5T^3/M_P$) (this term is also positive
in sign).  Observe that if $d(T_{res}^{ee'}-T)/dt > 0$, then the system 
passes through the resonance without significant destruction of
$L^{(e)}$. If on the other hand, $d(T_{res}^{ee'}-T)/dt 
\stackrel{<}{\sim} 0$, then the position of the resonance moves
to lower and lower temperatures and $L^{(e)} \to 0$.
Thus, a sufficient condition that $L^{(e)}$ survives without being destroyed
by $\nu_e - \nu'_e$ oscillations is that
\begin{equation}
{\partial T^{ee'}_{res} \over \partial L_{\nu_e}}{\partial L_{\nu_e} \over
\partial t}
+ {\partial T^{ee'}_{res} \over \partial L_{\nu_{\tau}}}{\partial 
L_{\nu_{\tau}} \over \partial t} > {dT \over dt}.
\label{yyy}
\end{equation}
To evaluate $\partial L_{\nu_{\tau}}/\partial t$, observe that
\begin{equation}
{\partial L^{(\tau)} \over \partial t} = 2{\partial L_{\nu_{\tau}} 
\over \partial t} + {\partial L_{\nu_e} \over \partial t} 
\simeq {-4L^{(\tau)} \over T}{dT \over dt}
\label{uu}
\end{equation}
where we have assumed that $L^{(\tau)} \sim T^{-4}$ 
for $T = T^{ee'}_{res}$.
Of course this latter assumption only holds provided that the
$\nu_e - \nu'_e$ resonance does not occur while $L_{\nu_{\tau}}$ 
is still growing exponentially.
However, for $\sin^2 2\theta_0^{\tau e'}$ sufficiently large,
the $\nu_e - \nu'_e$ resonance can occur during
the rapid exponential growth phase of $L_{\nu_{\tau}}$.
If this happens then the rate at which the $\nu_{\tau} 
- \nu'_e$ oscillations move the system
away from the $\nu_e - \nu'_e$ resonance is much more rapid.
Consequently, the region of parameter space where $L^{(e)}$
survives without being destroyed by $\nu_e - \nu'_e$ oscillations
is significantly larger in this case (this effect will be illustrated
later on when we study the system numerically).

Thus using Eq.(\ref{uu}), Eq.(\ref{yyy}) can be written in
the form
\begin{equation}
{3 \over 4}{\partial L_{\nu_{e}} \over \partial t} {\partial T^{ee'}_{res} 
\over \partial L_{\nu_e}} \stackrel{>}{\sim} 
\left[1 + {L^{(\tau)} \over T}{\partial T_{res}^{ee'} \over
\partial L_{\nu_e}}\right]{dT \over dt},
\label{jj}
\end{equation}
where we have used the relation $\partial T_{res}^{ee'}/
\partial L_{\nu_e} = 2\partial T_{res}^{ee'}/\partial L_{\nu_{\tau}}$ which
is easily obtainable from Eq.(\ref{mm}).

Note that the most stringent condition occurs at the $\nu_e - \nu'_e$
resonance temperature, Eq.(\ref{w}).
We are primarily interested in relatively large values of $|\delta 
m^2_{\tau e'}| \stackrel{>}{\sim} 10^{-1}\ eV^2$, which means
that $T^{ee'}_{res} \stackrel{>}{\sim} 8 \ MeV$. Thus, from section
III, we are in a region of parameter space where we expect
Eq.(\ref{eeqq}) to be valid. [In particular, note that since 
$T_{res}^{ee'}$ is not at the point
where the lepton number is initially created,
Eq.(\ref{hohum2}) should also be valid].
Thus, from Eq.(\ref{eeqq}) we can obtain the rate of change of
$L_{\nu_e}$ due to $\nu_e - \nu_e'$ oscillations, at
the $\nu_e - \nu'_e$ resonance (where $b -a -c = 0$). We find
\begin{equation}
{\partial L_{\nu_e} \over \partial t} \simeq -{3 \over 8} 
{\sin^2\theta_0^{ee'}  \over \Gamma_{\nu_e}T^2}
\left[{\delta m^2_{ee'} \over 6.3}\right]^2.
\label{kkk}
\end{equation}
Note that from Eq.(\ref{mm}), we have:
\begin{equation}
{\partial T_{res}^{ee'} \over \partial L_{\nu_e} } 
= {M_W^2 \over A_e T^{ee'}_{res}} = {T^{ee'}_{res} \over L^{(e)}}.
\label{kk}
\end{equation} 
Thus,
\begin{equation}
{L^{(\tau)} \over T_{res}^{ee'}} {\partial T_{res}^{ee'} \over
\partial L_{\nu_e}} = {L^{(\tau)} \over L^{(e)}} \simeq 2.
\label{hh}
\end{equation}  
Hence the sufficient condition that $L^{(e)}$ survives without
being destroyed by $\nu_e - \nu'_e$ oscillations can be
obtained by substituting Eqs.(\ref{kkk},\ref{kk},\ref{hh}) 
into Eq.(\ref{jj}). Doing this exercise we find 
\begin{equation}
|\delta m^2_{ee'}| \stackrel{<}{\sim} \lambda |\delta m^2_{\tau e'}|^{11
\over 12},
\label{888}
\end{equation}
where $\lambda$ is given by
\begin{eqnarray}
&\lambda \simeq 
{12.6 G_F\over \sqrt{\sin^2 2\theta_0^{ee'} }M_W}
\left({8y_e A_e 5.5 \over 3M_P}
\right)^{1\over 2}
\left( {T_{res}^{ee'} \over (|\delta m^2|)^{1/6}}\right)^{11 \over 2} 
\nonumber \\
&\simeq
{12.6 G_F\over \sqrt{\sin^2 2\theta_0^{ee'} }M_W}
\left({8y_e A_e 5.5 \over 3M_P}
\right)^{1\over 2}
\left( {4.1 M^2_W \over 6.3 \sqrt{2}G_F A_{\tau}}
{A_{\tau} \over A_{\tau} + 2A_e}\right)^{11 \over 12},
\label{677}
\end{eqnarray}
and we have used Eq.(\ref{w}).  Thus, putting the numbers in, we find
\begin{equation}
{ |\delta m^2_{ee'}| \over eV^2}
\stackrel{<}{\sim} 6 \times 10^{-7} 
\left({|\delta m^2_{\tau e'}|\over eV^2}\right)^{{11 \over 12}},
\label{q}
\end{equation}
where we have assumed maximal mixing (i.e. $\sin^2 2\theta_0^{ee'}
\simeq 1$).  Thus provided this condition holds, 
$L^{(e)}$ will not be destroyed significantly by $\nu_e 
-\nu'_e$ oscillations (under the assumption that the neutrino
thermal momentum distribution can be neglected; we will study the 
effects of the momentum distribution in a moment) and the system 
moves quickly away from the $\nu_e - \nu'_e$ resonance.
While this condition was derived as a sufficient condition,
it turns out to be a necessary one as well.
This is because if Eq.(\ref{yyy}) where not valid, then $\nu_e - 
\nu'_e$ oscillations would create $L_{\nu_e}$ rapidly enough such that
$\partial (T_{res}^{ee'} - T)/\partial t < 0$. This would mean 
that the $\nu_e - \nu'_e$ resonance temperature would move to lower 
and lower temperatures where the rate of change of $L_{\nu_e}$ 
[from Eq.(\ref{kkk})] would be even larger (as it is proportional 
to $1/T^7$) and the expansion rate slower.  Thus if the condition 
Eq.(\ref{yyy}) were not satisfied initially, it could certainly not 
be satisfied for lower temperatures.

In the above system consisting of $\nu_{\tau}, \nu_{e}$ and $\nu'_e$
(and their anti-particles) discussed above,
observe that we have neglected the effects of $\nu_{\tau} - \nu_e$
oscillations. As discussed in the previous section, the effect of 
these oscillations is to make $(L_{\nu_{\tau}} - L_{\nu_e})$ tend
to zero.  Since these oscillations cannot prevent $L_{\nu_{\tau}}$ from
being generated the effect of incorporating them should only increase
the allowed region of parameter space\cite{ie}. 

The effect of the muon neutrino can also only increase the allowed region
of parameter space.  The effect of $\nu_{\mu} - \nu_e'$ oscillations
will be to create $L_{\nu_{\mu}}$ provided that $\delta m^2_{\mu e'} < 0$.
The effects of $\nu_{\mu}$ are completely analogous to the effects of 
the $\nu_{\tau}$ neutrino, and we can replace $\nu_{\tau} $ with 
$\nu_{\mu}$ in the above analysis.  This means that it is only
necessary that either $\delta m^2_{\tau e'}$ or $\delta m^2_{\mu e'}$
(or both) satisfy Eq.(\ref{q}).

Hitherto we have examined the system neglecting the thermal 
distribution of neutrino momenta. We now study the realistic case 
where the thermal distribution of neutrino momenta is taken into 
consideration.  We first estimate approximately the effects of the 
momentum distribution analytically and then we will preform a 
more accurate numerical study.

The previous calculation assumes that all of the neutrinos
have a common momentum and thus they all enter the
resonance at the same time. In the realistic case, only
a small fraction (less than about 1 percent as we will show) 
of the neutrinos are at resonance at any given time. 
Note that the $\nu_{\tau} - \nu_e'$ oscillations are
not affected greatly by this consideration, since as we
showed in section V, the momentum spread does not
prevent $L_{\nu_{\tau}}$ from being created (and
it still satisfies approximately $L^{(\tau)} \sim T^4$ after it
is initially created).
On the other hand the effect of the neutrino momentum distribution
on the $\nu_e - \nu'_e$ oscillations is very important. This is 
because the $\nu_e  - \nu'_e$ oscillations cannot destroy $L^{(e)}$ as 
efficiently as before.  In fact Eq.(\ref{kkk}) will be reduced 
by a factor which is about equal to the fraction of neutrinos at 
the resonance.  In principle, one should solve
Eq.(\ref{jj}) at the point where electron neutrinos of
momentum $p = \gamma T$ are at resonance and then
calculate the minimum of the value of $\lambda$ [see Eqs.(\ref{888},
\ref{677})] over
the range of all possible values of $\gamma$.
For simplicity we will make a rough approximation and
assume that the minimum of $\lambda$ occcurs
when neutrinos of average momentum are at resonance (i.e.
assume $\gamma \simeq 3.15$).  [Note that later-on we will 
do a more accurate numerical calculation].

To calculate the fraction of $\nu_e$ neutrinos at the 
$\nu_e - \nu_e'$ resonance we need to calculate the
width of the resonance in momentum space.
We will denote this width by $\Delta p$.
From Eq.(\ref{eeqq}) it is easy to see that the width
of the resonance is governed approximately by the equation
\begin{equation}
\Delta p |{\partial (b^p_{ee'} - a^p_{ee'}) \over \partial p}| \simeq
2\sqrt{x_{ee'}},
\label{wid}
\end{equation}
where we have assumed maximal mixing (i.e. $\cos 2\theta_0^{ee'} 
\simeq 0$).
Note that from the momentum dependence of $b^p,a^p$ [see Eq.(\ref{rand2})],
it follows that
\begin{equation}
{\partial b^p_{ee'}\over \partial p} = {2b^p_{ee'} \over p},\ 
{\partial a^p_{ee'}\over \partial p} = {a^p_{ee'} \over p}, 
\label{wid1}
\end{equation}
and hence
\begin{equation}
{\partial (b^p_{ee'} - a^p_{ee'}) \over \partial p} = 
{2b^p_{ee'} \over p} - {a^p_{ee'} \over p} \simeq {a^p_{ee'}
\over p},
\label{ncc}
\end{equation}
where we have used the result that $b^p_{ee'} \simeq a^p_{ee'}$ at the
resonance (note that we have assumed that $L^{(e)} > 0$ for 
definiteness).  Note that we are essentially interested in evaluating 
the maximum value of the fraction of neutrinos at the resonance.
This maximum fraction should occur approximately when 
$p \sim \langle p \rangle$. Thus, from the previous analysis, 
the resonance for neutrinos of average momentum occurs when
\begin{equation}
a_{ee'} \simeq {a_{\tau e'} \over 2}
{\delta m^2_{\tau e'} \over \delta m^2_{ee'}}
\simeq {1 \over 2}
{\delta m^2_{\tau e'} \over \delta m^2_{ee'}}.
\label{wid2}
\end{equation}
Thus, from Eqs.(\ref{ncc}, \ref{wid2}) and Eq.(\ref{wid}) the
width of the resonance in momentum space becomes
\begin{equation}
\Delta p \approx 4p\sqrt{\langle x_{ee'}\rangle} 
\left[ {|\delta m^2_{ee'}| \over |\delta m^2_{\tau e'}|}\right].
\label{wid3}
\end{equation}
Recall that $\langle x_{ee'} \rangle$ is defined in Eq.(\ref{jjds}),
and is given by
\begin{equation}
\langle x_{ee'} \rangle = \sin^2 2\theta_0^{ee'} + 
\Gamma^2_{\nu_{e}} {
\langle p \rangle^2 \over (\delta m^2_{ee'})^2} \simeq
\sin^2 2\theta_0^{ee'} + {y_e^2 G_F^4 (3.15)^2 T^{12} \over 
( \delta m^2_{ee'})^2}.
\end{equation}
Expanding out $\langle x_{ee'} \rangle$ at the temperature 
$T \approx T^{ee'}_{res}$  [defined in Eq.(\ref{w})], we find
\begin{eqnarray}
\langle x_{ee'} \rangle = \sin^2 2\theta_0^{ee'} + \left[
{y_e G_F M^2_W 4.1 \over 2\sqrt{2} A_{\tau}}{A_{\tau} \over A_{\tau}
+ 2A_e}\right]^2 
\left[ {\delta m^2_{\tau e'} \over \delta m^2_{ee'}}\right]^2
\simeq
1.2\times 10^{-5}\left[
{\delta m^2_{\tau e'} \over \delta m^2_{ee'}}\right]^2,
\end{eqnarray}
where the last part follows provided that $|\delta m^2_{ee'}| 
\stackrel{<}{\sim} 10^{-3} |\delta m^2_{\tau e'}|$.
Thus, using the above equation, Eq.(\ref{wid3}) simplifies to
\begin{equation}
{\Delta p \over p} \simeq {2y_e G_F M^2_W 4.1 \over \sqrt{2}A_{\tau}}
{A_{\tau} \over A_{\tau} + A_e} \simeq 1.4 \times 10^{-2}.
\label{wid4}
\end{equation}
Thus it is clear that only a small fraction of neutrinos will be at 
the resonance. We denote the fraction of electron neutrinos at 
the $\nu_e - \nu'_e$ resonance by $\Delta n_{\nu}/n_{\nu}$. 
Note that $\Delta n_{\nu}/n_{\nu}$ is given approximately by 
the equation 
\begin{equation}
{\Delta n_{\nu} \over n_{\nu}} \simeq {\Delta p \over n_{\nu}}
{dn_{\nu}\over dp}.
\end{equation}
Using $n_{\nu} = {3\over 4}\zeta (3) T^3/\pi^2$, and 
\begin{equation}
{dn_{\nu} \over dp} =  {1 \over 2\pi^2} {p^2 \over 1 + e^{p/T}},
\end{equation}
we find
\begin{equation}
{\Delta n_{\nu} \over n_{\nu}}|_{max} \approx {2y_e G_F M^2_W
4.1 \over \sqrt{2}A_{\tau}}{A_{\tau} \over A_{\tau} + 2A_e}
{\langle p/T \rangle^3 \over 1.5\zeta (3) (1 + e^{\langle p \rangle/T})} 
\simeq {1.4\times 10^{-2} \over 1.5\zeta (3)}
{\langle p/T \rangle^3 \over 1 + e^{\langle p \rangle/T}} 
\simeq 1.0\times 10^{-2}.
\label{ttttt}
\end{equation}
The effect of the momentum spread is thus to reduce the
number of neutrinos at the resonance by the above factor.
Multiplying Eq.(\ref{kkk}) by this fraction
and repeating the same steps which lead to Eq.(\ref{q})
we find that Eq.(\ref{q}) is weakened by the
factor $\sqrt{\Delta n_{\nu}/n_{\nu}} \simeq 10^{-1}$.
In other words the effect of the neutrino momentum 
distribution is to increase the allowed region of parameter 
space for which $\nu_e - \nu'_e$ oscillations do not destroy 
the $L^{(e)}$ asymmetry created by $\nu_{\tau} - \nu'_e$ oscillations.
This region of parameter space is given approximately by
\begin{equation}
|\delta m^2_{ee'}| \stackrel{<}{\sim} {\lambda \over 
\sqrt{\Delta n_{\nu}/n_{\nu}}} |\delta m^2_{\tau e'}|^{11 \over 12},
\end{equation}
where $\lambda$ is given in Eq.(\ref{677}) and
$\sqrt{\Delta_{\nu}/n_{\nu}}$ is given in Eq.(\ref{ttttt}).
Putting the numbers in, the above condition can
be written in the form
\begin{equation}
{|\delta m^2_{ee'}| \over eV^2} \stackrel{<}{\sim} 6\times 10^{-6} 
\left({|\delta m^2_{\tau e'}|\over eV^2}\right)^{{11 \over 12}}.
\label{hello}
\end{equation}
We now check this result by doing a more accurate numerical study of 
this problem.

The rate of change of $L_{\nu_{e}}$ and $L_{\nu_{\tau}}$ due
to the $\nu_{\tau} - \nu'_{e}$, $\nu_{e} - \nu'_{e}$
oscillations can be obtained from Eq.(\ref{eeqq3}).
This leads to the following coupled differential equations
\begin{eqnarray}
& {dL_{\nu_{e}} \over dt} \simeq 
{1 \over 4\zeta(3)T^3} \int^{\infty}_{0} {\sin^2 2\theta_0^{e e'}
\Gamma^p_{\nu_e} a^p_{e e'}  (\cos2\theta_0^{e e'} -b^p_{e e'})\over
[x^p_{e e'} + (\cos2\theta_0^{e e'}- b^p_{e e'} + 
a^p_{e e'})^2][x^p_{e e'} + (\cos2\theta_0^{e e'} - 
b^p_{e e'}  - a^p_{e e'})^2]
} {(1 - z^+)p^2 dp \over (1 + e^{p/T})} \nonumber \\
& + {1 \over 8\zeta (3) T^3} \int^{\infty}_{0}
{\sin^2 2\theta_0^{e e'} \Gamma^p_{\nu_e} [x^p_{e e'} + 
(a^p_{e e'})^2 + (b^p_{e e'}  - \cos2\theta_0^{e e'})^2]\over
[x^p_{e e'} + (\cos2\theta_0^{e e'} - b^p_{e e'} + 
a^p_{e e'})^2][x^p_{e e'} + (\cos2\theta_0^{e e'}  - 
b^p_{e e'}  -a^p_{e e'})^2]}
{z^-p^2 dp \over 1 + e^{p/T}}, \nonumber \\
& {dL_{\nu_{\tau}} \over dt} \simeq 
{1 \over 4\zeta(3)T^3} \int^{\infty}_{0} {\sin^2 2\theta_0^{\tau e'}
\Gamma^p_{\nu_{\tau}} 
a^p_{\tau e'}  (\cos2\theta_0^{\tau e'} -b^p_{\tau e'})\over
[x^p_{\tau e'} + (\cos2\theta_0^{\tau e'}- b^p_{\tau e'} + 
a^p_{\tau e'})^2][x^p_{\tau e'} + (\cos2\theta_0^{\tau e'} - 
b^p_{\tau e'}  - a^p_{\tau e'})^2]
} {(1 - z^+)p^2 dp \over (1 + e^{p/T})} \nonumber\\
& + {1 \over 8\zeta (3) T^3 } \int^{\infty}_{0}
{\sin^2 2\theta_0^{\tau e'} \Gamma^p_{\nu_{\tau}} [x^p_{\tau e'} + 
(a^p_{\tau e'})^2 + (b^p_{\tau e'}  - \cos2\theta_0^{\tau e'})^2]\over
[x^p_{\tau e'} + (\cos2\theta_0^{\tau e'} - b^p_{\tau e'} + 
a^p_{\tau e'})^2][x^p_{\tau e'} + (\cos2\theta_0^{\tau e'}  - 
b^p_{\tau e'}  -a^p_{\tau e'})^2]
} {z^- p^2 dp \over 1 + e^{p/T}}. 
\label{longeq2}
\end{eqnarray}
These equations are coupled differential equations because
$a^p_{ee'}$ and $a^p_{\tau e'}$ depend on both $L_{\nu_e}$
and $L_{\nu_{\tau}}$. Recall that $z^{\pm} = (z \pm \bar z)/2$. 
From Eq.(\ref{rate2}) the $z$ parameter, which is related
to the number of sterile neutrinos produced, is governed by:
\begin{equation}
{dz \over dt} =  
     {1 \over 4}(1 - z)\left[ {\Gamma^p_{\nu_e} \sin^2 2\theta_0^{ee'}\over 
[x^p_{ee'} + (\cos 2\theta_0^{ee'} - b^p_{ee'} + a^p_{ee'})^2] }
+ {\Gamma^p_{\nu_{\tau}} \sin^2 2\theta_0^{\tau e'}\over 
[x^p_{\tau e'} + (\cos 2\theta_0^{\tau e'} - b^p_{\tau e'} + 
a^p_{\tau e'})^2]} \right] ,
\label{monn3}
\end{equation}
and the evolution of $\bar z$ is governed by an equation
similar to the above (but with $a^p \to -a^p$).

The above equations can be integrated numerically (following
the proceedure mentioned in section V). 
Doing this, we can find the region of parameter space 
where the $L^{(e)}$ asymmetry created by the 
$\nu_{\tau} - \nu'_{e}$ oscillations 
does not get destroyed by the $\nu_{e} - \nu'_{e}$ oscillations.
We will solve Eq.(\ref{longeq2}) and Eq.(\ref{monn3}) under 
the assumption that $\sin^2 2\theta_0^{e e'} \simeq 1$ (i.e. 
the $\nu_e-\nu'_{e}$ oscillations are approximately maximal).
Performing the necessary numerical work, we find that $L^{(e)}$ is
created by $\nu_{\tau}-\nu'_{e}$ oscillations and not 
subsequently destroyed by $\nu_{e} - \nu'_{e}$ oscillations 
for the region of parameter space shown in Figure 5.
For definiteness we have taken two illustrative choices for
$\sin^2 2\theta_0^{\tau e'}$, 
$\sin^2 2\theta_0^{\tau e'} = 10^{-8}, \ 10^{-6}$. 
Note that in our numerical work, we have studied the region 
$10^{-1} \stackrel{<}{\sim}
|\delta m^2_{\tau e'}|/eV^2 \stackrel{<}{\sim} 10^2$.
Of course, there will be parameter space outside this region
where the $L^{(e)}$ created by $\nu_{\tau} - \nu'_e$ oscillations
is not destroyed by $\nu_e - \nu'_e$ oscillations. However,
one should keep in mind that there is a rather stringent
cosmology bound, $m_{\nu_{\tau}} \stackrel{<}{\sim} 40\ eV$\cite{cbbb} 
(which implies that $|\delta m^2_{\tau e'}| \stackrel{<}{\sim}
1600 \ eV^2$).  This bound assumes that the neutrino is
approximately stable, which is expected given the standard
model interactions. Of course, if there are new 
interactions beyond the standard model, then it is possible
to evade this cosmology bound\cite{cmp}.

Observe that the region of parameter space where $L^{(e)}$
survives is somewhat larger than our analytical estimate
Eq.(\ref{hello}). This is partly because the point where
$L_{\nu_{\tau}}$ is created occurs at a significantly higher
temperature than the analytical estimate (see section V
for some discussion about this point). Note that the
quantity $\lambda/\sqrt{\Delta n_{\nu}/n_{\nu}} \ \alpha
\ T_{res}^{11/2}/T_{res}^3 \sim T_{res}^{5/2}$. Thus, the
result that the lepton number is created at a higher 
temperature than our analytic estimate can easily
lead to a significant increase in the parameter space. 
Also, for large $\sin^2 2\theta_0^{\tau e'}$,
the magnitude of $L_{\nu_{\tau}}$ created 
by $\nu_{\tau} - \nu'_e$ oscillations is considerably
larger before the growth of $L_{\nu_{\tau}}$ is cut
off by the non-linearity of the differential equation
governing its evolution (compare the solid line with the  dashed or
dashed-dotted lines in Figures 3 or Figure 4).
Recall that our analytical estimate assumed that the creation of
$L^{(e)}$ due to $\nu_{\tau} - \nu'_e$ oscillations
had already passed the rapid exponential growth phase
at the point where the destruction of $L^{(e)}$ due
to $\nu_e - \nu'_e$ oscillations reached a maximum.
While this latter assumption is generally true for
small values of $\sin^2 2\theta_0^{\tau e'}$,
it is not true for larger values.
In this case, the rate of change of $T^{ee'}_{res}$ due
to $\nu_{\tau} - \nu_e'$ oscillations will be much larger
than our analytical estimate.
Consequently, the allowed region of parameter space 
is increased.
Thus the result that the allowed region of parameter space for $\sin^2
2\theta_0^{\tau e'} = 10^{-6}$ is significantly larger
than the allowed region for $\sin^2 2\theta_0^{\tau e'} = 10^{-8}$
is not unexpected.

Having established the condition that the $\nu_e - \nu'_e$ oscillations
do not destroy the $L^{(e)}$ which is created by the 
$\nu_{\tau} - \nu_e'$ oscillations (or $\nu_{\mu} - \nu'_e$
oscillations), we must also check that the magnitude of
$L^{(e)}$ is large enough to invalidate the bound in Eq.(\ref{naive}).

For $\delta m^2_{ee'}$ in the range
$|\delta m^2_{ee'}| \stackrel{>}{\sim} 10^{-6} \ eV^2$,
the bound Eq.(\ref{naive}) arises by requiring that the $\nu_{e}-
\nu'_{e}$ oscillations do not bring the $\nu_{e}'$ sterile neutrino
into equilibrium above the kinetic decoupling temperature ($\sim 3$
MeV). The sterile neutrino $\nu_{e}'$ will not be brought into 
equilibrium provided that the rate of $\nu_{e}'$ production is 
approximately less than the expansion rate $H$, i.e. 
\begin{equation}
\Gamma (\nu_e \to \nu'_e)/H \simeq
{1 \over 4}\Gamma_{\nu_{e}}\sin^2 2\theta_m^{e e'}/H 
\stackrel{<}{\sim}\ 1,
\label{bb}
\end{equation}
where we have used Eq.(\ref{Gamma}) with $\langle \sin^2 
\tau/2L_m \rangle \simeq 1/2$\cite{fn79}.
Recall that we are primarily interested in the region 
$1$ MeV $\stackrel{<}{\sim} T \stackrel{<}{\sim} 100 $ MeV, where
$H \simeq 5.5 T^2/M_P$.  Using Eq.(\ref{bel}) with $a \simeq 0$, 
the above equation can be re-written in the form
\begin{equation}
{y_{e}G_F^2 M_P \sin^2 2\theta_0^{e e'} T^3 \over 22\left[
b_{e e'}^2 + 1\right]}
\stackrel{<}{\sim} 1,
\label{hhn}
\end{equation}
where we have assumed large mixing i.e. $\cos2\theta_0^{ee'} 
\ll 1$. Recall that $b_{ee'}$ can be obtained from Eq.(\ref{ab3}).
Obtaining the maximum of the left-hand side of Eq.(\ref{hhn})
leads approximately to the bound $|\delta m^2| \stackrel{<}{\sim}
10^{-6}\ eV^2$.

In the case where $L_{\nu_{\tau}}$ is created by $\nu_{\tau} - \nu_{e}'$ 
oscillations, the situation is very different. The
lepton number $L_{\nu_{\tau}}$ is created at the
temperature when $b_{\tau e'} \approx 1$ (assuming that
$\cos2\theta_0^{\tau e'} \sim 1$). Denoting this
temperature by $T_c^{\tau e'}$, then as per Eq.(\ref{critT})
\begin{equation}
T_c^{\tau e'} \approx 16 \left({|\delta m^2_{\tau e'}| \over eV^2}
\right)^{1 \over 6} \ {\rm MeV}.
\label{fd}
\end{equation}
The evolution of this
system can be divided into two regions, the region before
lepton number creation (i.e. $T > T_c^{\tau e'}$), and the
region after the lepton number creation (i.e. $T < T_c^{\tau e'}$).
In the region before the lepton number is created, $a_{\tau e'}
\simeq 0$ and Eq.(\ref{hhn}) holds. We will obviously be interested
in the parameter space where $\delta m^2_{\tau e'}$ is sufficiently
large (recall that $\delta m^2_{\tau e'}$ is related to
$T_c^{\tau e'}$ by Eq.(\ref{fd}) above) so that $L_{\nu_{\tau}}$ 
is created at some point above the kinetic decoupling temperature 
$T_{dec} \simeq 3$ MeV, of $\nu_{e}$. Let us assume that 
$|\delta m^2_{\tau e'}|$ is large enough so that 
$b_{e e'}^2 \gg 1$ for temperatures $T > T_c^{\tau e'}$
(which corresponds approximately to, $|\delta m^2_{\tau e'}| > 
|\delta m^2_{ee'}|$).  In this case, $b_{e e'}^2 + 1 
\simeq b_{e e'}^2$ and Eq.(\ref{hhn}) can be re-written in the form
\begin{equation}
T^9 \stackrel{>}{\sim} \left( {
4.1\delta m^2_{ee'}M_W^2 \over 6.3\sqrt{2}G_F A_e}
\right)^2
{ \sin^2 2\theta_0^{e e'} y_{e}
G_F^2 M_{P} \over 22}.
\label{aa}
\end{equation}
where we have used Eq.(\ref{ab3}) with $n_{\gamma} \simeq T^3/4.1$.
Observe that the most stringent condition occurs for $T = T_c^{\tau e'}$.
Thus taking $T = T_c^{\tau e'}$, using Eq.(\ref{fd}), we find that
\begin{equation}
{|\delta m^2_{\tau e'}| \over eV^2} \stackrel{>}{\sim} 
15 \left[{|\delta m^2_{e e'}| \over eV^2}\right]^{4\over 3}
\left[\sin^2 2\theta_0^{e e'}\right]^{2 \over 3}.
\label{friday}
\end{equation}
Assuming maximal $\nu_{e} -\nu'_{e}$ oscillations 
(i.e. $\sin 2\theta_0^{e e'} \simeq 1$) and assuming
$|\delta m^2_{e e'}| \stackrel{<}{\sim} 10^{-3}\ eV^2$\cite{fn1}, 
Eq.(\ref{friday}) implies that 
\begin{equation}
|\delta m^2_{\tau e'}| \stackrel{>}{\sim}\ 10^{-3}\  eV^2.
\label{bound} 
\end{equation}
Thus provided that this constraint is satisfied, the sterile neutrino, 
$\nu_e'$ will not come into equilibrium for temperatures
greater than the temperature where $L_{\nu_{\tau}}$ is 
created, $T_c^{\tau e'}$.

We now need to check that the lepton number created is
sufficient to suppress $\nu_{e} - \nu'_{e}$ oscillations for
temperatures less than $T_c^{\tau e'}$.
Demanding that the interactions do not bring the sterile neutrino
into equilibrium with the muon neutrino, that is again imposing the
inequality Eq.(\ref{bb}), but this time for $T < T_c^{\tau e'}$
where there is significant creation of $L^{(e)}$ \cite{fv1}, we
find that
\begin{equation}
{y_{e}G_F^2M_P\sin^2 2\theta_0^{e e'} T^3 \over
22\left[(b_{e e'} \pm a_{e e'})^2 + 1\right]}
\stackrel{<}{\sim} 1,
\label{Lbound}
\end{equation}
where the $-(+)$ signs correspond to $\nu_{e}-\nu_{e}'$ (
$\bar \nu_{e}-\bar \nu'_{e}$ ) oscillations.
Note that Eq.(\ref{Lbound}) is only required to be satisfied
for $T > T_{dec} \simeq 3 $ MeV (since we only need to 
require that the sterile neutrinos do not come into equilibrium
before kinetic decoupling of the electron neutrinos occurs).
Once $L^{(e)}$ is created at $T = T_c^{\tau e'}$ (
where $b_{\tau e'} = \cos 2\theta_0^{\tau e'} \simeq 1$), 
its magnitude will rise according to the constraint $a_{\tau e'} 
\stackrel{>}{\sim} 1$ (assuming for definiteness that $L^{(e)} > 0$).
Note that the quantities $b_{e e'}, a_{e e'}$ are related
to $b_{\tau e'}, a_{\tau e'}$ as follows:
\begin{equation}
{b_{e e'}\over b_{\tau e'}} = {A_e \over A_{\tau}}
{\delta m^2_{\tau e'}\over
\delta m^2_{e e'}},\ {a_{e e'}\over a_{\tau e'}} \simeq 
{1 \over 2} {\delta m^2_{\tau e'}\over \delta m^2_{e e'}}.
\label{bbba}
\end{equation}
After the initial resonance $a_{\tau e'} \stackrel{>}{\sim} 1$ while
$b_{\tau e'} \le 1$ (and quickly becomes much less than one).
Thus very soon after the resonance,
$a_{\tau e'} \gg b_{\tau e'}$ and hence from Eq.(\ref{bbba}),
$a_{e e'} \gg b_{e e'}$. As before the most stringent bound
occurs when $T \simeq T_c^{\tau e'}$, and Eq.(\ref{Lbound}) leads 
to approximately the same bound as before [i.e. Eq.(\ref{bound})],
since at the point $T = T_c^{\tau e'},\ a_{\tau e'} 
\approx b_{\tau e'}$.

Finally, we need to check that the oscillations of the
$\nu_e$, $\nu'_e$ neutrinos do not significantly
deplete the number of electron neutrinos for
the temperature range, 
\begin{equation}
0.7\ {\rm MeV}\stackrel{<}{\sim} T
\stackrel{<}{\sim} T_{dec} \simeq 3 \ {\rm MeV}.
\label{wed4}
\end{equation}
Neutrino oscillations in this temperature range can affect
BBN because they will deplete electron neutrinos (and anti-neutrinos)
and thus modify the temperature when the neutron/proton ratio 
freezes out. This effect is generally small unless 
$\sin^2 2\theta_m \stackrel{>}{\sim} 10^{-2}$\cite{B2,B4}.
If we demand that $\sin^2 2\theta_m \stackrel{<}{\sim} 10^{-2}$ for
this temperature range, then from Eq.(\ref{bel}) we 
require $|a| \stackrel{>}{\sim} 10$ (for the most stringent case of 
maximal mixing) for this temperature range (or  $|\delta m^2| 
\stackrel{<}{\sim} 10^{-8}\ eV^2$).  Thus, from Eq.(\ref{ab3}), 
$|a| \stackrel{>}{\sim} 10$ implies
\begin{equation}
|L^{(e)}|
\stackrel{>}{\sim} 2\left( {\delta m^2_{ee'}\over eV^2} \right).
\label{wed5}
\end{equation}
Recall that for temperatures $T \stackrel{>}{\sim} T_f$ (where $T_f$ is the 
temperature where the change in $a$ due to the expansion is 
larger in magnitude to the change in $a$ due to oscillations, see
the earlier comments around Eq.(\ref{sat5}) for some discussion
about this), $L_{\nu_{\tau}}$ is created such that $a_{\tau e'} \simeq
1$, from this it follows that
\begin{equation}
L^{(e)} \simeq L_{\nu_{\tau}} \simeq 2\times 10^{-2}
\left({|\delta m^2_{\tau e'}|
\over eV^2}\right) \left({{\rm MeV} \over T_f}\right)^4.
\label{wed6}
\end{equation}
Combining Eq.(\ref{wed5}) and Eq.(\ref{wed6}), 
sufficient lepton number will be generated to suppress the oscillations
in the temperature range Eq.(\ref{wed4}) provided that
\begin{equation}
|\delta m^2_{e e'}| \stackrel{<}{\sim} 10^{-2} |\delta m^2_{\tau e'}|
\left({{\rm MeV} \over T_f}\right)^4.
\label{wed7}
\end{equation} 
Note that the temperature $T_f$ is generally less than about
$4$ MeV [see Eq.(\ref{sat5}) for a discussion about this].
Thus, Eq.(\ref{wed7}) will be easily satisfied given the condition 
Eq.(\ref{hello}).

In summary, a consequence of the creation of $L_{\nu_{\tau}}$ by 
$\nu_{\tau} - \nu'_e$ oscillations is that the
large angle or maximal $\nu_e - \nu'_e$ oscillations
will not significantly modify BBN provided that 
$L^{(e)}$ does not get destroyed by $\nu_e - \nu'_e$ oscillations
(see Figure 5 for some of this region of parameter 
space) and the condition Eq.(\ref{bound}) holds.
Thus, it is clear that the oscillation generated neutrino asymmetry
can weaken the rather stringent BBN bound ($|\delta m^2_{e e'}|
\stackrel{<}{\sim} 10^{-8} \ eV^2$ for maximal mixing) by
many orders of magnitude.  A consequence of this is that
the maximal ordinary-sterile neutrino oscillation solution 
to the solar neutrino problem does not significantly modify
BBN for a large range of parameters.

While we have focussed on a particular scenario,
our analysis will be relevant to other models with sterile neutrinos.
For example, assume that there is a sterile
neutrino which mixes with parameters corresponding
to the large angle MSW solution to the solar neutrino 
problem, that is $\delta m^2 \sim 10^{-5}\ eV^2$ and $\sin^2 2\theta_0
\sim 0.7$\cite{hl}.  This scenario has been ``ruled out'' (assuming
negligible lepton number asymmetry) in Refs.\cite{B2,B4}.
However, if the sterile neutrino also mixes slightly
with the mu and/or tau neutrino (and such mixing
would be expected), then these BBN bounds can be evaded provided that
$|\delta m^2_{\tau e'}|$ and/or $|\delta m^2_{\mu e'}| 
\stackrel{>}{\sim} 0.1 - 1 \ eV^2$.
Note that the evidence for $\nu_{\mu} - \nu_e$ oscillations
found by the LSND collaboration suggests that 
$|\delta m^2_{\mu e}| \stackrel{>}{\sim} \ 0.3\ eV^2$\cite{lsnd, fn345}.
If this is the case then the large angle MSW solution
will not lead to a significant modification to BBN
for a large range of values for  $\sin^2 2\theta_0^{\mu e'}$.

We now discuss the possibility that the atmospheric neutrino anomaly 
is due to large angle or maximal muon neutrino - sterile
neutrino oscillations. Here, we will denote the sterile
neutrino by $\nu'_{\mu}$ (this neutrino is expected to be distinct
from $\nu'_e$).  Note that the possibility that the atmospheric 
neutrino anomaly is due to large angle or maximal $\nu_{\mu}-
\nu'_{\mu}$ oscillations can be well motivated.
For example, the exact parity model\cite{flv} predicts that
all three ordinary neutrinos mix maximally with mirror neutrinos
if neutrinos have mass.  [See also Ref.\cite{oth} for some
other interesting models which can solve the atmospheric 
neutrino anomaly through maximal ordinary - sterile neutrino
oscillations]. The deficit of atmospheric 
muon neutrinos can be explained if there are $\nu_{\mu}-\nu'_{\mu}$
oscillations with $\sin^2 2\theta_0 \stackrel{>}{\sim} 0.5$
and $10^{-3} \stackrel{<}{\sim} |\delta m^2_{\mu \mu'}|/eV^2 
\stackrel{<}{\sim} 10^{-1}$\cite{ana, barpak}. The best fit occurs for
$\delta m^2_{\mu \mu'} \simeq 10^{-2} \ eV^2$ and 
$\sin^2 2\theta_0 \simeq 1$ \cite{ana}.  However, this parameter range 
is naively inconsistent with BBN [see Eq.(\ref{negl})] if the lepton 
number asymmetries are neglected.  Can the generation of lepton number 
by ordinary-sterile neutrino oscillations reconcile this solution to 
the atmospheric neutrino anomaly with BBN?

To study this issue, consider the system consisting of
$\nu_{\tau}, \nu_{\mu}, \nu'_{\mu}$.
This system is similar to the $\nu_{\tau}, \nu_e, \nu'_e$ 
system that we have discussed above.
Doing a similar analysis to the above (i.e. replacing
$\nu_e$ and $\nu'_e$ by $\nu_{\mu}$ and $\nu'_{\mu}$), we 
find that the $L^{(\mu)}$ asymmetry created by $\nu_{\tau} - \nu'_{\mu}$ 
oscillations will not be destroyed by $\nu_{\mu} - \nu'_{\mu}$ 
oscillations provided that
\begin{equation}
|\delta m^2_{\mu \mu'}| \stackrel{<}{\sim} {\lambda \over
\sqrt{\Delta n_{\nu}/n_{\nu}}} |\delta m^2_{\tau \mu'}|^{11 \over 12},
\end{equation}
where $\lambda$ and $\Delta n_{\nu}/n_{\nu}$ are given by
equations similar to Eq.(\ref{677}) and Eq.(\ref{ttttt}) except
that the replacements $y_e \to y_{\mu}, \ A_e \to A_{\mu}$ have
to be made.  Thus, evaluating the resulting expressions for
$\lambda$ and $\Delta n_{\nu}/n_{\nu}$, we find
\begin{equation}
{|\delta m^2_{\mu \mu'}| \over eV^2} \stackrel{<}{\sim} 5\times 10^{-6} 
\left({|\delta m^2_{\tau \mu'}|\over eV^2}\right)^{{11 \over 12}}.
\label{qqq2}
\end{equation}
As before, we have made a more accurate numerical study of this
problem. If we solve the system of equations Eq.(\ref{longeq2})
and Eq.(\ref{monn3}),
with the replacements $\nu_e, \nu_e' \to \nu_{\mu}, \nu'_{\mu}$,
then we can obtain the region of parameter space where the
$L^{(\mu)}$ created by $\nu_{\tau} - \nu'_{\mu}$ oscillations
does not get destroyed by $\nu_{\mu} - \nu'_{\mu}$ oscillations.
We show some of this parameter space in Figure 6.

If we assume the best fit of the atmospheric neutrino
data, then $|\delta m^2_{\mu \mu'}| \simeq 10^{-2}\ eV^2$ 
and $\sin^2 2\theta_0^{\mu \mu'} \simeq 1$.
Numerically solving the Eqs.(\ref{longeq2}) and Eq.(\ref{monn3})
(with the replacement of $\nu_e, \nu'_e$ with $\nu_{\mu},
\nu'_{\mu}$) assuming the best fit parameters, $|\delta m^2_{\mu \mu'}|
\simeq 10^{-2}\ eV^2$ and $\sin^2 2\theta_0^{\mu \mu'} \simeq 1$,
we again obtain the region of parameter space where the $L^{(\mu)}$
asymmetry is created by $\nu_{\tau} - \nu_{\mu}'$ oscillations
and does not get destroyed subsequently by $\nu_{\mu} - 
\nu'_{\mu}$ oscillations. Our results are shown in
Figure 7.  As the figure shows, 
the asymmetry $L^{(\mu)}$ created by
$\nu_{\tau} - \nu_{\mu}'$ oscillations will not be destroyed
by $\nu_{\mu} - \nu'_{\mu}$ oscillations provided 
that $\delta m^2_{\tau \mu'}$ is quite large, i.e., 
\begin{equation}
|\delta m^2_{\tau \mu'}| \stackrel{>}{\sim} 30 \ eV^2.
\label{friday3}
\end{equation}
Recall that our analysis neglects the possible effects of
$\nu_{\tau} - \nu_{\mu}$ oscillations. It may be possible
that smaller $\delta m^2_{\tau \mu'}$ 
are allowed if the $\nu_{\tau} - \nu_{\mu}$ 
mixing parameters are large enough.

The requirement that $\nu_{\tau} - \nu'_{\mu}$ oscillations do
not produce too many sterile states implies an upper limit on
$\sin^2 2\theta_0^{\tau \mu'}$ [see Eq.(\ref{gh})].
This upper limit has been shown in the Figure (dashed-dotted line).
Also shown in Figure 7 (dashed line) is the cosmological 
energy density bound $|\delta m^2_{\tau \mu'}| 
\stackrel{<}{\sim} 1600 \ eV^2$\cite{cbbb}.

Recall that the differential equations, Eq.(\ref{longeq2}) are only
valid provided that Eq.(\ref{hohum2}) holds. [We also require
Eq.(\ref{76}) to hold for $\delta m^2 = \delta m^2_{\tau e'}$, which
is clearly valid for the region of parameter space studied].
Note that in our numerical work we found that the condition
Eq.(\ref{hohum2}) was approximately valid for the points in the
allowed region of the figures except for the
region with relatively large values of $\sin^2 2\theta_0 
\stackrel{>}{\sim} 10^{-6}$.
It would be a useful exercise to check our analysis by
preforming a more accurate study using the density matrix
equations, Eq.(\ref{jk}), modified to incorporate the
neutrino momentum distribution.

The $\sin^2 2\theta_0^{\tau \mu'}$ dependence shown in Figures 6,7 
can be understood qualitatively as follows. For small 
$\sin^2 2\theta_0^{\tau \mu'}$ ($\stackrel{<}{\sim} 10^{-8}$), the creation 
of $L^{(\mu)}$ is sluggish which has the effect of delaying the point 
where the destruction of $L^{(\mu)}$ by $\nu_{\mu} -\nu'_{\mu}$ 
oscillations reaches its maximum rate.  As mentioned earlier, for 
lower temperatures the rate at which $\nu_{\mu} - \nu'_{\mu}$ 
oscillations destroy $L^{(\mu)}$ increases, which has the effect 
of reducing the allowed parameter space. For larger values 
of $\sin^2 2\theta_0^{\tau \mu'}$ ($\stackrel{>}{\sim} 10^{-8}$), the
maximum rate at which the $\nu_{\mu} - \nu'_{\mu}$ oscillations
destroy $L^{(\mu)}$ occurs during the time when
$L^{(\mu)}$ is still growing exponentially. In this case
the system moves rapidly away from the $\nu_{\mu} - \nu_{\mu} '$
resonance region. Consequently, the allowed region of parameter
space is significantly increased.  

Observe that from Eq.(\ref{friday}), this lepton number will easily 
be sufficiently large and created early enough to prevent the 
$\nu'_{\mu}$ sterile neutrino from coming into equilibrium given
Eq.(\ref{friday3}).  Thus, the large angle or maximal muon - sterile 
neutrino oscillation solution to the atmospheric neutrino anomaly is 
in fact consistent with BBN for a significant range of parameters.  
Note that the condition Eq.(\ref{friday3}) 
implies quite large tau neutrino masses,
$m_{\nu_{\tau}} \stackrel{>}{\sim} 6\ eV$. Note that if the 
neutrinos are approximately stable (which would be
expected unless some new interactions exist\cite{cmp})
then there is a stringent cosmology bound 
of $m_{\nu_{\tau}} \stackrel{<}{\sim} 40  \ eV$\cite{cbbb}.
Although this parameter space is not so big, it can be
well motivated from the point of view of dark matter (since
stable tau neutrinos with masses in the range
$6 \ eV\stackrel{<}{\sim} m_{\nu_{\tau}} \stackrel{<}{\sim} 40\ eV$
could provide a significant fraction of the matter in the Universe).

If we add the $\nu'_{\mu}$ sterile neutrino to the 
$\nu_{\tau}, \nu_e , \nu'_e$ 
system we considered earlier (in connection to the 
large angle ordinary - sterile neutrino oscillation solution
to the solar neutrino problem), then $\nu_{\tau} - \nu_{\mu}'$ 
oscillations will also generate $L^{(e)}$ in a similar 
manner to the way in which $\nu_{\tau} - \nu_e'$ oscillations 
generated $L^{(e)}$.  Consequently, the bounds on 
$\delta m^2_{\tau e'}, \ \sin^2 2\theta_0^{\tau e'}$, can 
alternatively be considered as bounds on 
$\delta m^2_{\tau \mu'}, \ \sin^2 2\theta_0^{\tau \mu'}$.
Of course, we only need to require that either $\delta m^2_{\tau e'},
\ \sin^2 2\theta_0^{\tau e'}$ or $\delta m^2_{\tau \mu'},\ 
\sin^2 2\theta_0^{\tau \mu'}$ satisfy the bounds derived.
Similarly, we can add the $\nu'_e$ sterile neutrino to the
$\nu_{\tau}, \nu_{\mu}, \nu'_{\mu}$ system and analogous 
reasoning leads to the conclusion that the bounds
on $\delta m^2_{\tau e'}, \ \sin^2 2\theta_0^{\tau e'}$ can 
alternatively be considered as bounds on $\delta m^2_{\tau \mu'},
\ \sin^2 2\theta_0^{\tau \mu'}$.  Observe that with 
$\delta m^2_{\tau e'},\ \sin^2 2\theta_0^{\tau e'}$ or 
$\delta m^2_{\tau \mu'}, \ \sin^2 2\theta_0^{\tau \mu'}$ in the 
range identified in Figure 7 (where the atmospheric neutrino 
anomaly is explained by large angle $\nu_{\mu} -\nu'_{\mu}$ 
oscillations without significantly modifying BBN) the solar 
neutrino problem can also be solved for the entire parameter space 
[Eq.(\ref{solar})], without significantly modifying BBN.
Alternatively one can argue that the present data may allow
the $\nu'_{\mu}$ to come into equilibrium with
the ordinary neutrinos and still be consistent with BBN
\cite{ks} and thus we only require the less stringent bounds given 
in Figure 5.  Clearly this is a possibility at the moment. 
Note however, that for the case of the exact parity symmetric
model\cite{flv}, where the mirror neutrinos interact with
themselves, this way out is not possible. This is because
if the mirror muon neutrino is brought into equilibrium above
the kinetic decoupling temperature (which is about $5$ MeV for
muon neutrinos) then the mirror weak interactions will bring all
three mirror neutrinos together with the mirror photon and
mirror electron-positron into equilibrium (which would lead to about 
9 effective neutrino degrees of freedom during nucleosynthesis).
For the case of mirror neutrinos it seems to be necessary
to ensure that the mirror muon neutrino is not brought into
equilibrium in the first place.

Note that in our previous analysis, we have assumed that
the sterile neutrino is truly sterile and does not interact 
with the background. In the special case of mirror neutrinos,
the mirror neutrinos are expected to interact with
the background because they interact with themselves\cite{br}. 
In general the effective potential describing 
coherent forward scattering of the neutrino with the background
has the form $V = V_{\alpha} - V_{s }'$.
For truly sterile neutrinos, $V_s' = 0$ (as has been
assumed hitherto).  For mirror neutrinos $V_{s}'$ is non-zero.
Denoting the mirror neutrinos by $\nu'_{\beta}$,
then for the case of  $\nu_{\alpha} - \nu_{\beta}'$ oscillations 
we will denote the effective potential by 
\begin{equation}
V = V_{\alpha} - V'_{\beta},
\end{equation}
where $V_{\alpha}$ is given by Eq.(\ref{V}) and $V_{\beta}'$ is 
the effective potential due to the interactions of the mirror 
neutrinos with the background. The mirror effective potential $V'_{\beta}$ 
can be expressed in an analogous way to $V_{\alpha}$, that is
there is a part which is proportional to
mirror lepton number and a part which is independent of mirror lepton
number, 
\begin{equation}
V'_{\beta} = (-a'^p + b'^p)\Delta_0^p.
\end{equation}
If the number of mirror neutrinos is much less than the
number of ordinary neutrinos then $b' \simeq 0$.
[Note that the $b$-part of the effective potential is
proportional to the number densities of the background particles.
This dependence is not given in Eq.(\ref{ab}) since for this
equation the number densities were set equal to their
equilibrium values].  The parameter $a'$ has the form
\begin{equation}
a'^p \equiv {-\sqrt{2}G_F n_{\gamma}L'^{(\beta)} \over \Delta_0^p},  
\end{equation}
where
$L'^{(\beta)}$ is given 
\begin{equation}
L'^{(\beta)}  = L_{\nu'_{\beta}} + L_{\nu'_e} + L_{\nu'_{\mu}}
+ L_{\nu'_{\tau}} + \eta',
\end{equation}
where $L_{\nu'_{\beta}}$ are are the mirror lepton numbers, which
are defined by $L_{\nu'_{\beta}} \equiv  (n_{\nu_{\beta}'} 
- n_{\bar \nu'_{\beta}})/n_{\gamma}$ (note that $n_{\gamma}$ 
is the number density of {\it ordinary} photons) and 
$\eta'$ is a function of the mirror baryon/electron number asymmetries
[which is defined analogous to Eq.(\ref{toomanyeq})]. We 
will assume that $\eta'$ is small and can be approximately neglected.
Since ordinary + mirror lepton number is conserved (and we will
assume that it is zero), it follows that
\begin{equation}
L_{\nu_e} + L_{\nu_{\mu}} + L_{\nu_{\tau}} + 
L_{\nu'_e} + L_{\nu'_{\mu}} + L_{\nu'_{\tau}} = 0.
\label{145}
\end{equation}
From the above equation, it follows that $a'$ is expected to be
of the same order of magnitude as $a$.
In the case of the $\nu_{\tau}, \nu_{\mu}, \nu'_{\mu}$
system, the effect of the mirror - neutrino effective
potential can be accounted for by simply replacing $L^{(\mu,\tau)}$ 
in $V_{\alpha}$ by
\begin{eqnarray}
& L^{(\mu)} \to L^{(\mu)} - L'^{(\mu)} \simeq 
2L_{\nu_{\mu}} + L_{\nu_{\tau}} - 2L_{\nu'_{\mu}}
\simeq 4L_{\nu_{\mu}} + 3L_{\nu_{\tau}},\nonumber \\
& L^{(\tau)} \to L^{(\tau)} - L'^{(\mu)} \simeq 
2L_{\nu_{\tau}} + L_{\nu_{\mu}} - 2L_{\nu'_{\mu}}
\simeq 4L_{\nu_{\tau}} + 3L_{\nu_{\mu}},
\end{eqnarray}
where we have used $L_{\nu_e'} \simeq L_{\nu_{\tau}'} 
\simeq 0$ and Eq.(\ref{145}).  Thus, from the above equation, 
assuming that negligible $L_{\nu_{\mu}}$ is produced, we see that 
$|a_{\mu \mu'}| \simeq {3 \over 4}R |a_{\tau \mu'}|$ (where 
$R \equiv |\delta m^2_{\tau \mu'}/\delta m^2_{\mu \mu'}|$).
The factor of $3/4$ replaces the factor of $1/2$ that we had
earlier (for the case where $\nu_{\mu}'$ or $\nu'_e$ were
sterile neutrinos). This difference will increase the region of
allowed parameter space, because it will make $T^{\mu \mu'}_{res}$
closer to the point where $L_{\nu_{\tau}}$ is initially
created. At this point $\partial L_{\nu_{\tau}}/\partial t$ can
be significantly enhanced because it is very close to the resonance
(also note that $\partial L_{\nu_{\mu}}/\partial t$ will be
suppressed because it is proportional to $1/T^7$).

Finally, observe that another important feature of mirror neutrinos
is that the mirror interactions can potentially bring all three
of the mirror neutrinos into equilibrium with themselves as well
as the mirror photon and mirror electron positron. [However
the temperature of the mirror particles will generally be
less than the temperature of the ordinary particles if
the oscillations satisfy Eq.(\ref{gh})].
Detailed studies involving mirror neutrinos will need to 
incorporate this. We leave a more detailed study of mirror
neutrinos to the future\cite{fv3}.
\vskip 0.4cm
\noindent
{\bf VII. Conclusion}
\vskip 0.5cm
In summary, we have studied the phenomenon of neutrino oscillation 
generated lepton number asymmetries in the early Universe in detail. 
This extended study clarifies the origin of the approximations 
adopted in the earlier work\cite{ftv}.  We have also studied the effects 
of the thermal distribution of the neutrino momenta and non-negligible 
sterile neutrino number densities. 

In the unrealistic case where the neutrino momentum distribution is
neglected, the evolution of $L_{\nu_{\alpha}}$ can
be approximately described by seven coupled differential equations
[Eqs.(\ref{jk})], which can be obtained from the density matrix.
We showed in section III that these equations
can be reduced to a single integro-differential
equation (we show in the appendix that the same equation can be obtained
from the Hamiltonian formalism). In general, the density matrix
equations cannot be solved analytically, and must be solved numerically. 
However, if the system is sufficiently smooth (the static limit),
then the integro-differential equation reduces to a
relatively simple first order ordinary differential equation 
[Eq.(\ref{eeqq})].  This equation gives quite a reasonable description 
of the evolution of $L_{\nu_{\alpha}}$, except possibly at the initial
resonance where significant generation of $L_{\nu_{\alpha}}$ occurs.
We show that when the thermal distribution of the neutrino momenta is 
incorporated several important effects occur. One
of these effects is that the creation of lepton number
is much smoother. This allows a considerable computational simplification,
because it means that the static approximation can be a reasonably good
approximation, even at the resonance for a much larger range of
parameters.
This means that $L_{\nu_{\alpha}}$ can be accurately described by the 
relatively simple first order differential equation (modified to
incorporate the neutrino momentum distribution). This equation is 
given by Eq.(\ref{eeqq2}), expressed as a function of the number
distribution of sterile states.
In section V, we showed that the number distribution of sterile
neutrino states satisfied a first order differential equation [
Eq.(\ref{rate2})] which must be integrated for each momentum step.

We first applied our analysis to obtain the region of parameter
space where large neutrino asymmetries are generated. This
region of parameter space is given in Eq.(\ref{ghh}).
This analysis included the effects of the neutrino momentum
distribution which was neglected in earlier studies\cite{ftv, shi}.
We also examined the implications of lepton number
generation for the BBN bounds for $\delta m^2, \sin^2 2\theta_0$ for
ordinary-sterile neutrino mixing. There are two ways in which
the creation of lepton number can modify the BBN bounds.
One way is where the $\nu_{\alpha} - \nu_s$ oscillations themselves
produce $L_{\nu_{\alpha}}$ thereby suppressing the number of
sterile neutrinos produced from the same oscillations.
The other way is where the $\nu_{\beta} - \nu_s$ oscillations
create $L_{\nu_{\beta}}$ which thereby suppresses $\nu_s$ production
from $\nu_{\alpha} - \nu_s$ oscillations.
The bound for the former case is given in Eq.(\ref{gh}), while
the latter case studied in section VI, in the
context of the maximal vacuum oscillation solutions to the solar
and atmospheric neutrino problems.  The maximal vacuum oscillation
solution of the solar neutrino problem 
assumes that the electron neutrino is approximately maximally
mixed with a sterile neutrino.
For a large range of parameter space, the maximal mixing leads
to an energy independent factor of two reduction in the solar
neutrino fluxes. This leads to a reasonably simple predictive
solution to the solar neutrino problem which is supported by
the experiments. However, most of the parameter space 
for this solution is inconsistent with standard big bang 
nucleosynthesis (BBN) if the lepton numbers are assumed 
to be negligible\cite{B,B2,B3,B4}.  We showed that there is a large
region of parameter space where the oscillations generated
lepton number in such a way so as to allow the maximal vacuum 
oscillation solution to the solar neutrino problem 
to be solved without significantly modifying BBN.  
The allowed parameter space is given in Figure 5. 
We also showed that there is a range of parameters where
the lepton number is generated so that the large angle muon
- sterile neutrino oscillation solution to the atmospheric neutrino
anomaly does not lead to any significant modification of BBN.
This parameter space is illustrated in Figure 6 and Figure 7.

We finish with a speculation.
One of the mysteries of cosmology is the origin of the
observed baryon asymmetry of the early Universe. In principle, 
it may be possible that the baryon asymmetry arises from 
a lepton number asymmetry. The lepton number asymmetry can
be converted into a baryon number asymmetry through sphaleron
transitions at or above the weak phase transition.
It may be possible that a small lepton number asymmetry
arises from the mechanism of ordinary-sterile neutrino
oscillations, which is seeded by statistical
fluctuations of the background.
One interesting feature of this possibility is that the baryon number
asymmetry would not be related to the CP asymmetry of the
Lagrangian. Instead the origin of matter over anti-matter would
be due to a statistical fluctuation which is then amplified
by neutrino oscillations.
However before this speculation can be checked, it would
be necessary to work out the effective potential at 
high temperatures ($T \sim 250 \ GeV$) and study the phase
transition region.
\vskip 0.5cm
\noindent
{\bf Acknowledgements}
\vskip 0.4cm
\noindent
We would like to thank A. Ignatiev and M. Thomson for useful discussions, 
and one of us (R.R.V) would like to thank K.Enqvist and K. Kainulainen
for a useful discussion.  This work was supported by the 
Australian Research Council.
\vskip 0.5cm
\noindent
{\bf Appendix}
\vskip 0.4cm
\noindent
The purpose of this appendix is to show that Eq.(\ref{wh1}) can be
derived from the Hamiltonian formalism. In applying this formalism,
we will assume that the rate at which collisions collapse the
wavefunction (i.e. the rate of measurement of whether the
state is a weak or sterile eigenstate) is given by the
damping frequency which is half of the collision frequency.
For further discussion of this point see section II and Ref.\cite{thomo}.

The expectation value that an initial weak-eigenstate neutrino 
$\nu_{\alpha}$ has oscillated into a sterile state $\nu_{s}$ after $\tau$
seconds will be denoted by $|\psi'_s(t,\tau)|^2$ (where $t$ is
the age of the Universe). The average probability that an initial
weak eigenstate has oscillated into a sterile state can be obtained
by averaging the quantity $|\psi'_s(t,\tau)|^2$ over all
possible times $\tau$ (weighted by the probability that the neutrino
has survived $\tau $ seconds since its last ``measurement''). 
This average has the form
\begin{equation}
\langle |\psi'_s(t)|^2\rangle =
{1 \over \omega_0} \int^{t}_{0} e^{-\tau/\omega_0}|\psi'_s(t,\tau)|^2 d\tau,
\end{equation}
where $\omega_0$ is the mean time between measurements.
According to Ref.\cite{thomo}, $\omega_0 = 1/D = 2/\Gamma_{\nu_{\alpha}}$.
If we denote the analogous quantity for anti-neutrinos by 
$\langle |\stackrel{\sim}{\psi}'_s(t,\tau)|^2\rangle$,
then the rate of change of lepton number can be expressed as
\begin{equation}
{dL_{\nu_{\alpha}} \over dt} \simeq 
- {3 \over 8}\langle \Omega (t) \rangle {\Gamma_{\nu_{\alpha}} \over 2}
 -{3 \over 8}{\partial \langle \Omega (t) \rangle \over
\partial t}, 
\label{ww1}
\end{equation}
where
\begin{equation}
\langle \Omega (t)\rangle = \langle |\psi'_s(t)|^2 \rangle -
\langle |\stackrel{\sim}{\psi}'_s(t)|^2\rangle.
\end{equation}
Note that the first term in Eq.(\ref{ww1}) represents the rate
of change of lepton number due to collisions (which produce sterile
neutrino states). The second term represents the rate of change
of lepton number due to the oscillations between collisions. 

In the adiabatic limit, the transformation $\theta_0 \to
\theta_m$ and $L_0 \to L_m$ diagonalizes the
Hamiltonian. In this limit, the mean probability
$\langle |\psi'_s(t)|^2\rangle$ is given by:
\begin{equation}
\langle |\psi'_s(t)|^2\rangle = \sin^2 2\theta_m
\langle \sin^2 {\tau \over 2L_m}\rangle.
\end{equation}
Note that in the static limit, $\partial \langle \Omega (t) 
\rangle/\partial t = 0$ and hence Eq.(\ref{dLdt1}) results.
However, in the expanding Universe which is non-static, the above
equation is not generally valid (although it turns out that it is a good
approximation for oscillations away from resonance where
the system changes sufficiently slowly and even at some 
resonance regions which are sufficiently smooth).
To calculate the probability $\langle |\psi'_s(t)|^2\rangle$
in the general case, we go back to the fundamental Hamiltonian
equations
\begin{equation}
i{d\over dt}\left(\begin{array}{c}
\psi_{\alpha}\\
\psi_{s}'
\end{array}\right) = {1 \over 2p}{\cal M}^2 \left(\begin{array}{c}
\psi_{\alpha} \\
\psi_{s}'
\end{array}
\right),
\label{a1}
\end{equation}
where
\begin{equation}
{\cal M}^2 = {1 \over 2}\left[
R_{\theta}\left(\begin{array}{cc}
-\delta m^2&0\\
0&\delta m^2
\end{array}\right)
R_{\theta}^{T} + 4p\left(\begin{array}{cc}
\langle V \rangle&0\\
0&0
\end{array}\right)
\right], 
\end{equation}
and
\begin{equation}
R_{\theta} = \left(
\begin{array}{cc}
\cos\theta_0&\sin\theta_0\\
-\sin\theta_0&\cos\theta_0
\end{array}
\right),\  \langle H \rangle = {(b \pm a)\delta m^2
\over 2p},
\end{equation}
where the $-(+)$ sign corresponds to neutrino (anti-neutrino) 
oscillations.  Expanding out Eq.(\ref{a1}), we find:
\begin{equation}
i{d\psi_{\alpha}\over dt} = \alpha \psi_{\alpha} + 
{\beta \over 2}\psi_{s}',\ 
i{d\psi_{s}'\over dt} = {\beta \over 2} \psi_{\alpha} 
+ \gamma \psi_{s}',
\label{a2}
\end{equation}
where 
\begin{equation}
\alpha = {\delta m^2 \over 4p}(2b \pm 2a - \cos2\theta_0),
\ \beta = {\delta m^2 \over 2p} \sin2\theta_0, \
\gamma = {\delta m^2 \over 4p }\cos2\theta_0.
\label{xx99}\end{equation}
If we divide the equations Eqs.(\ref{a2}) by $\psi_{\alpha}$ and
$\psi'_{s}$ respectively, then they can be combined into
the single differential equation:
\begin{equation}
i{d\Psi \over dt} = \lambda\Psi 
+ {\beta \over 2}(1 - \Psi^2),
\label{gg}
\end{equation}
where $\Psi \equiv \psi_{s}'/\psi_{\alpha}$ and 
\begin{equation}
\lambda = \gamma - \alpha = {\delta m^2 \over 2p}(\cos2\theta_0 - 
b \pm a).
\label{de}
\end{equation}
The $+(-)$ sign in the above equation corresponds to $\nu_{\alpha} -
\nu_s$ ($\bar \nu_{\alpha} - \bar \nu_s$) oscillations.
Note that $|\psi_{s}'|^2 = {|\Psi|^2 \over 1 + |\Psi|^2}$.
If the non-linear term ($\Psi^2$) can be neglected, then
the solution for constant $\alpha, \beta, \gamma$ is
\begin{equation}
\Psi (t) = {-\beta \over 2\lambda}
\left[ 1 - e^{-i(\lambda)(t-t^*)}\right],
\end{equation}
with boundary condition $\Psi(t^*) = 0$.  Introducing the 
variable $\tau \equiv t-t^*$, and evaluating $|\Psi(t,\tau)|^2$ we find
\begin{equation}
|\Psi(t,\tau)|^2 = {\beta^2 \over  \lambda^2}
\sin^2 \left[{\lambda \tau \over 2} 
\right],
\end{equation}
which is approximately $\sin^2 2\theta_m \sin^2 \tau/2L_m$
provided that $|\Psi|^2 \ll 1$.

In the general case where $\alpha, \beta$ and $\gamma$ are not constant,
the general solution is (where we have again neglected the
non-linear $\Psi^2$ term):
\begin{equation}
\Psi (t) = {-i\over 2}\int^t_{t^*} e^{i\stackrel{\sim}{\lambda}(t')}
\beta(t') dt',
\label{fff}
\end{equation}
where 
\begin{equation}
\stackrel{\sim}{\lambda}(t') \equiv \int^{t'}_{t}  \lambda dt'',
\end{equation}
and the boundary condition $\Psi (t^*) = 0$ has again 
been taken. 
One may easily verify that Eq.(\ref{fff}) is indeed the solution
by directly substituting it into Eq.(\ref{gg}).  
The probability that a weak-eigenstate at $t=t^*$ has oscillated
into a sterile eigenstate at time t is thus
\begin{equation}
|\Psi (t)|^2 \simeq {\beta^2 \over 4}\int^t_{t^*}\int^t_{t^*}
\cos\left[ \int^{t_1}_{t_2}\lambda dt'\right] 
dt_1 dt_2,
\end{equation}
where we have assumed that $\beta$ is approximately constant
over the interaction time scale $t - t^*$, so that it can
be taken  outside the integral. This step is a good
approximation provided $T \stackrel{>}{\sim} 2\ {\rm MeV}$\cite{last}.
Again defining the quantity $\tau \equiv t - t^*$ (recall that $\tau$ 
is the time between measurements), and averaging $|\Psi(t,\tau)|^2$ 
over $\tau$, with the appropriate weighting factor, we find that
\begin{equation}
\langle |\Psi(t)|^2 \rangle
\simeq {\beta^2 \over 4\omega_0}
\int^{t}_{0}e^{-\tau/\omega_0}\int^t_{t-\tau}
\int^t_{t-\tau} \cos\left[\int^{t_1}_{t_2}\lambda dt'
\right] dt_1 dt_2 d\tau. 
\end{equation}
Integrating this equation by parts (with respect to the $\tau$ 
integration), we find:
\begin{equation}
\langle |\Psi(t)|^2 \rangle
\simeq {\beta^2 \over 2}\int^{t}_{0}\int^t_{t-\tau}
e^{-\tau/\omega_0} \cos\left[\int^{t_1}_{t-\tau}\lambda dt'
\right] dt_1 d\tau, 
\label{unc}
\end{equation}
where we have used the fact that $e^{-t/\omega_0} \simeq 0$\cite{kk}.
The analogous quantity for anti-neutrinos,
$\langle |\stackrel{\sim}{\Psi}(t)|^2 \rangle$, can 
similarly be defined. 
Recall that the functions $\langle |\Psi(t)|^2 \rangle$,
$\langle |\stackrel{\sim}{\Psi}(t)|^2 \rangle$ are
related to the rate of change of lepton number through
Eq.(\ref{ww1}):
\begin{equation}
{dL_{\nu_{\alpha}} \over dt} \simeq
-{3 \over 8}{\langle |\Psi(t)|^2 \rangle \over \omega_0}
-{3 \over 8}{\partial \langle |\Psi(t)|^2 \rangle \over \partial t}
- \left[\Psi \to \stackrel{\sim}{\Psi} \right].
\end{equation}
Evaluating $\partial \langle |\Psi(t)|^2\rangle/\partial t$ we find:
\begin{eqnarray}
&{\partial \langle |\Psi(t)|^2 \rangle \over \partial t} =
{\beta^2\over 2}\int^t_0 e^{-\tau/\omega_0}
\left(\cos\left[\int^{t}_{t-\tau}\lambda dt'\right] - 1
\right)d\tau \nonumber \\
& + {\beta^2 \over 2}\int^t_0 e^{-\tau/\omega_0}
\lambda_{(t-\tau)} \int^t_{t-\tau}
\sin\left[\int^{t_1}_{t-\tau}\lambda dt'\right]
dt_1 d\tau, 
\end{eqnarray}
where we use the notation that $\lambda_{(t-\tau)}$ denotes $\lambda$ 
evaluated at the point $(t - \tau)$.  Dividing Eq.(\ref{unc}) by 
$\omega_0$ and integrating Eq.(\ref{unc}) by parts we find (with
respect to the $\tau$ integration), we find
\begin{equation}
{1 \over \omega_0}\langle |\Psi(t)|^2\rangle =
{\beta^2 \over 2}\int^t_0 e^{-\tau/\omega_0} 
\left(1 - \lambda_{(t-\tau)}\int^t_{t-\tau}\sin
\left[\int^{t_1}_{t-\tau}\lambda dt'\right] 
dt_1 \right)d\tau.
\end{equation}
Adding the above two equations and subtracting the analogous term
for anti-neutrinos, we obtain the following rather
compact expression for the rate of change of lepton number:
\begin{equation}
{dL_{\nu_{\alpha}} \over dt} =
{-3\beta^2 \over 16}\int^t_0 e^{-\tau/\omega_0} 
\left(\cos\left[\int^t_{t-\tau} \lambda 
dt'\right]
- \cos\left[\int^t_{t-\tau} \bar \lambda
dt'\right] \right) d\tau,
\label{wh}
\end{equation}
where $\bar \lambda$ is defined similarly to $\lambda$ except that 
$a \to -a$.  Note that the total contribution to the rate of change 
of lepton number is in fact simpler than either of the two separate 
contributions coming from collisions and oscillations between collisions.
Equation (\ref{wh}) can be re-written (using a trigonometric
identity)
\begin{equation}
{dL_{\nu_{\alpha}} \over dt} =
{3\beta^2 \over 8}\int^t_0 e^{-\tau/\omega_0} 
\sin\left[\int^t_{t-\tau} \lambda^+ 
dt'\right]
\sin\left[\int^t_{t-\tau} \lambda^-
dt''\right] d\tau,
\label{wh2}
\end{equation}
where $\lambda^{\pm} = (\lambda \pm \bar \lambda)/2$.
Note that this is exactly the same equation that we derived 
in section III [Eq.(\ref{wh1})] from the density matrix
equations.

\newpage

\vskip 0.5cm
\noindent
{\large \bf Table Caption}
\vskip 0.6cm
\noindent
Summary of the predictions for the chlorine and gallium 
experiments assuming 1) standard electro-weak theory (i.e. no new 
physics) 2) that the electron neutrino oscillates maximally into 
a sterile state (maximal mixing model) and 3) the experimental 
measurements. All numbers are in units of SNU.

\vskip 1cm
\noindent
{\large \bf Figure Captions}
\vskip 0.5cm
\noindent
Figure 1.  The evolution of the $\nu_{\mu} - \nu_s$ (or
$\nu_{\tau} -\nu_s$) oscillation generated lepton number
asymmetry, $L_{\nu_{\mu}}$ (or $L_{\nu_{\tau}}$).
We have taken by way of example, the parameter choices $\delta m^2 = -1\ eV^2,
\sin^2 2\theta_0 = 10^{-8}$ ($\sin^2 2\theta_0 = 10^{-4}$) for 
the bottom two curves (top two curves).
The solid lines represent the results of the numerical
integration of the density matrix equations [Eq.(\ref{jk})],
while the dashed lines result from the numerical integration
of Eq.(\ref{eeqq}).
\vskip 0.5cm
\noindent
Figure 2.  
The evolution of the $\nu_{\mu} - \nu_s$ (or
$\nu_{\tau} - \nu_s$) oscillation generated lepton number
asymmetry, $L_{\nu_{\mu}}$ (or $L_{\nu_{\tau}}$).
In this example we have taken the parameter choices,
$\delta m^2 = -1000 \ eV^2$,
$\sin^2 2\theta_0 = 10^{-9}$ ($\sin^2 2\theta_0 = 10^{-6}$) for 
the bottom two curves (top two curves).  As in Figure 1, the
solid lines represent the results of the numerical
integration of the density matrix equations [Eq.(\ref{jk})],
while the dashed lines result from the numerical integration
of Eq.(\ref{eeqq}).
\vskip 0.5cm
\noindent
Figure 3. The evolution of the $\nu_{\mu} - \nu_s$ (or
$\nu_{\tau} - \nu_s$) oscillation generated lepton number
asymmetry, $L_{\nu_{\mu}}$ (or $L_{\nu_{\tau}}$).
In this example we have taken the parameter choices 
$\delta m^2 = -1\ eV^2,
\sin^2 2\theta_0 = 10^{-4}$ (dashed line), $\sin^2 2\theta_0 = 10^{-6}$
(dashed-dotted line) and $\sin^2 2\theta_0 = 10^{-8}$ (solid line).
These curves result from integrating the coupled differential
equations Eq.(\ref{eeqq3}, \ref{rate2}), which in contrast to Figures 1,2,
incorporate the momentum distribution of the neutrino.
They also incorporate the effect of the non-zero number density of 
the sterile neutrinos which are produced by the oscillations.
\vskip 0.5cm
\noindent
Figure 4. Same as Figure 3 except that $\delta m^2 = -1000\ eV^2,
\sin^2 2\theta_0 = 10^{-6}$ (dashed line),
$\sin^2 2\theta_0 = 10^{-7}$ (dashed-dotted line)
and $\sin^2 2\theta_0 = 10^{-9}$ (solid line).
\vskip 0.5cm
\noindent
Figure 5.  Region of parameter space in the $-\delta m^2_{\tau e'}$, 
$|\delta m^2_{ee'}|$, plane (assuming $\sin^2 2\theta_0^{ee'} \simeq 1$)
where the $L^{(e)}$ created by $\nu_{\tau} - \nu'_e$ oscillations
does not get destroyed by $\nu_e - \nu_e'$ oscillations.
The solid line corresponds to  $\sin^2 2\theta_0^{\tau e'}
= 10^{-6}$, while the dashed line
corresponds to $\sin^2 2\theta_0^{\tau e'} = 10^{-8}$.
Note that similar results hold for $\nu_{\mu} - \nu'_e$ oscillations
by replacing $\nu_{\tau} \to \nu_{\mu}$.
\vskip 0.5cm
\noindent
Figure 6. Region of parameter space
in the $-\delta m^2_{\tau \mu'}$, $|\delta m^2_{\mu \mu'}|$ plane
where the $L^{(\mu)}$ created by $\nu_{\tau} - \nu_{\mu}'$ oscillations
does not get destroyed by $\nu_{\mu} - \nu'_{\mu}$ oscillations
(assuming $\sin^2 2\theta_0^{\mu \mu'} \simeq 1$). 
The solid line corresponds to  $\sin^2 2\theta_0^{\tau e'}
= 10^{-6}$, while the dashed line
corresponds to $\sin^2 2\theta_0^{\tau e'} = 10^{-8}$.
Note that similar results hold for $\nu_{\tau} - \nu'_e$
oscillations if both $\nu_{\mu}'$ and $\nu'_e$ exist.
\vskip 0.5cm
\noindent
Figure 7. Region of parameter space 
($\sin^2 2\theta_0^{\tau \mu'}$, $ - \delta m^2_{\tau \mu'}$)
where the $L^{(\mu)}$ created
by $\nu_{\tau} - \nu'_{\mu}$ oscillations does not get
destroyed by $\nu_{\mu} - \nu'_{\mu}$ oscillations.
This region which in the figure is denoted by the ``Allowed region'' 
is all of the parameter space above the solid line.
We have assumed that $\sin^2 2\theta_0^{\mu \mu'} \simeq 1$ and
$|\delta m^2_{\mu \mu'}| = 10^{-2}\ eV^2$ (which is the
best fit to the atmospheric neutrino data).
Also shown (the dashed line) is the cosmology bound 
$m_{\nu_{\tau}} \stackrel{<}{\sim} 40\ eV$ (which
implies $|\delta m^2_{\tau \mu'}| \stackrel{<}{\sim}
1600\ eV^2$), which is required if the neutrino is sufficiently
long lived.  The dashed-dotted line is the BBN bound, Eq.(\ref{gh}).

\newpage
\vskip 2cm
\noindent
{\large \bf Table 1}
\vskip 1.4cm
\tabskip=0pt \halign to \hsize{
\vrule#\tabskip=0pt plus 1fil\strut&
\hfil#\hfil& \vrule#& \hfil#\hfil& \vrule#& \hfil#\hfil&
\tabskip=0pt\vrule#\cr
\noalign{\hrule}
&Prediction/Expt&&Chlorine&&Gallium &\cr
\noalign{\hrule}
&Standard Electro-weak theory&&$4.5\pm0.5$&&$123^{+8}_{-6}$&\cr
&Maximal mixing model&&$3.7
\pm 0.4 $&&$65^{+7}_{-4}$&\cr
&Experiment&&$2.78 \pm 0.35 $&&$71 \pm 7$&\cr
\noalign{\hrule}
}


\begin{thebibliography}{99}
\bibitem{shif}
We have implicitly assumed that the collision rate for
neutrinos, $\Gamma_{\nu_{\alpha}}$, is equal to the
collision rate for anti-neutrinos, $\bar \Gamma_{\nu_{\alpha}}$.
Of course, in a $CP$ asymmetric background the collision rate 
for neutrinos will be slightly different to the collision rate 
for anti-neutrinos. However as shown in Ref.{\cite{shi}},
$\Gamma_{\nu_{\alpha}} - \bar \Gamma_{\nu_{\alpha}}
\ll L_{\nu_{\alpha}}\Gamma_{\nu_{\alpha}}$ and hence
to a good approximation $\Gamma_{\nu_{\alpha}} =
\bar \Gamma_{\nu_{\alpha}}$.
\bibitem{last}
The quantity $\beta$ can be taken outside the integral (to
a good approximation) provided that 
$|\omega_0 {\partial \beta \over \partial t}|
\stackrel{<}{\sim} |\beta|$. This equation implies that
$$ |\omega_0 {\partial \beta \over \partial t}| = 
{2 \over \Gamma_{\nu_{\alpha}}} |{\partial T \over \partial t}
{\partial \beta \over \partial T}| \stackrel{<}{\sim} |\beta|$$
Using $\partial \beta/\partial T \simeq -\beta/T$, and
$\partial T/\partial t = -HT$ (where $H$ is the Hubble parameter),
it follows that the above condition holds provided that
$\Gamma_{\nu_{\alpha}} \stackrel{>}{\sim} 2H$, i.e. 
for $T \stackrel{>}{\sim} 2 \ {\rm MeV}$.
\bibitem{fn345}
Note that the LSND experiment\cite{lsnd} suggests that
$|\delta m^2_{\mu e}| \stackrel{>}{\sim} 0.3\ eV^2$.
This also implies that $|\delta m^2_{\mu e'}| \stackrel{>}{\sim}
0.3\ eV^2$ because $\delta m^2_{\mu e'} = \delta m^2_{\mu e} + 
\delta m^2_{ee'}$ and $|\delta m^2_{ee'}| \stackrel{<}{\sim}
10^{-3} \ eV^2$.
\bibitem{ie}
In the case where 
$\nu_e - \nu_{\tau}$ oscillations are very significant (this
should happen when the mixing angle $\sin^2 2\theta_0^{\tau e} $
is relatively large),
it may be possible that the rate of change of $(L_{\nu_{\tau}} -
L_{\nu_e})$ can be large enough to make
$L_{\nu_{\tau}} \simeq L_{\nu_e}$.
In this case one then has approximately only one differential
equation rather than two coupled differential equations.
Even in this case, it is possible to show that 
the $L^{(e)}$ created by $\nu_{\tau} - \nu'_e$ oscillations cannot get
destroyed by $\nu_e - \nu_e'$ oscillations provided 
that $\sin^2 2\theta_0^{\tau e'} \stackrel{>}{\sim}
(\delta m^2_{ee'}/\delta m^2_{\tau e'})^2$.
\bibitem{kk}
From Eq.(\ref{unc}), using $\cos[\int blah] \le 1$, it
is straightforward to show that $|\Psi |^2 < 1$ provided that 
$\sin^2 2\theta_0 \stackrel{<}{\sim}  10^{-3}$.
\end{thebibliography}
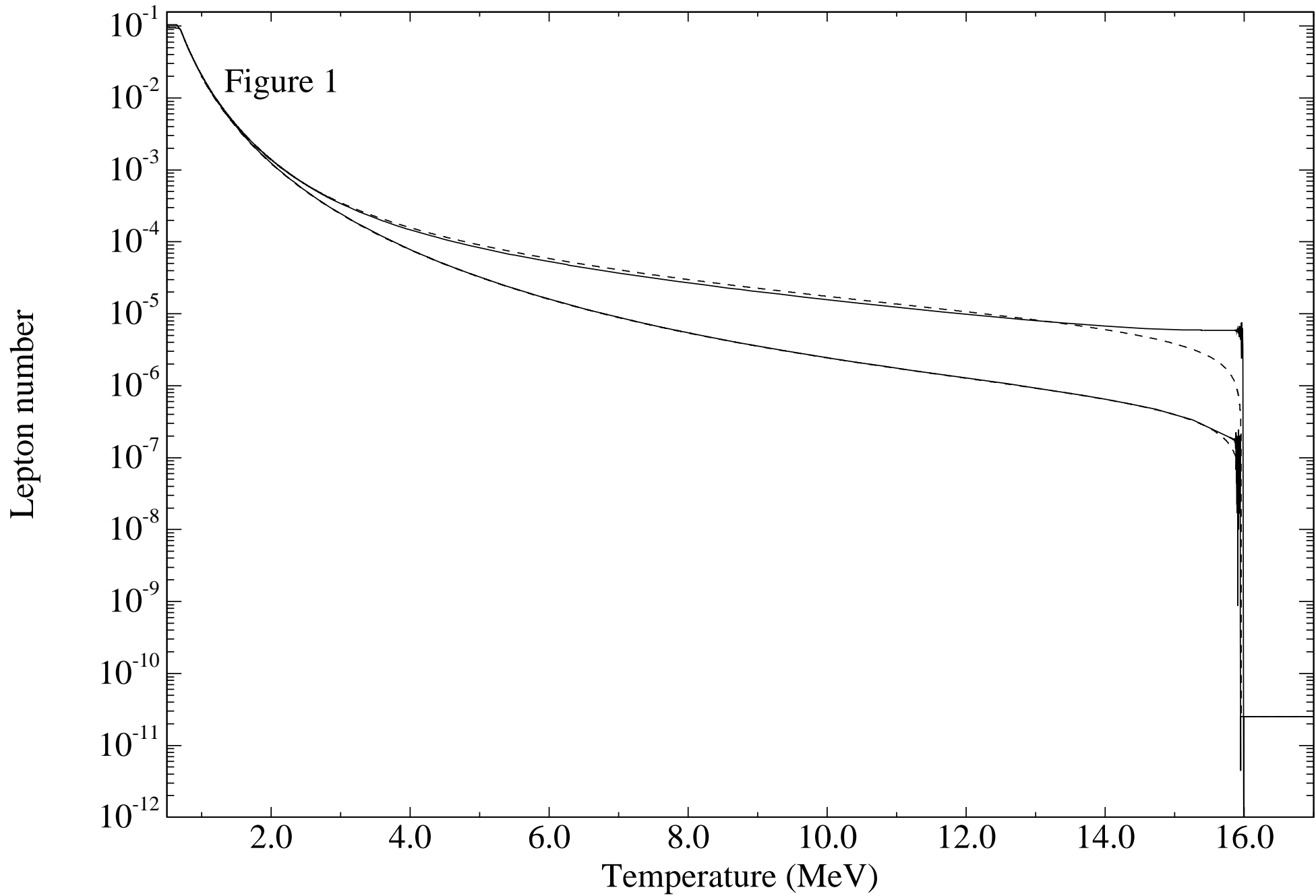
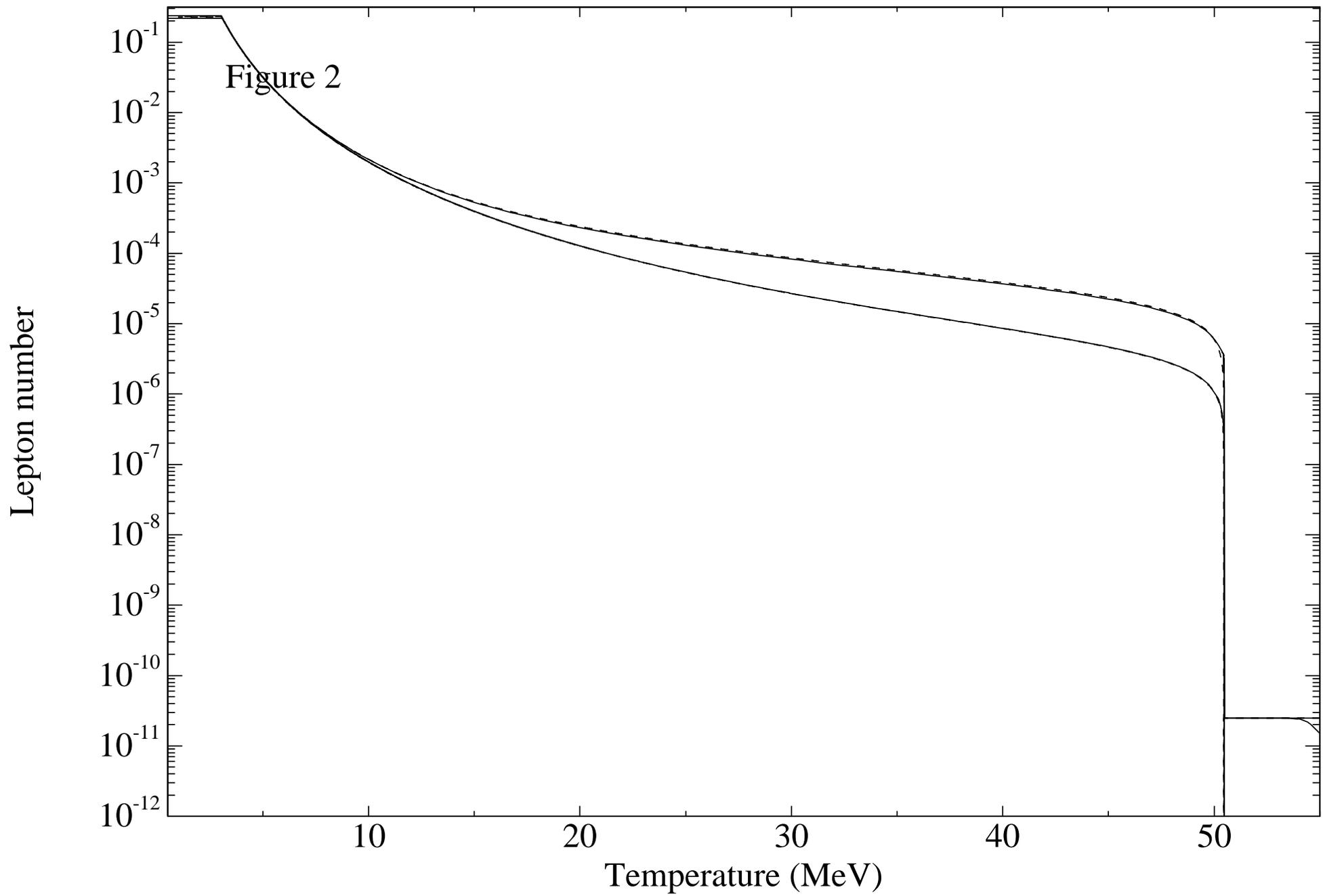
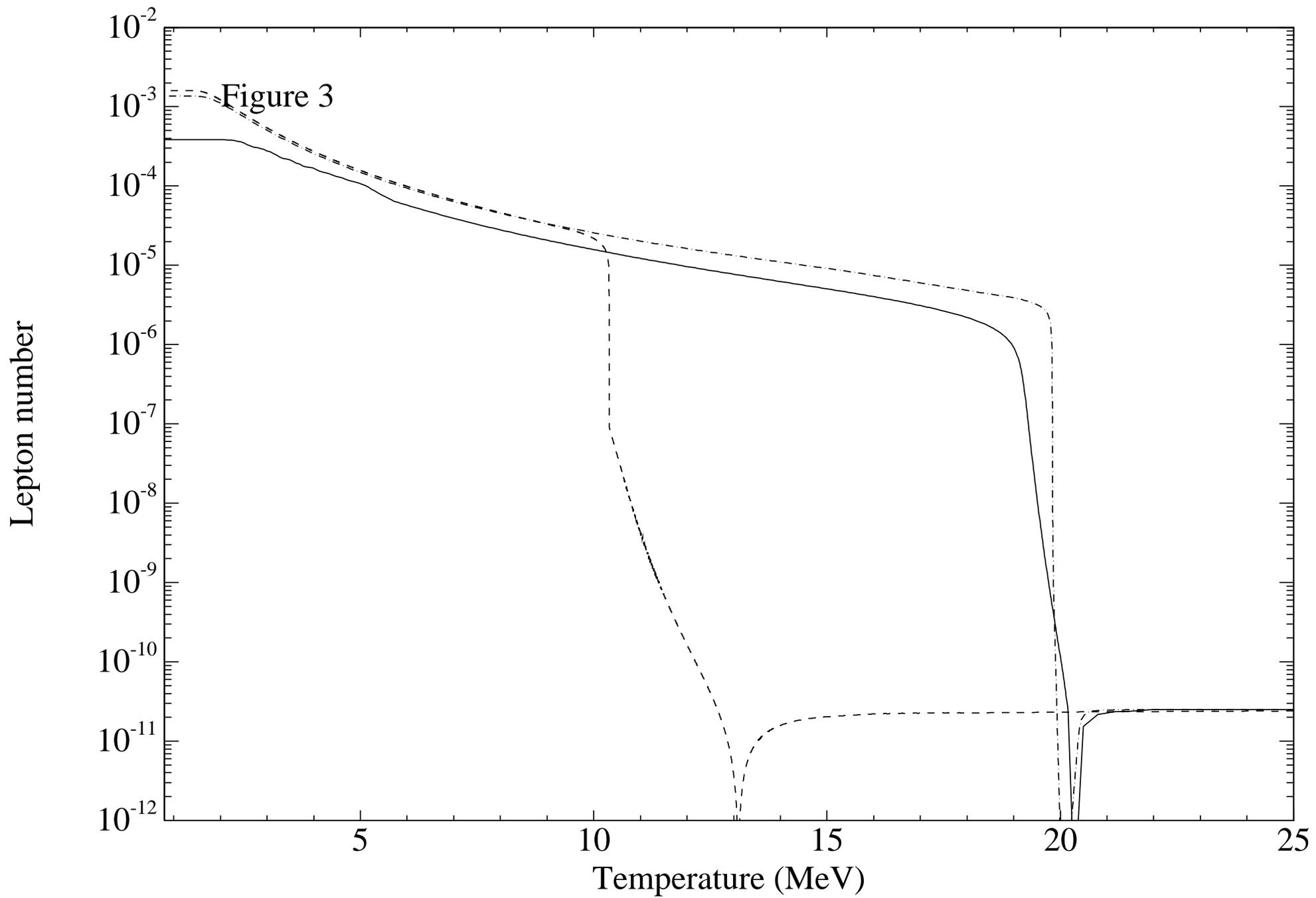
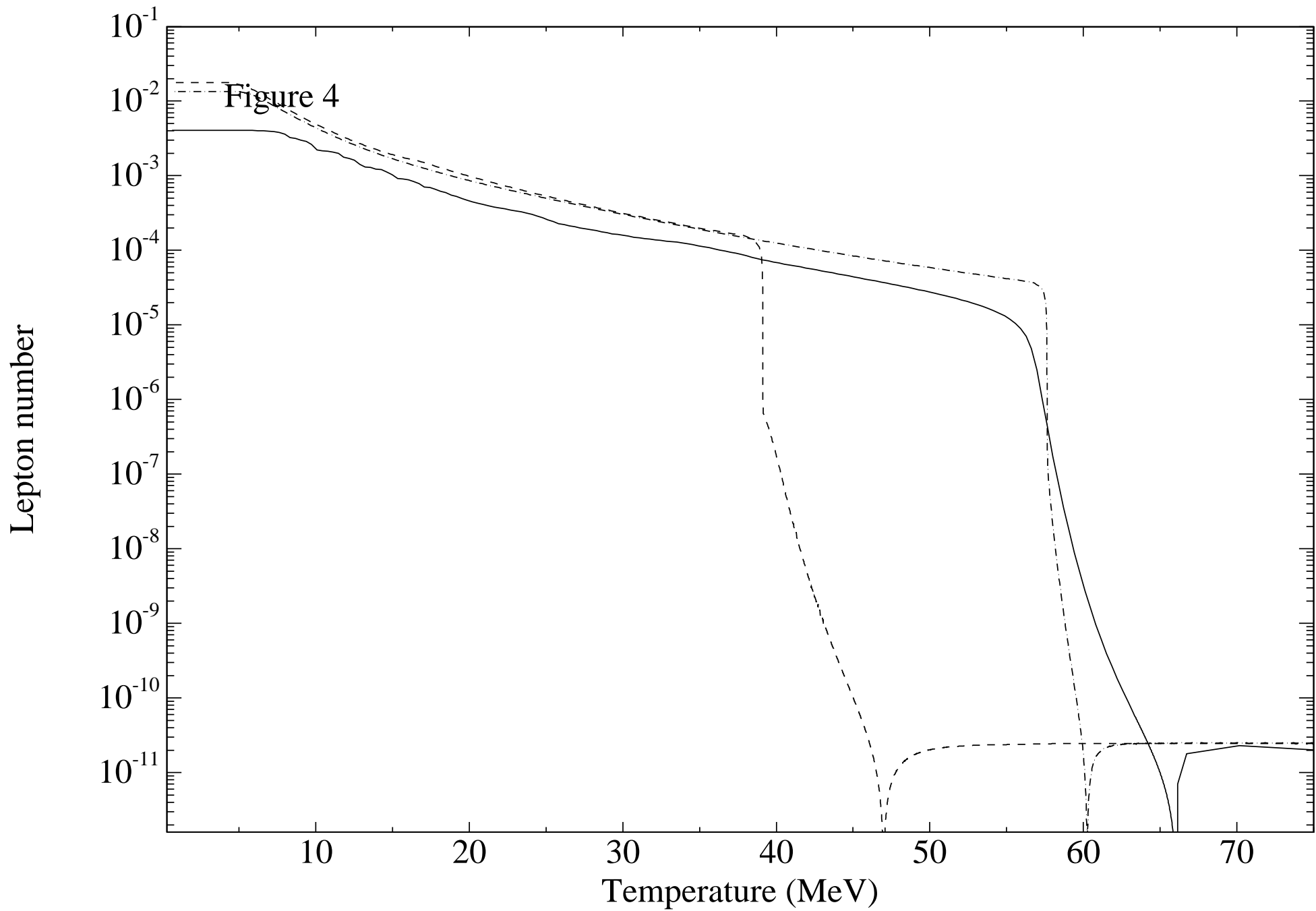
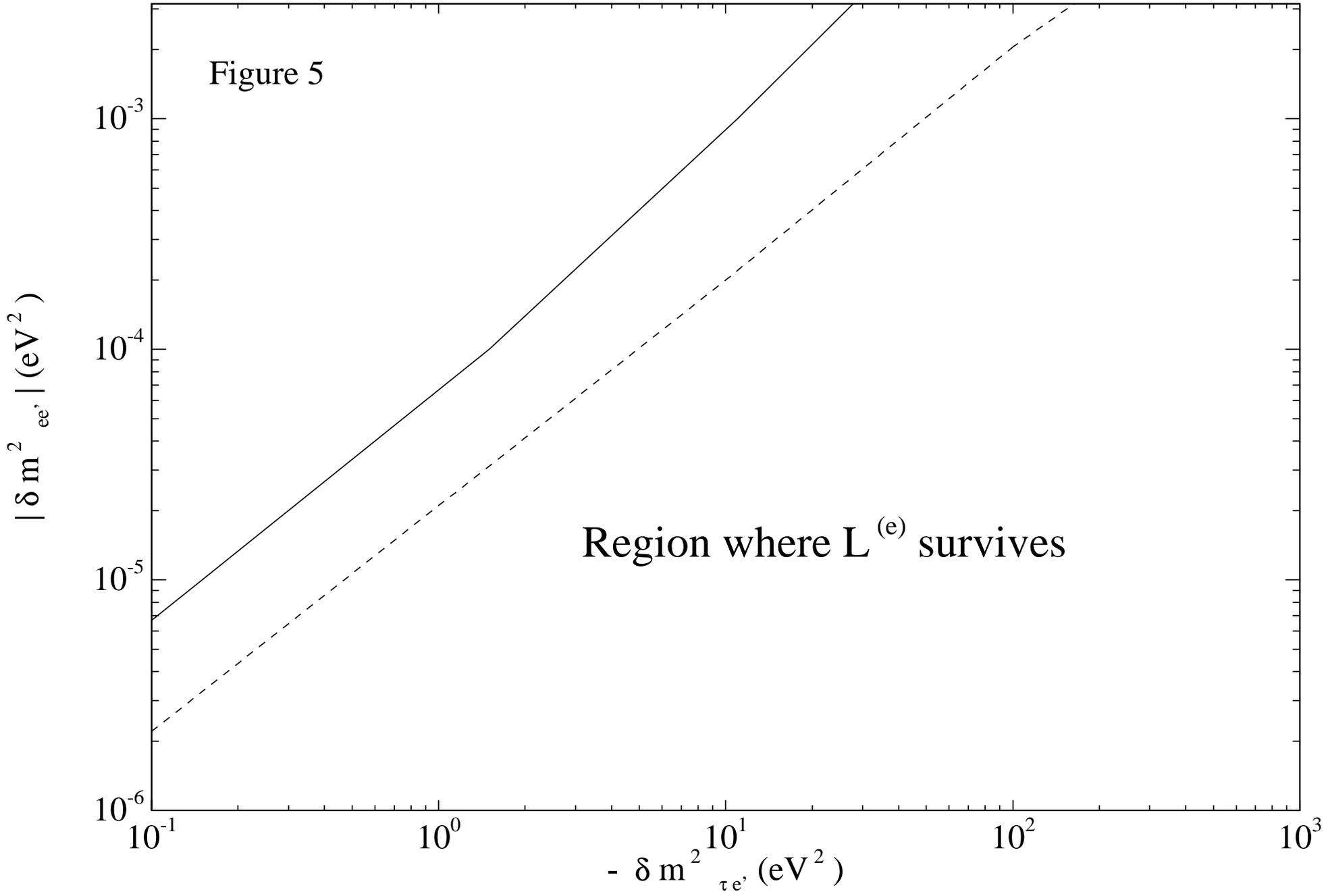
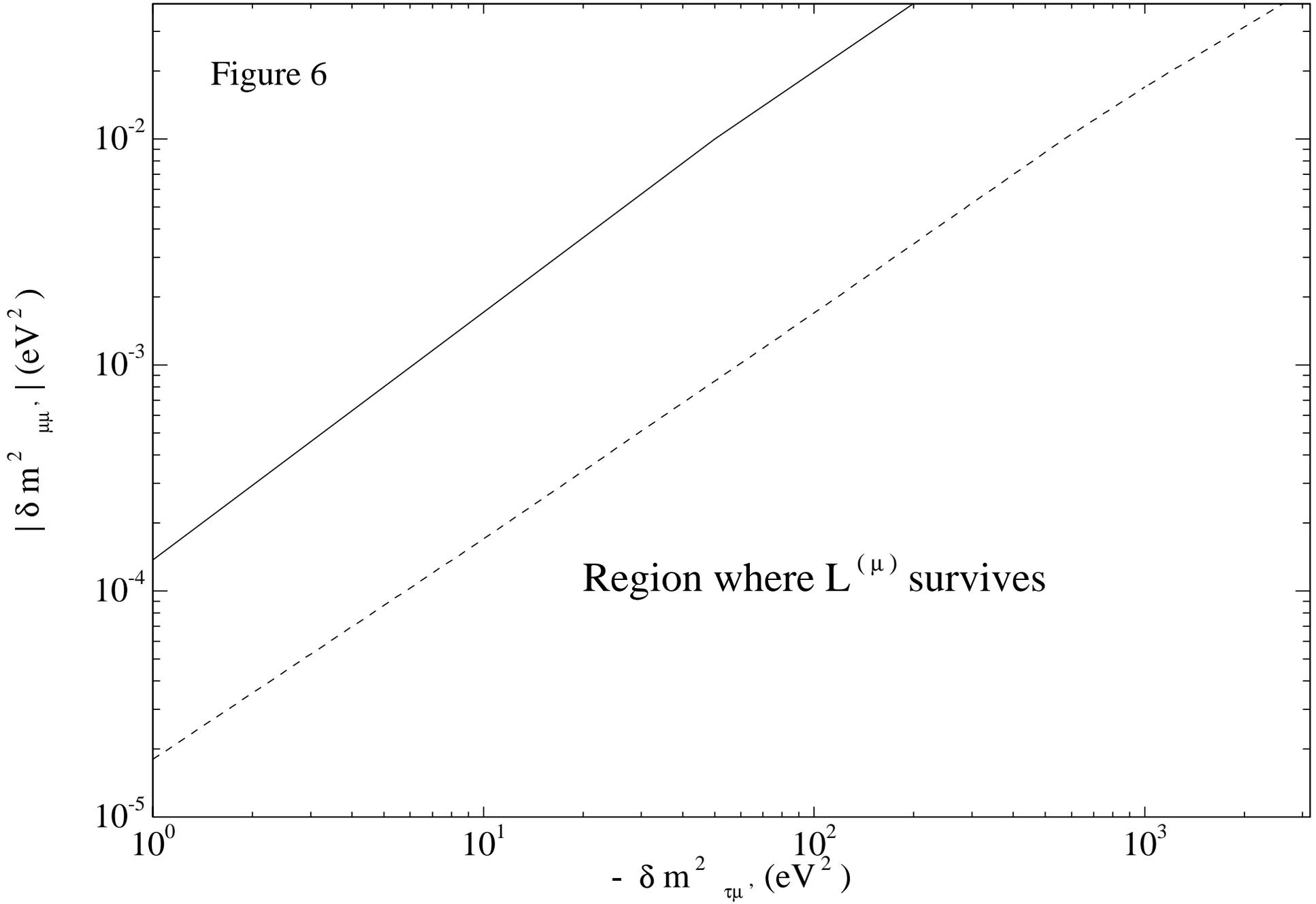
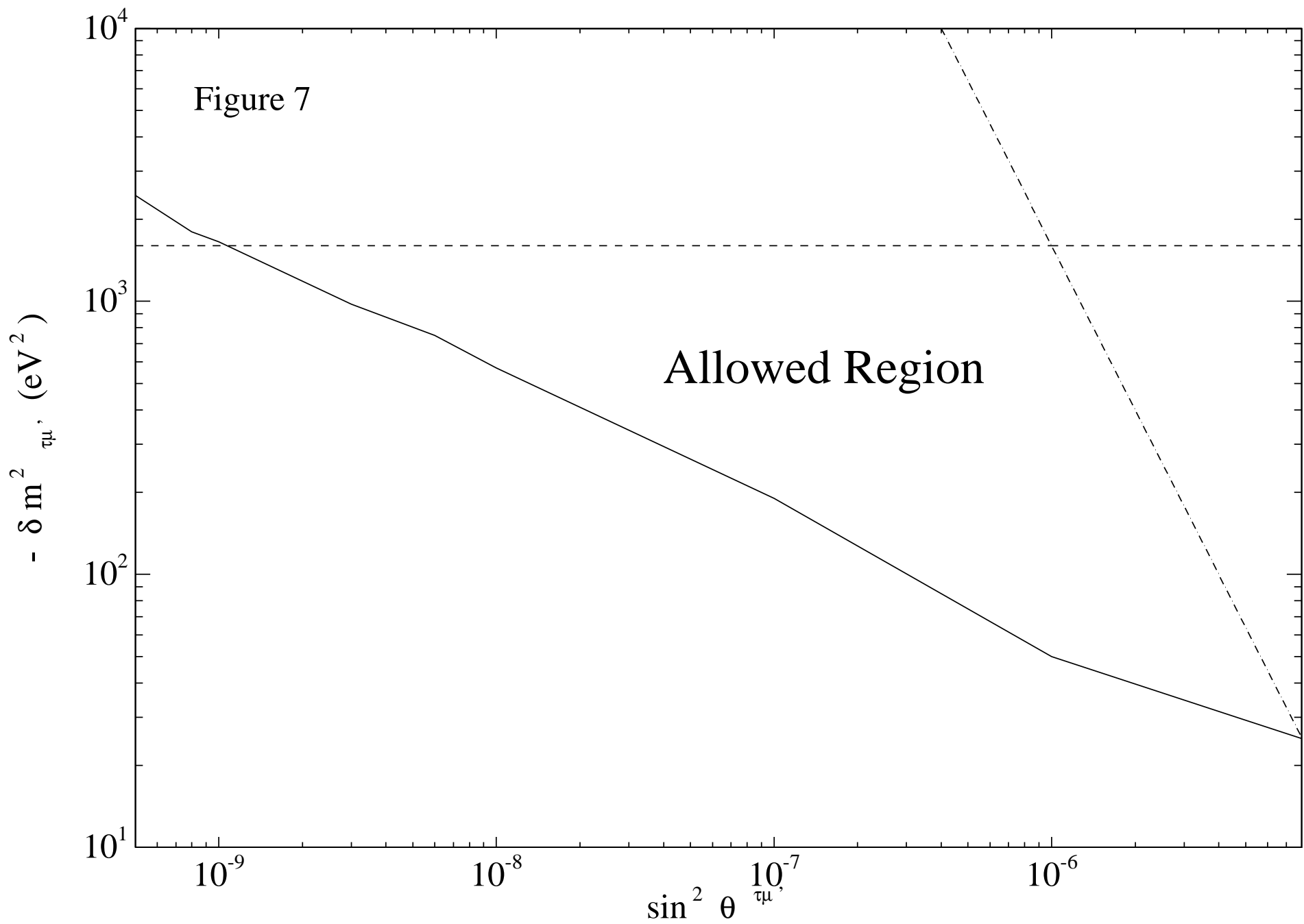
\end{document}